\documentclass[preprint,superscriptaddress,nofootinbib]{revtex4}
   
   \usepackage{amssymb,amsmath,bbm,graphicx,subfigure,color}
   \usepackage{rotating,multirow}
   \usepackage[colorlinks]{hyperref}
   \allowdisplaybreaks[4]
   \begin{document}

  \title{Reinvestigating $B$ ${\to}$ $PV$ decays by including
         contributions from ${\phi}_{B2}$ with the perturbative
         QCD approach}
  \author{Yueling Yang}
  \affiliation{Institute of Particle and Nuclear Physics,
              Henan Normal University, Xinxiang 453007, China}
  \author{Xule Zhao}
  \affiliation{Institute of Particle and Nuclear Physics,
              Henan Normal University, Xinxiang 453007, China}
  \author{Lan Lang}
  \affiliation{Institute of Particle and Nuclear Physics,
              Henan Normal University, Xinxiang 453007, China}
  \author{Jinshu Huang}
  \affiliation{School of Physics and Electronic Engineering,
              Nanyang Normal University, Nanyang 473061, China}
  \author{Junfeng Sun}
  \affiliation{Institute of Particle and Nuclear Physics,
              Henan Normal University, Xinxiang 453007, China}

  \begin{abstract}
  Considering the $B$ mesonic wave function ${\phi}_{B2}$,
  the $B$ ${\to}$ $PV$ decays are restudied at the leading order
  for three scenarios using the perturbative QCD  approach
  within the standard model, where $P$ $=$ ${\pi}$ and $K$, and
  $V$ denotes the ground $SU(3)$ vector mesons.
  It is found that contributions from ${\phi}_{B2}$ can enhance
  most branching ratios, and are helpful for improving the
  overall consistency of branching ratios between the updated
  calculations and available data, although there are still
  several discrepancies between the experimental and theoretical
  results.

  \href{https://doi.org/10.1088/1674-1137/ac6573}{Chin. Phys. C 46, 083103 (2022)}
  \end{abstract}

   \maketitle

   \section{Introduction}
   \label{sec:introduction}
   $B$ meson physics is an important and popular area
   of particle physics because of continuous impetus
   from experimental and theoretical efforts and pursuits.
   With the running of the Belle-II and LHCb experiments, an increasing
   number of $B$ meson events will be accumulated, with an expected goal
   of the integrated luminosity of $50$ $ab^{-1}$ by the Belle-II
   detector at the $e^{+}e^{-}$ SuperKEKB collider \cite{1808.10567}
   and approximately $300$ $fb^{-1}$ by the LHCb detector at the future
   High Luminosity LHC (HL-LHC) hadron collider \cite{1808.08865}.
   More than $10^{11}$ $B_{u,d}$ mesons are expected to become available
   at the future CEPC \cite{1811.10545} and FCC-ee \cite{epjc.79.474}
   experiments based on about $10^{12}$ $Z^{0}$ boson decays with
   a branching ratio ${\cal B}(Z^{0}{\to}b\bar{b})$ ${\approx}$
   $15.12\%$ \cite{pdg2020} and a fragmentation fraction
   $f(b{\to}B_{u})$ ${\approx}$ $f(b{\to}B_{d})$ ${\approx}$ $41.8\%$
   \cite{plb.576.29}.
   With the gradual improvement of data processing technology,
   besides numerous new and unforeseen phenomena, higher precision
   measurements of $B$ meson weak decays will be achieved.
   The experimental study of $B$ meson decays is stepping into a
   golden age of big data and high precision.
   Higher requirements have been placed on the
   accuracy of theoretical calculations for $B$ meson decays,
   which is the fundamental premise behind rigorous testing of
   the standard model (SM) of elementary particles, finding a
   solution to the discrepancies between data and theoretical
   expectations, and searching for new physics beyond the SM.

   Owing to an inadequate understanding of the dynamic
   mechanisms of hadronization and quantum chromodynamics (QCD)
   behavior at low energy scales, the main uncertainties on
   various theoretical estimations for $B$ meson decays arise from
   the hadronic matrix elements (HMEs) describing the transition
   from the quarks to hadrons involved.
   The calculation of the nonleptonic $B$ meson decays is especially
   complicated, because both the initial and final states are hadrons.
   Additionally, nonleptonic $B$ decay modes are rich, and the
   study of these is very interesting and significant.
   The measurement of nonleptonic $B$ meson decays has been
   providing abundant information and various constraints on the SM,
   for example, the angles and sides of the commonly termed
   unitarity triangle, $V_{ud}V_{ub}^{\ast}$ $+$ $V_{cd}V_{cb}^{\ast}$
   $+$ $V_{td}V_{tb}^{\ast}$ $=$ $0$, arising from the
   Cabibbo-Kobayashi-Maskawa (CKM) matrix
   \cite{PhysRevLett.10.531,PTP.49.652}.
   How to deal with HMEs as reasonably and reliably as possible
   is now a central and urgent issue in the theoretical
   calculation of nonleptonic $B$ meson decays.

   Based on the widely used Lepage-Brodsky procedure for
   exclusive processes with a large momentum transfer between
   hadrons \cite{PhysRevD.22.2157}
   and the power counting rules in the heavy quark limits,
   several attractive QCD-inspired methods, such as
   the perturbative QCD (PQCD) approach \cite{PhysRevLett.74.4388,
   plb.348.597,PhysRevD.52.3958,PhysRevD.63.074006,
   PhysRevD.63.054008,PhysRevD.63.074009,plb.555.197},
   QCD factorization (QCDF) approach \cite{PhysRevLett.83.1914,
   npb.591.313,npb.606.245,plb.488.46,plb.509.263,PhysRevD.64.014036,
   epjc.36.365,PhysRevD.69.054009,npb.774.64,PhysRevD.77.074013},
   soft and collinear effective theory \cite{PhysRevD.63.014006,
   PhysRevD.63.114020,plb.516.134,PhysRevD.65.054022,
   PhysRevD.66.014017,npb.643.431,plb.553.267,npb.685.249}
   and so on,
   have been fully developed to evaluate HMEs for nonleptonic
   $B$ meson decays, where HMEs are phenomenologically
   expressed as the convolution integral of the scattering
   amplitudes at the quark level and wave functions (or
   distribution amplitudes) at the hadronic level.
   The calculation accuracy of HMEs may be improved via the
   following two aspects together. The first is the scattering amplitudes,
   and the second is the hadronic wave functions (WFs) or distribution
   amplitudes (DAs).
   Owing to the asymptotic freedom of QCD,
   the scattering amplitudes describing hard interactions among
   quarks are calculable, in principle, order by order with the
   perturbative quantum field theory.
   The higher order radiative corrections to HMEs are necessary
   and important for approaching the true values, reducing the
   dependence of theoretical results on the renormalization scale,
   obtaining strong phases closely related to the $CP$
   violation, verifying models, and perfecting methods.
   In recent years, the next-to-next-to-leading order (NNLO)
   corrections to HMEs have become available and have shown a large
   model sensitivity to the hadronic distribution amplitudes,
   for example, in Refs. \cite{npb.794.154,npb.832.109,
   plb.750.348,jhep.2016.09.112,jhep.2020.04.055}.
   The influences of WFs on HMEs are also significant; however,
   they have attracted relatively insufficient attention compared with
   the scattering amplitudes. There are numerous studies on
   nonleptonic $B$ decays using the PQCD approach, which show that
   the theoretical uncertainties mainly originate from the
   parameters of WFs or DAs, for example, in Refs.
   \cite{PhysRevD.71.034018,PhysRevD.103.056006,
   PhysRevD.74.014027,epjc.28.515,npb.625.239,npb.642.263,
   PhysRevD.89.094004,front.phys.16.24201,
   PhysRevD.64.112002,epjc.23.275,epjc.72.1923,
   PhysRevD.90.074018,
   PhysRevD.75.014019,epjc.59.49,PhysRevD.104.016025,
   PhysRevD.74.094020}, and the actual contributions from the
   higher twist (for example, twist-3) DAs to the hadronic
   transition form factors
   are as important as those from the leading twist (twist-2) DAs
   \cite{PhysRevD.74.014027,epjc.28.515,npb.625.239,npb.642.263,
   PhysRevD.89.094004,front.phys.16.24201}  and
   those from the next-to-leading order (NLO)
   QCD radiative corrections to the scattering
   amplitudes \cite{PhysRevD.103.056006,PhysRevD.89.094004}.
   It has already been recognized from the numerical perspective
   that the effects from the higher twist hadronic DAs are
   considerably large rather than formally power suppressed.

   In our recent study \cite{PhysRevD.103.056006}, the $B$ ${\to}$
   $PP$ decays were systemically reinvestigated using the PQCD approach
   by considering contributions from $B$ mesonic subleading
   twist WFs and the updated DAs of the final pseudoscalar mesons.
   It was found that the contribution from the $B$ mesonic WFs ${\phi}_{B2}$,
   which are usually paid less attention in the previous calculations,
   have certain influences over HMEs and branching ratios, and are
   comparable with those from the NLO corrections.
   In this work, a comprehensive study of the effects of the WFs
   ${\phi}_{B2}$ and updated DAs of final states using the PQCD
   approach is extended to charmless $B$ ${\to}$ $PV$ decays to
   match the precision improvement of theoretical and experimental
   results, where $V$ denote the ground $SU(3)$ vector mesons.
   Because of our inadequate understanding of the flavor mixing
   and possible glueball components, the final states of ${\eta}$
   and ${\eta}^{\prime}$ mesons are not considered here for the
   moment, {\em i.e.}, $P$ $=$ ${\pi}$ and $K$.

   This paper is organized as follows.
   In Section \ref{sec:hamiltonian}, the theoretical framework
   is briefly described.
   Definitions of kinematic variables and expressions for
   the WFs involved are presented in Section
   \ref{sec:kinematics} and \ref{sec:wave-function}, respectively.
   The contributions from different twist WFs
   to the form factors of $B$ ${\to}$ $PV$ decays are quantitatively
   analyzed in Section \ref{sec:formfactor}.
   In Section \ref{sec:branch}, the branching ratios and $CP$ asymmetries
   of $B$ ${\to}$ $PV$ decays are reevaluated by taking the $B$
   mesonic WFs ${\phi}_{B2}$ into consideration.
   We conclude with a summary in Section \ref{sec:summary}.
   The decay amplitudes and amplitude building blocks for the
   $B$ ${\to}$ $PV$ decays are displayed in Appendices \ref{sec:mode}
   and \ref{sec:block}, respectively.

   \section{The effective Hamiltonian}
   \label{sec:hamiltonian}
   It is widely accepted that charmless nonleptonic $B$ ${\to}$
   $PV$ decays are predominantly induced by the heavy $b$ quark
   weak decays within the SM, {\em i.e.}, $b$ ${\to}$ $W^{{\ast}-}$ $+$ $u$.
   There are at least three energy scales, the mass of the $W^{\pm}$
   gauge boson $m_{W}$, the mass of the $b$ quark $m_{b}$, and the
   QCD characteristic scale ${\Lambda}_{\rm QCD}$, with the
   hierarchical relationship $m_{W}$ ${\gg}$ $m_{b}$ ${\gg}$
   ${\Lambda}_{\rm QCD}$ and each energy scale corresponding to a
   different interaction dynamics.
   Based on the operator product expansion and renormalization
   group (RG) method, the effective Hamiltonian in charge
   of charmless $B$ ${\to}$ $PV$ decays can be factorized
   by the renormalization scale ${\mu}$ into three parts,
   the Wilson coefficients $C_{i}$, four-quark operators $Q_{i}$,
   and the ${\mu}$-independent couplings of weak interactions,
   including the Fermi constant  $G_{F}$ ${\approx}$
   $1.166{\times}10^{-5}$ ${\rm GeV}^{-2}$  \cite{pdg2020} and
   CKM factors, and written as \cite{RevModPhys.68.1125},
   \begin{equation}
  {\cal H}_{\rm eff}\, =\,
   \frac{G_{F}}{\sqrt{2}}\, \sum\limits_{q=d,s}
   \Big\{ V_{ub}\,V_{uq}^{\ast} \sum\limits_{i=1}^{2}C_{i}\,Q_{i}
   - V_{tb}\,V_{tq}^{\ast} \sum\limits_{j=3}^{10}C_{j}\,O_{j} \Big\}
   +{\rm h.c.}
   \label{hamilton}
   \end{equation}
   With the phenomenological Wolfenstein parametrization and up to
   ${\cal O}({\lambda}^{8})$, the CKM factors involved
   are expressed as,
    \begin{eqnarray}
    V_{ub}\,V_{ud}^{\ast} &=&
    A\,{\lambda}^{3}\,({\rho}-i\,{\eta})\,
    (1-\frac{1}{2}\,{\lambda}^{2}-\frac{1}{8}\,{\lambda}^{4})
    +{\cal O}({\lambda}^{8})
    \label{ckm-vub-vud}, \\
    V_{tb}\,V_{td}^{\ast} &=&
    A\,{\lambda}^{3}
    +A^{3}\,{\lambda}^{7}\,({\rho}-i\,{\eta}-\frac{1}{2})
    -V_{ub}\,V_{ud}^{\ast}
    +{\cal O}({\lambda}^{8})
    \label{ckm-vtb-vtd}, \\
    V_{ub}\,V_{us}^{\ast} &=&
    A\,{\lambda}^{4}\,({\rho}-i\,{\eta})
    +{\cal O}({\lambda}^{8})
    \label{ckm-vub-vus}, \\
    V_{tb}\,V_{ts}^{\ast} &=&
    -A\,{\lambda}^{2}\,(1-\frac{1}{2}\,{\lambda}^{2}-\frac{1}{8}\,{\lambda}^{4})
    +\frac{1}{2}\,A^{3}\,{\lambda}^{6}
    -V_{ub}\,V_{us}^{\ast}
    +{\cal O}({\lambda}^{8})
    \label{ckm-vtb-vts},
    \end{eqnarray}
   where $A$, ${\lambda}$, ${\rho}$ and ${\eta}$ are the
   Wolfenstein parameters; their latest fitted values
   can be found in Ref. \cite{pdg2020}.
   The Wilson coefficients $C_{i}$ summarize the physical
   contributions above the energy scale ${\mu}$, and are computable
   using the RG-assisted perturbative theory.
   Their explicit expressions, including the NLO corrections,
   can be found in Ref. \cite{RevModPhys.68.1125}.
   The local four-quark operators are defined as follows.
    \begin{equation}
    Q_{1} \, =\,
    \big[\, \bar{u}_{\alpha}\,{\gamma}_{\mu}\,(1-{\gamma}_{5})\,b_{\alpha} \,\big]\
    \big[\, \bar{q}_{\beta}\,{\gamma}^{\mu}\,(1-{\gamma}_{5})\,u_{\beta}   \,\big]
    \label{operator:q1},
    \end{equation}
    \begin{equation}
    Q_{2} \, =\,
    \big[\, \bar{u}_{\alpha}\,{\gamma}_{\mu}\,(1-{\gamma}_{5})\,b_{\beta} \,\big]\
    \big[\, \bar{q}_{\beta}\,{\gamma}^{\mu}\,(1-{\gamma}_{5})\,u_{\alpha} \,\big]
    \label{operator:q2},
    \end{equation}
    \begin{equation}
    Q_{3} \, =\, \sum\limits_{q^{\prime}}\,
    \big[\, \bar{q}_{\alpha}\,{\gamma}_{\mu}\,(1-{\gamma}_{5})\,b_{\alpha} \,\big]\
    \big[\, \bar{q}^{\prime}_{\beta}\,{\gamma}^{\mu}\,(1-{\gamma}_{5})\,q^{\prime}_{\beta} \,\big]
    \label{operator:q3},
    \end{equation}
    \begin{equation}
    Q_{4} \, =\, \sum\limits_{q^{\prime}}\,
    \big[\, \bar{q}_{\alpha}\,{\gamma}_{\mu}\,(1-{\gamma}_{5})\,b_{\beta} \,\big]\
    \big[\, \bar{q}^{\prime}_{\beta}\,{\gamma}^{\mu}\,(1-{\gamma}_{5})\,q^{\prime}_{\alpha} \,\big]
    \label{operator:q4},
   \end{equation}
    \begin{equation}
    Q_{5} \, =\, \sum\limits_{q^{\prime}} \,
    \big[\, \bar{q}_{\alpha}\,{\gamma}_{\mu}\,(1-{\gamma}_{5})\,b_{\alpha} \,\big]\
    \big[\, \bar{q}^{\prime}_{\beta}\,{\gamma}^{\mu}\,(1+{\gamma}_{5})\,q^{\prime}_{\beta} \,\big]
    \label{operator:q5},
    \end{equation}
    \begin{equation}
    Q_{6} \, =\, \sum\limits_{q^{\prime}}
    \big[\, \bar{q}_{\alpha}\,{\gamma}_{\mu}\,(1-{\gamma}_{5})\,b_{\beta} \,\big]\
    \big[\, \bar{q}^{\prime}_{\beta}\,{\gamma}^{\mu}\,(1+{\gamma}_{5})\,q^{\prime}_{\alpha} \,\big]
    \label{operator:q6},
    \end{equation}
    \begin{equation}
    Q_{7} \, =\, \sum\limits_{q^{\prime}}\,\frac{3}{2}\,Q_{q^{\prime}}\,
    \big[\, \bar{q}_{\alpha}\,{\gamma}_{\mu}\,(1-{\gamma}_{5})\,b_{\alpha} \,\big]\
    \big[\, \bar{q}^{\prime}_{\beta}\,{\gamma}^{\mu}\,(1+{\gamma}_{5})\,q^{\prime}_{\beta} \,\big]
    \label{operator:q7},
    \end{equation}
    \begin{equation}
    Q_{8} \, =\, \sum\limits_{q^{\prime}}\,\frac{3}{2}\,Q_{q^{\prime}}\,
    \big[\, \bar{q}_{\alpha}\,{\gamma}_{\mu}\,(1-{\gamma}_{5})\,b_{\beta} \,\big]\
    \big[\, \bar{q}^{\prime}_{\beta}\,{\gamma}^{\mu}\,(1+{\gamma}_{5})\,q^{\prime}_{\alpha} \,\big]
    \label{operator:q8},
    \end{equation}
    \begin{equation}
    Q_{9} \, =\, \sum\limits_{q^{\prime}}\,\frac{3}{2}\,Q_{q^{\prime}}\,
    \big[\, \bar{q}_{\alpha}\,{\gamma}_{\mu}\,(1-{\gamma}_{5})\,b_{\alpha} \,\big]\
    \big[\, \bar{q}^{\prime}_{\beta}\,{\gamma}^{\mu}\,(1-{\gamma}_{5})\,q^{\prime}_{\beta} \,\big]
    \label{operator:q9},
    \end{equation}
    \begin{equation}
    Q_{10} \, =\, \sum\limits_{q^{\prime}}\,\frac{3}{2}\,Q_{q^{\prime}}\,
    \big[\, \bar{q}_{\alpha}\,{\gamma}_{\mu}\,(1-{\gamma}_{5})\,b_{\beta} \,\big]\
    \big[\, \bar{q}^{\prime}_{\beta}\,{\gamma}^{\mu}\,(1-{\gamma}_{5})\,q^{\prime}_{\alpha} \,\big]
    \label{operator:q10},
    \end{equation}
   where ${\alpha}$ and ${\beta}$ are the color indices;
   $q^{\prime}$ ${\in}$ \{$u$, $d$, $c$, $s$, $b$\}, and
   $Q_{q^{\prime}}$ is the electric charge of quark
   $q^{\prime}$ in the unit of ${\vert}e{\vert}$.
   The physical contributions below the energy scale ${\mu}$ are
   contained in the HMEs ${\langle}Q_{i}{\rangle}$ $=$
   ${\langle}PV{\vert}Q_{i}{\vert}B{\rangle}$, which are the
   focus of the current theoretical calculation.

   The various treatments on HMEs depend on the different
   phenomenological approaches corresponding to the understanding
   of the perturbative and nonperturbative contributions.
   The joint effort of the transverse momentum for quarks and
   the Sudakov factors for all participant WFs is considered
   within the PQCD approach to settle the soft endpoint
   contributions from the collinear approximation.
   The master formula for HMEs with the PQCD approach is
   generally written as
   \begin{eqnarray}
  {\langle}PV{\vert}Q_{i}{\vert}B{\rangle} &{\propto}&
  {\int}\,dx_{1}\,dx_{2}\,dx_{3}\,db_{1}\,db_{2}\,db_{3}\,
   H_{i}(t_{i},x_{1},b_{1},x_{2},b_{2},x_{3},b_{3})
   \nonumber \\ & & \quad
  {\Phi}_{B}(x_{1},b_{1})\,e^{-S_{B}}\,
  {\Phi}_{P}(x_{2},b_{2})\,e^{-S_{P}}\,
  {\Phi}_{V}(x_{3},b_{3})\,e^{-S_{V}}
   \label{eq:HMEs},
   \end{eqnarray}
   where $b_{i}$ is the conjugate variable of the transverse
   momentum $\vec{k}_{i{\perp}}$ of the valence quarks;
   $H_{i}$ is the scattering amplitudes for hard gluon exchange
   interactions among quarks; $e^{-S_{i}}$ is the Sudakov factor.
   Other variables and inputs are described below.

   \section{Kinematics}
   \label{sec:kinematics}
   It is usually assumed that in the heavy quark limit, the light
   quarks rapidly move away from the $b$ quark decaying point at
   near the speed of light.
   The light cone variables are generally used in expressions.
   The relations between the four-dimensional space-time
   coordinates $x^{\mu}$ $=$ ($x^{0}$, $x^{1}$, $x^{2}$, $x^{3}$)
   $=$ ($t$, $x$, $y$, $z$) and the light-cone coordinates $x^{\mu}$
   $=$ ($x^{+}$, $x^{-}$, $\vec{x}_{\perp}$) are defined as
   $x^{\pm}$ $=$ $(x^{0}{\pm}x^{3})/\sqrt{2}$ and
   $\vec{x}_{\perp}$ $=$ ($x^{1}$, $x^{2}$).
   The light cone planes correspond to $x^{\pm}$ $=$ $0$.
   The scalar product of any two vectors is given by
   $a{\cdot}b$ $=$ $a_{\mu}b^{\mu}$ $=$ $a^{+}b^{-}$ $+$
   $a^{-}b^{+}$ $-$ $\vec{a}_{\perp}{\cdot}\vec{b}_{\perp}$.

   In the rest frame of the $B$ meson, the light cone kinematic
   variables are defined as
   \begin{equation}
    p_{B}\, =\, p_{1}\, =\, \frac{m_{B}}{\sqrt{2}}(1,1,0)
   \label{kine-p1},
   \end{equation}
   \begin{equation}
    p_{P}\, =\, p_{2}\, =\, \frac{m_{B}}{\sqrt{2}}(0,1-r_{V}^{2},0)
   \label{kine-p2},
   \end{equation}
   \begin{equation}
    p_{V}\, =\, p_{3}\, =\, \frac{m_{B}}{\sqrt{2}}(1,r_{V}^{2},0)
   \label{kine-p3},
   \end{equation}
   \begin{equation}
   e^{\parallel}_{V}\, =\, \frac{p_{V}}{m_{V}}-
   \frac{m_{V}}{p_{V}{\cdot}n_{-}}n_{-}
   \label{kine-polarization-vector},
   \end{equation}
   \begin{equation}
    k_{1}\, =\, x_{1}\,p_{1}+(0,0,\vec{k}_{1{\perp}})
   \label{kine-k1},
   \end{equation}
   \begin{equation}
   k_{2}\, =\, \frac{m_{B}}{\sqrt{2}}(0,x_{2},\vec{k}_{2{\perp}})
   \label{kine-k2},
   \end{equation}
   \begin{equation}
   k_{3}\, =\, \frac{m_{B}}{\sqrt{2}}(x_{3},0,\vec{k}_{3{\perp}})
   \label{kine-k3},
   \end{equation}
   where the mass ratio $r_{V}$ $=$ $m_{V}/m_{B}$.
   $e^{\parallel}_{V}$ is the longitudinal polarization vector.
   The variables $x_{1}$ and $\vec{k}_{1{\perp}}$ are the longitudinal
   momentum fraction and transverse momentum of the light quark
   in the $B$ meson, respectively.
   The variables $x_{i}$ and $\vec{k}_{i{\perp}}$ for $i$ $=$ $2$
   and $3$ are the longitudinal momentum fractions and transverse
   momentum of the antiquarks in the final pseudoscalar and
   vector mesons, respectively.

   \section{Hadronic wave functions and distribution amplitudes}
   \label{sec:wave-function}
   The $B$ mesonic WFs are generally defined as
   \cite{PhysRevD.103.056006,PhysRevD.74.014027,epjc.28.515,
   PhysRevD.55.272,npb.592.3},
   \begin{eqnarray} & &
  {\langle}\,0\,{\vert}\, \bar{q}_{\alpha}(z)\,b_{\beta}(0)\,{\vert}\,
   \overline{B}(p_{1})\,{\rangle}
   \nonumber \\ &=&
  +\frac{i}{4}\,f_{B}\,{\int} \mathbbm{d}^{4}k\,e^{-i\,k_{1}{\cdot}z}\,
   \Big\{ \big(\!\!\not{p}_{1}+m_{B}\big)\, {\gamma}_{5}\,
   \Big[\,\frac{\not{n}_{-}}{\sqrt{2}}\,{\phi}_{B}^{+}
  +\frac{\not{n}_{+}}{\sqrt{2}}\,{\phi}_{B}^{-} \Big] \Big\}_{{\beta}{\alpha}}
   \nonumber \\ &=&
   -\frac{i}{4}\,f_{B}\,{\int}\mathbbm{d}^{4}k\,e^{-i\,k_{1}{\cdot}z}\,
    \Big\{ \big(\!\!\not{p}_{1}+m_{B}\big)\,
   {\gamma}_{5}\, \Big[{\phi}^{+}+\frac{\not{n}_{+}}{\sqrt{2}}\,\big(
   {\phi}_{B}^{+}-{\phi}_{B}^{-} \big) \Big] \Big\}_{{\beta}{\alpha}}
    \nonumber \\ &=&
   -\frac{i}{4}\,f_{B}\,{\int}\mathbbm{d}^{4}k\,e^{-i\,k_{1}{\cdot}z}\,
    \Big\{ \big(\!\!\not{p}_{1}+m_{B}\big)\,
   {\gamma}_{5}\, \Big({\phi}_{B1}+\frac{\not{n}_{+}}{\sqrt{2}}\,
   {\phi}_{B2} \Big) \Big\}_{{\beta}{\alpha}}
   \label{B-meson-WF-definition},
   \end{eqnarray}
   where $f_{B}$ is the decay constant.
   The coordinate $z$ of the light quark, and
   the vectors $n_{+}$ $=$ ($1$, $0$, $\vec{0}$) and
   $n_{-}$ $=$ ($0$, $1$, $\vec{0}$) are on the light cone,
   {\em i.e.}, $z^{2}$ $=$ $0$ and $n_{\pm}^{2}$ $=$ $0$.
   The scalar functions ${\phi}_{B}^{+}$
   and ${\phi}_{B}^{-}$ are the leading and subleading twist WFs,
   respectively.
   ${\phi}_{B}^{+}$ and ${\phi}_{B}^{-}$ have different asymptotic
   behaviors as the longitudinal momentum fraction of the light quark
   $x_{1}$ ${\to}$ $0$. Their relations are
    \begin{equation}
   {\phi}_{B}^{+}(x_{1}) + x_{1}\, {\phi}_{B}^{-{\prime}}(x_{1})\, =\, 0
    \label{wf-b-meson-equation-of-motion},
    \end{equation}
    \begin{equation}
   {\phi}_{B1}\, =\, {\phi}_{B}^{+}
    \label{wf-b-meson-b1},
    \end{equation}
    \begin{equation}
   {\phi}_{B2}\, =\, {\phi}_{B}^{+}-{\phi}_{B}^{-}
    \label{wf-b-meson-b2}.
    \end{equation}

   Although the expressions of ${\phi}_{B}^{+}$ are generally
   different from those of ${\phi}_{B}^{-}$ with the equation
   of motion Eq.(\ref{wf-b-meson-equation-of-motion}),
   an approximation of ${\phi}_{B}^{+}$ $=$ ${\phi}_{B}^{-}$
   is often used in many phenomenological studies of nonleptonic
   $B$ meson decays, {\em i.e.}, only contributions from
   ${\phi}_{B1}$ are considered, and those from
   ${\phi}_{B2}$ are absent. However, it has been shown in Refs.
   \cite{PhysRevD.71.034018,PhysRevD.103.056006,PhysRevD.74.014027,
   epjc.28.515,npb.625.239,npb.642.263,PhysRevD.89.094004} that ${\phi}_{B2}$
   is necessary to HMEs rather than a negligible factor, and its
   contributions to the form factors $F_{0}^{B{\to}{\pi}}$ with
   the PQCD approach can even reach up to $30\,\%$ in certain cases
   \cite{PhysRevD.74.014027,epjc.28.515}.
   Additionally, its share of the
   branching ratio could be as large as those from NLO corrections
   \cite{PhysRevD.103.056006}.
   The possible influence of ${\phi}_{B2}$ on $B$ ${\to}$ $PV$
   decays with the PQCD approach is a focus of this paper.
   One candidate for the most commonly used leading $B$ mesonic WFs
   ${\phi}_{B}^{+}$ in actual calculations with the PQCD approach
   is written as \cite{PhysRevD.63.054008},
    \begin{equation}
   {\phi}_{B}^{+}(x_{1},b_{1})\, =\, N\, x_{1}^{2}\,\bar{x}_{1}^{2}\,
   {\exp}\Big\{ -\Big( \frac{x_{1}\,m_{B}}{\sqrt{2}\,{\omega}_{B}}
    \Big)^{2} -\frac{1}{2} {\omega}_{B}^{2}\,b_{1}^{2} \Big\}
    \label{wf-b-meson-plus},
    \end{equation}
   and the corresponding $B$ mesonic WFs ${\phi}_{B}^{-}$
   is written as \cite{PhysRevD.103.056006,PhysRevD.74.014027},
    \begin{eqnarray}
   {\phi}_{B}^{-}(x_{1},b_{1}) &=& N\,
    \frac{2\,{\omega}_{B}^{4}}{m_{B}^{4}}\,
   {\exp}\Big(-\frac{1}{2} {\omega}_{B}^{2}\,b_{1}^{2} \Big)\,\Big\{
    \sqrt{{\pi}}\,\frac{m_{B}}{\sqrt{2}\,{\omega}_{B}}
   {\rm Erf}\Big( \frac{m_{B}}{\sqrt{2}\,{\omega}_{B}},
    \frac{x_{1}\,m_{B}}{\sqrt{2}\,{\omega}_{B}}\Big)
    \nonumber \\ & &+
    \Big[1+\Big(\frac{m_{B}\,\bar{x}_{1}}{\sqrt{2}\,{\omega}_{B}}\Big)^{2}\Big]
   {\exp}\Big[-\Big(\frac{x_{1}\,m_{B}}{\sqrt{2}\,{\omega}_{B}}\Big)^{2} \Big]
   -{\exp}\Big(-\frac{m_{B}^{2}}{2\,{\omega}_{B}^{2}} \Big) \Big\}
    \label{wf-b-meson-minus},
    \end{eqnarray}
   where ${\omega}_{B}$ is the shape parameter, and
   $\bar{x}_{1}$ $=$ $1$ $-$ $x_{1}$.
   The normalization constant $N$ is determined by,
    \begin{equation}
   {\int}_{0}^{1}dx_{1}\, {\phi}_{B}^{\pm}(x_{1},0)\, =\, 1
    \label{wf-b-meson-normalization}.
    \end{equation}

   The WFs of the final states including the light pseudoscalar mesons
   and longitudinally polarized vector mesons are respectively
   defined as \cite{jhep.1999.01.010,jhep.2006.05.004,
   PhysRevD.65.014007,jhep.2007.03.069},
   \begin{eqnarray} & &
  {\langle}\,P(p_{2})\,{\vert}\,\bar{q}_{i}(0)\,q_{j}(z)\,
  {\vert}\,0\,{\rangle}
   \nonumber \\ &=&
  -i\,\frac{f_P}{4}\,{\int}_{0}^{1}\mathbbm{d}x_{2}\,
   e^{+i\,k_{2}{\cdot}z}\,  \Big\{ {\gamma}_{5}\,
   \Big[ \!\!\not{p}_{2}\,{\phi}_{P}^{a}+
  {\mu}_{P}\,{\phi}_{P}^{p}-{\mu}_{P}\,
  (\not{n}_{-}\!\not{n}_{+}-1)\, {\phi}_{P}^{t}
   \Big] \Big\}_{ji}
   \label{pseudoscalar-meson-WF-definition},
   \end{eqnarray}
   \begin{eqnarray} & &
  {\langle}\,V(p_{3},e_{\parallel})\,{\vert}\,
   \bar{q}_{i}(0)\,q_{j}(z)\,{\vert}\,0\,{\rangle}
   \nonumber \\ &=&
   \frac{1}{4}\, {\int}_{0}^{1}\mathbbm{d}x_{3}\, e^{+i\,k_{3}{\cdot}z}\,
   \big\{ \!\not{e}_{{\parallel}}\,m_{V}\,f_{V}^{\parallel}\,
  {\phi}_{V}^{v}\,+ \!\not{e}_{{\parallel}}
   \!\not{p}_{3}\, f_{V}^{\perp}\, {\phi}_{V}^{t}- m_{V}\,
   f_{V}^{\perp}\, {\phi}_{V}^{s} \big\}_{ji}
   \label{vector-meson-WF-definition},
   \end{eqnarray}
   where $f_{P}$, $f_{V}^{\parallel}$, and $f_{V}^{\perp}$
   are the decay constants.
   ${\phi}_{P}^{a}$ and ${\phi}_{V}^{v}$ are the twist-2 WFs;
   ${\phi}_{P}^{p,t}$ and ${\phi}_{V}^{t,s}$ are the twist-3 WFs.
   It has been previously shown that the numerical values of the
   formfactor $F_{0,1}^{B{\to}{\pi}}$ were highly dependent on the
   models for pionic WFs \cite{PhysRevD.74.014027,npb.625.239,
   npb.642.263,PhysRevD.89.094004},
   and the contributions from the twist-3
   pionic DAs to $F_{0,1}^{B{\to}{\pi}}$ were larger than those
   from twist-2 pionic DAs
   \cite{PhysRevD.74.014027,npb.642.263,PhysRevD.89.094004}.
   According to the convention of Refs.
   \cite{jhep.2006.05.004,jhep.2007.03.069}
   and taking the pseudoscalar $P$ $=$ $K$ meson and vector $V$
   $=$ $K^{\ast}$ meson as an example,
   their DAs are written as,
    \begin{equation}
   {\phi}_{K}^{a}(x)\, =\, 6\,x\,\bar{x}\,\big\{
    1+a_{1}^{K}\,C_{1}^{3/2}({\xi})
     +a_{2}^{K}\,C_{2}^{3/2}({\xi})\big\}
    \label{pseudoscalar-twsit-2-a},
    \end{equation}
    \begin{eqnarray}
   {\phi}_{K}^{p}(x) &=& 1+3\,{\rho}_{+}^{K}
   -9\, {\rho}_{-}^{K}\,a_{1}^{K}
   +18\,{\rho}_{+}^{K}\,a_{2}^{K}
    \nonumber \\ &+&
    \frac{3}{2}\,({\rho}_{+}^{K}+{\rho}_{-}^{K})\,
    (1-3\,a_{1}^{K}+6\,a_{2}^{K})\,{\ln}(x)
    \nonumber \\ &+&
    \frac{3}{2}\,({\rho}_{+}^{K}-{\rho}_{-}^{K})\,
    (1+3\,a_{1}^{K}+6\,a_{2}^{K})\,{\ln}(\bar{x})
    \nonumber \\ &-&
    (\frac{3}{2}\, {\rho}_{-}^{K}
    -\frac{27}{2}\,{\rho}_{+}^{K}\,a_{1}^{K}
    +27\,{\rho}_{-}^{K}\,a_{2}^{K})\,C_{1}^{1/2}(\xi)
    \nonumber \\ &+&
    ( 30\,{\eta}_{K}-3\,{\rho}_{-}^{K}\,a_{1}^{K}
    +15\, {\rho}_{+}^{K}\,a_{2}^{K})\,C_{2}^{1/2}(\xi)
    \label{pseudoscalar-twsit-3-p},
    \end{eqnarray}
    \begin{eqnarray}
   {\phi}_{K}^{t}(x)  &=&
    \frac{3}{2}\,({\rho}_{-}^{K}-3\,{\rho}_{+}^{K}\,a_{1}^{K}
    +6\,{\rho}_{-}^{K}\,a_{2}^{K})
    \nonumber \\ &-&
    C_{1}^{1/2}(\xi)\big\{
    1+3\,{\rho}_{+}^{K}-12\,{\rho}_{-}^{K}\,a_{1}^{K}
   +24\,{\rho}_{+}^{K}\,a_{2}^{K}
    \nonumber \\ & & \quad +
    \frac{3}{2}\,({\rho}_{+}^{K}+{\rho}_{-}^{K})\,
    (1-3\,a_{1}^{K}+6\,a_{2}^{K})\,{\ln}(x)
    \nonumber \\ & & \quad +
    \frac{3}{2}\,({\rho}_{+}^{K}-{\rho}_{-}^{K})\,
    (1+3\,a_{1}^{K}+6\,a_{2}^{K})\, {\ln}(\bar{x}) \big\}
    \nonumber \\ &-&
    3\,(3\,{\rho}_{+}^{K}\,a_{1}^{K}
    -\frac{15}{2}\,{\rho}_{-}^{K}\,a_{2}^{K})\,C_{2}^{1/2}(\xi)
    \label{pseudoscalar-twsit-3-t},
    \end{eqnarray}
   \begin{equation}
  {\phi}_{K^{\ast}}^{v}(x) \,= \,
   6\,x\,\bar{x}\, \big\{ 1
   + a_{1}^{{\parallel},K^{\ast}}\,C_{1}^{3/2}(\xi)
   + a_{2}^{{\parallel},K^{\ast}}\,C_{2}^{3/2}(\xi) \big\}
   \label{vector-twist-2-v},
   \end{equation}
   \begin{eqnarray}
  {\phi}_{K^{\ast}}^{t}(x) &=&
    3\,{\xi}\,\big\{ C_{1}^{1/2}({\xi})
   +a_{1}^{{\perp},K^{\ast}}\,C_{2}^{1/2}({\xi})
   +a_{2}^{{\perp},K^{\ast}}\,C_{3}^{1/2}({\xi}) \big\}
   \nonumber \\ &+&
    \frac{3}{2}\, \frac{ m_{s}+m_{q} }{ m_{K^{\ast}} }\,
    \frac{ f_{K^{\ast}}^{\parallel} }{ f_{K^{\ast}}^{\perp} }\,
    \big\{ 1 + 8\,{\xi}\,a_{1}^{{\parallel},K^{\ast}}
   +(21-90\,x\,\bar{x})\,a_{2}^{{\parallel},K^{\ast}}
   \nonumber \\ & &
   +{\xi}\,{\ln}\bar{x}\,(1+3\,a_{1}^{{\parallel},K^{\ast}}
      +6\,a_{2}^{{\parallel},K^{\ast}})
   -{\xi}\,{\ln}x\,(1-3\,a_{1}^{{\parallel},K^{\ast}}
      +6\,a_{2}^{{\parallel},K^{\ast}}) \big\}
    \nonumber \\ &-&
    \frac{3}{2}\, \frac{ m_{s}-m_{q} }{ m_{K^{\ast}} }\,
    \frac{ f_{K^{\ast}}^{\parallel} }{ f_{K^{\ast}}^{\perp} }\,
    {\xi}\, \big\{ 2 + 9\,{\xi}\,a_{1}^{{\parallel},K^{\ast}}
   +(22-60\,x\,\bar{x})\,a_{2}^{{\parallel},K^{\ast}}
   \nonumber \\ & &
   +{\ln}\bar{x}\,(1+3\,a_{1}^{{\parallel},K^{\ast}}
   +6\,a_{2}^{{\parallel},K^{\ast}})
   +{\ln}x\,(1-3\,a_{1}^{{\parallel},K^{\ast}}
   +6\,a_{2}^{{\parallel},K^{\ast}}) \big\}
   \label{vector-twist-3-t},
   \end{eqnarray}
   \begin{eqnarray}
  {\phi}_{K^{\ast}}^{s}(x) &=&
    - 3\,C_{1}^{1/2}({\xi})
    - 3\,C_{2}^{1/2}({\xi})\,a_{1}^{{\perp},K^{\ast}}
    - 3\,C_{3}^{1/2}({\xi})\,a_{2}^{{\perp},K^{\ast}}
   \nonumber \\ &-&
   \frac{3}{2}\, \frac{ m_{s}+m_{q} }{ m_{K^{\ast}} }\,
   \frac{ f_{K^{\ast}}^{\parallel} }{ f_{K^{\ast}}^{\perp} }\,
   \big\{ C_{1}^{1/2}({\xi})
   +2\,C_{2}^{1/2}({\xi})\,a_{1}^{{\parallel},K^{\ast}}
   \nonumber \\ & &
    + \big[ 3\,C_{3}^{1/2}({\xi})
    +18\, C_{1}^{1/2}({\xi}) \big]\,a_{2}^{{\parallel},K^{\ast}}
   \nonumber \\ & &
   + ({\ln}\bar{x}+1)\,
    (1+3\,a_{1}^{{\parallel},K^{\ast}}
    +6\,a_{2}^{{\parallel},K^{\ast}})
   \nonumber \\ & &
   - ({\ln}x+1)\,
    (1-3\,a_{1}^{{\parallel},K^{\ast}}
    +6\,a_{2}^{{\parallel},K^{\ast}}) \big\}
    \nonumber \\ &+&
    \frac{3}{2}\, \frac{ m_{s}-m_{q} }{ m_{K^{\ast}} }\,
    \frac{ f_{K^{\ast}}^{\parallel} }{ f_{K^{\ast}}^{\perp} }\,
    \big\{ 9\,C_{1}^{1/2}({\xi})\,a_{1}^{{\parallel},K^{\ast}}
    +10\, C_{2}^{1/2}({\xi})\,a_{2}^{{\parallel},K^{\ast}}
   \nonumber \\ & &
   +({\ln}\bar{x}+1)\,
    (1+3\,a_{1}^{{\parallel},K^{\ast}}
    +6\,a_{2}^{{\parallel},K^{\ast}})
   \nonumber \\ & &
   +({\ln}x+1)\,
    (1-3\,a_{1}^{{\parallel},K^{\ast}}
    +6\,a_{2}^{{\parallel},K^{\ast}}) \big\}
   \label{vector-twist-3-s},
   \end{eqnarray}
   where $x$ is the longitudinal momentum fraction of the strange
   quark, and ${\xi}$ $=$ $x$ $-$ $\bar{x}$ $=$ $2\,x$ $-$ $1$.
   $C_{n}^{m}$ is the Gegenbauer polynomials.
   $a_{n}^{K}$, $a_{n}^{{\parallel},K^{\ast}}$ and
   $a_{n}^{{\perp},K^{\ast}}$ are the Gegenbauer moments.
   The dimensionless parameters ${\rho}_{+}^{K}$ $=$
   $(m_{s}+m_{q})^{2}/m_{K}^{2}$ and ${\rho}_{-}^{K}$ $=$
   $(m_{s}^{2}-m_{q}^{2})/m_{K}^{2}$ \cite{jhep.2006.05.004}.

   \begin{table}[ht]
   \caption{The values of input parameters, where their central
   values are regarded as the default inputs unless
   otherwise specified.}
   \label{tab:input}
   \begin{ruledtabular}
   \begin{tabular}{cccc}
   \multicolumn{4}{c}{Wolfenstein parameters of the CKM matrix
   \cite{pdg2020} } \\ \hline
     $A$ $=$ $0.790^{+0.017}_{-0.012}$
   & ${\lambda}$ $=$ $0.22650{\pm}0.00048$
   & $\bar{\rho}$ $=$ $0.141^{+0.016}_{-0.017}$
   & $\bar{\eta}$ $=$ $0.357{\pm}0.011$ \\ \hline
   \multicolumn{4}{c}{mass of particle (in the unit of MeV)
   \cite{pdg2020} } \\ \hline
     $m_{{\pi}^{\pm}}$ $=$ $139.57$
   & $m_{K^{\pm}}$ $=$ $493.677{\pm}0.016$
   & $m_{\rho}$ $=$ $775.26{\pm}0.25$
   & $m_{K^{{\ast}{\pm}}}$ $=$ $895.5{\pm}0.8$  \\
     $m_{{\pi}^{0}}$ $=$ $134.98$
   & $m_{K^{0}}$ $=$ $497.611{\pm}0.013$
   & $m_{\omega}$ $=$ $782.65{\pm}0.12$
   & $m_{K^{{\ast}0}}$ $=$ $895.55{\pm}0.20$  \\
     $m_{B_{u}}$ $=$ $5279.34{\pm}0.12$
   & $m_{B_{d}}$ $=$ $5279.65{\pm}0.12$
   & $m_{\phi}$ $=$ $1019.461{\pm}0.016$ \\ \hline
   \multicolumn{4}{c}{decay constants (in the unit of MeV)}
   \\ \hline
     $f_{\rho}^{\parallel}$ $=$ $216{\pm}3$   \cite{jhep.2007.03.069}
   & $f_{\omega}^{\parallel}$ $=$ $187{\pm}5$ \cite{jhep.2007.03.069}
   & $f_{\phi}^{\parallel}$ $=$ $215{\pm}5$   \cite{jhep.2007.03.069}
   & $f_{K^{\ast}}^{\parallel}$ $=$ $220{\pm}5$ \cite{jhep.2007.03.069} \\
     $f_{\rho}^{\perp}$ $=$ $165{\pm}9$ \cite{jhep.2007.03.069}
   & $f_{\omega}^{\perp}$ $=$ $151{\pm}9$ \cite{jhep.2007.03.069}
   & $f_{\phi}^{\perp}$ $=$ $186{\pm}9$ \cite{jhep.2007.03.069}
   & $f_{K^{\ast}}^{\perp}$ $=$ $185{\pm}10$ \cite{jhep.2007.03.069} \\
     $f_{B}$ $=$ $190.0{\pm}1.3$ \cite{pdg2020}
   & $f_{{\pi}}$ $=$ $130.2{\pm}1.2$ \cite{pdg2020}
   & $f_{K}$ $=$ $155.7{\pm}0.3$ \cite{pdg2020} \\ \hline
   \multicolumn{4}{c}{Gegenbauer moments at the scale of
    ${\mu}$ $=$ $1$ GeV \cite{jhep.2006.05.004,jhep.2007.03.069} }
   \\ \hline
     $a_{2}^{{\parallel},{\rho},{\omega}}$ $=$ $0.15{\pm}0.07$
   & $a_{2}^{{\parallel},{\phi}}$ $=$ $0.18{\pm}0.08$
   & $a_{1}^{{\parallel},K^{\ast}}$ $=$ $0.03{\pm}0.02$
   & $a_{2}^{{\parallel},K^{\ast}}$ $=$ $0.11{\pm}0.09$ \\
     $a_{2}^{{\perp},{\rho},{\omega}}$ $=$ $0.14{\pm}0.06$
   & $a_{2}^{{\perp},{\phi}}$ $=$ $0.14{\pm}0.07$
   & $a_{1}^{{\perp},K^{\ast}}$ $=$ $0.04{\pm}0.03$
   & $a_{2}^{{\perp},K^{\ast}}$ $=$ $0.10{\pm}0.08$ \\
     $a_{1}^{{\pi},{\rho},{\omega},{\phi}}$ $=$ $0$
   & $a_{2}^{\pi}$ $=$ $0.25{\pm}0.15$
   & $a_{1}^{K}$ $=$ $0.06{\pm}0.03$
   & $a_{2}^{K}$ $=$ $0.25{\pm}0.15$
   \end{tabular}
   \end{ruledtabular}
   \end{table}
   \begin{figure}[ht]
   \includegraphics[width=0.31\textwidth]{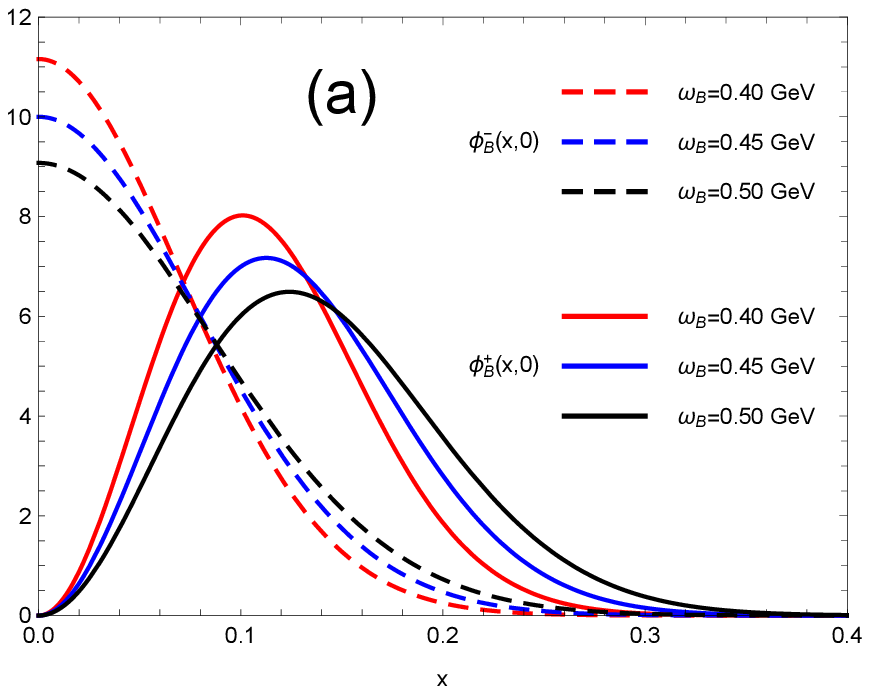}    \quad
   \includegraphics[width=0.31\textwidth]{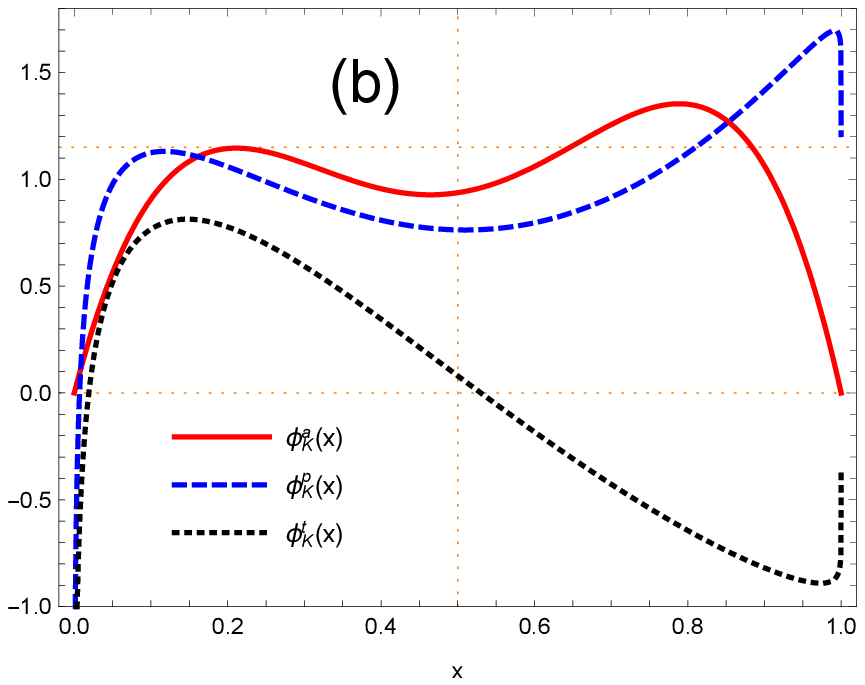} \quad
   \includegraphics[width=0.31\textwidth]{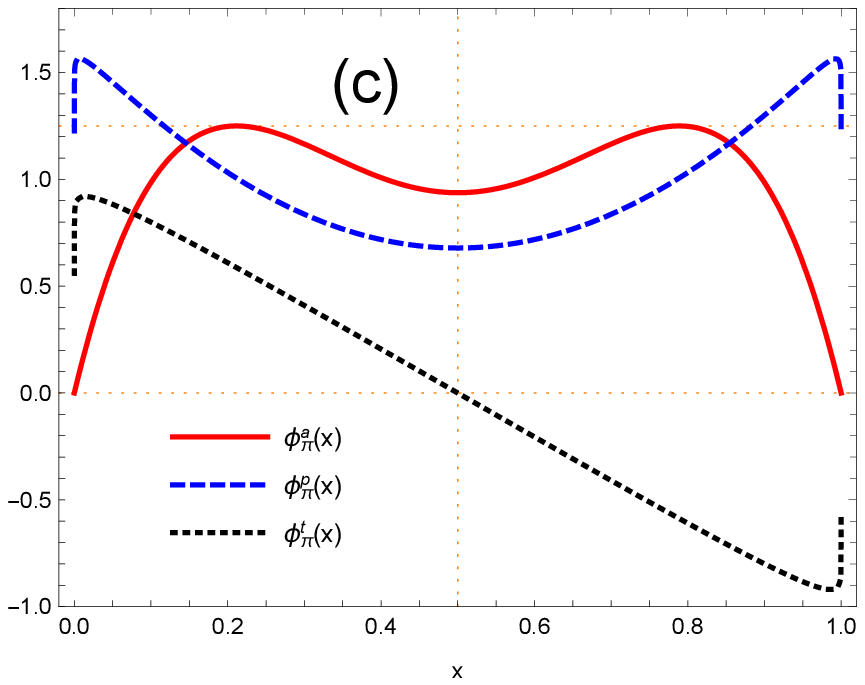}
   \\  \vspace{4mm}
   \includegraphics[width=0.31\textwidth]{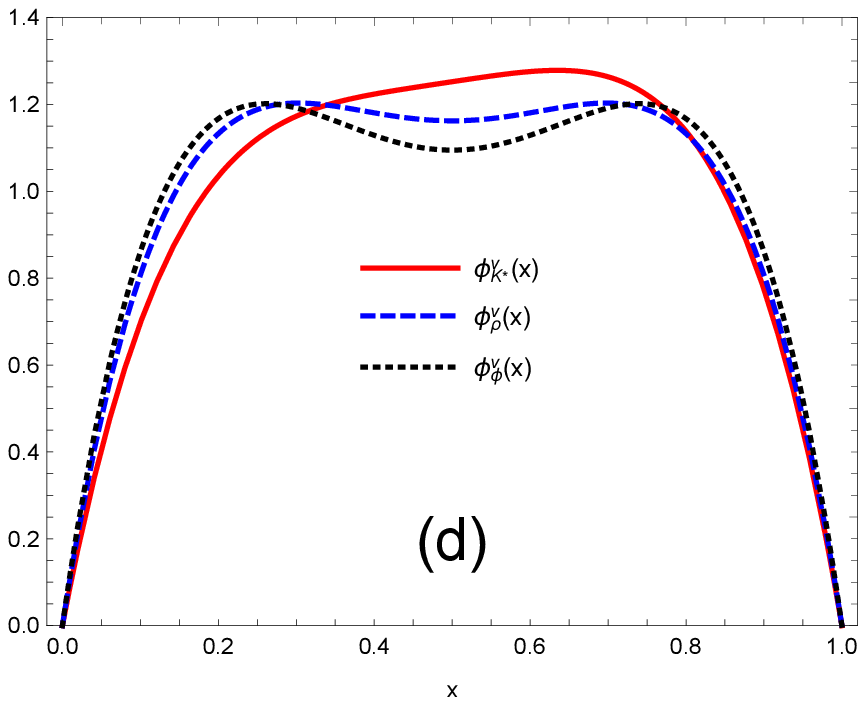} \quad
   \includegraphics[width=0.31\textwidth]{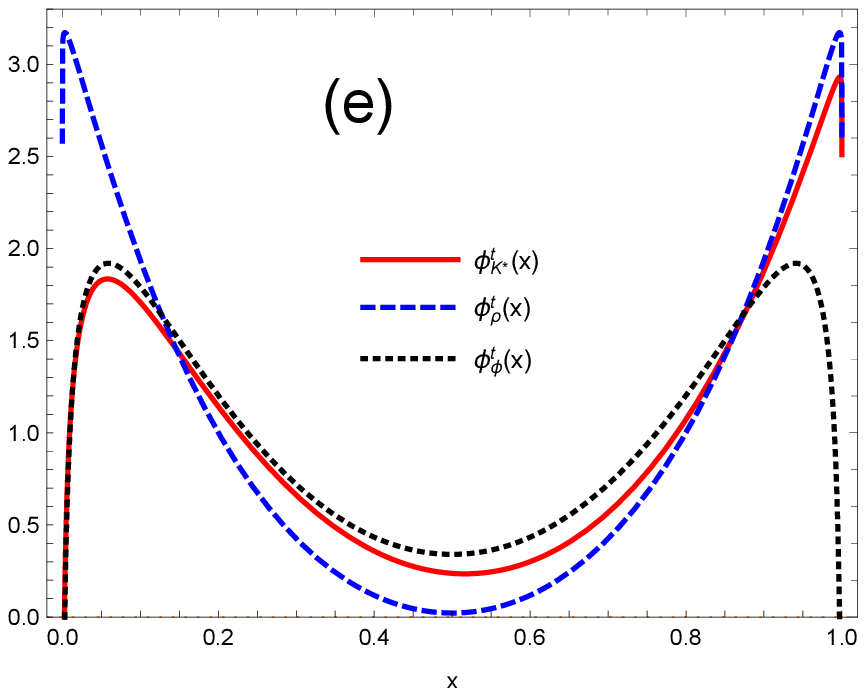} \quad
   \includegraphics[width=0.31\textwidth]{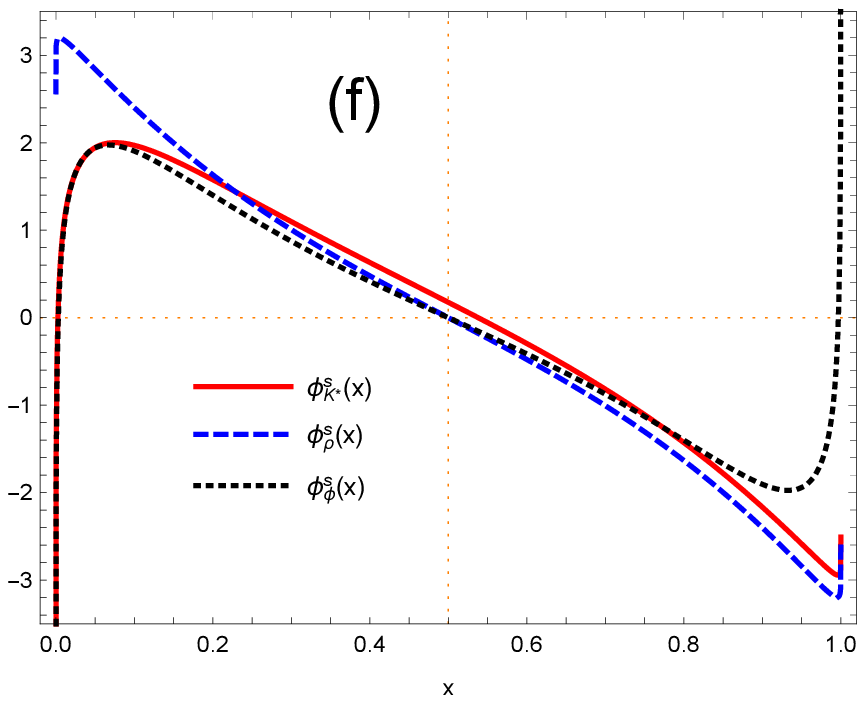}
   \caption{Shape lines of the DAs ${\phi}_{B}^{\pm}$
   (a), ${\phi}_{K}^{a,p,t}$ (b),
   ${\phi}_{\pi}^{a,p,t}$ (c),
   ${\phi}_{V}^{v}$ (d),
   ${\phi}_{V}^{t}$ (e) and
   ${\phi}_{V}^{s}$ (f) versus the longitudinal
   momentum fraction $x$ (horizontal axis).}
   \label{fig-wf}
   \end{figure}

   The shape lines of mesonic DAs with the inputs in Table
   \ref{tab:input} are displayed in Fig. \ref{fig-wf}.
   It is clearly seen that
   (1) the nonzero distributions of ${\phi}_{B}^{\pm}$ are mainly
   located in the small $x$ regions, and ${\phi}_{B}^{\pm}$ vanishes
   as ${x}$ ${\to}$ $1$.
   This fact is basically consistent with the intuitive
   expectation that the light quark shares a small longitudinal
   momentum fraction in the $B$ meson.
   (2) The shape lines of ${\phi}_{B}^{-}$ differ apparently from
   those of ${\phi}_{B}^{+}$ in the small $x$ regions.
   It is particularly noticeable that the DAs ${\phi}_{B}^{-}$ and
   ${\phi}_{B}^{+}$ exhibit different endpoint behaviors
   at $x$ $=$ $0$. Thus, it is clear that ${\phi}_{B2}$ $=$
   ${\phi}_{B}^{+}$ $-$ ${\phi}_{B}^{-}$ ${\ne}$ $0$, and the
   approximation ${\phi}_{B2}$ $=$ $0$
   in previous studies might be inappropriate and insufficient.
   (3) The integral ${\int}dx\frac{{\phi}_{B}^{-}}{x}$ will appear
   in the scattering amplitudes, for example, the form
   factors for the transition from the $B$ meson to final hadrons.
   The value of ${\phi}_{B}^{-}$ increases with the decrease of $x$,
   which implies that the integrals ${\int}dx\frac{{\phi}_{B}^{-}}{x}$
   and ${\int}dx\frac{{\phi}_{B2}}{x}$ may be significant in the
   small $x$ regions.
   The potential contributions from the subleading DAs ${\phi}_{B}^{-}$
   could be greatly enhanced when $x$ approaches to zero and
   should be given due consideration in the calculation.
   (4) The values of ${\phi}_{B}^{-}$ and ${\phi}_{B2}$ are nonzero
   at the endpoint $x$ $=$ $0$, so the integral
   ${\int}dx\frac{{\phi}_{B2}}{x}$ will be infrared divergent at
   the endpoint with the collinear approximation.
   This fact indicates that it may be reasonable and necessary
   for the PQCD approach to conciliate the nonperturbative
   contributions by considering the effects of the transverse
   momentum of valence quarks and the Sudakov factors.
   (5) The distributions of ${\phi}_{B}^{\pm}$ are sensitive to
   the shape parameter ${\omega}_{B}$.
   The larger the value of ${\omega}_{B}$, the wider distributions
   of ${\phi}_{B}^{\pm}$.
   The theoretical results with the PQCD approach will depend
   on the choice of ${\omega}_{B}$.
   (6) The expressions of DAs ${\phi}_{P}^{a,p,t}$ and
   ${\phi}_{V}^{v,t,s}$ are different from their asymptotic forms.
   With respect to the exchange $x$ ${\leftrightarrow}$
   $\bar{x}$, the DAs ${\phi}_{\pi}^{a,p}$ and
   ${\phi}_{{\rho},{\phi},{\omega}}^{v,t}$ are entirely symmetric,
   and the twist-3 DAs ${\phi}_{\pi}^{t}$ and
   ${\phi}_{{\rho},{\phi},{\omega}}^{s}$ are entirely antisymmetric,
   whereas the kaonic DAs ${\phi}_{K}^{a,p,t}$ and
   ${\phi}_{K^{\ast}}^{v,t,s}$ are asymmetric.

   \section{Form factors}
   \label{sec:formfactor}
\begin{figure}[ht]
   \includegraphics[width=0.25\textwidth,bb=180 520 320 605]{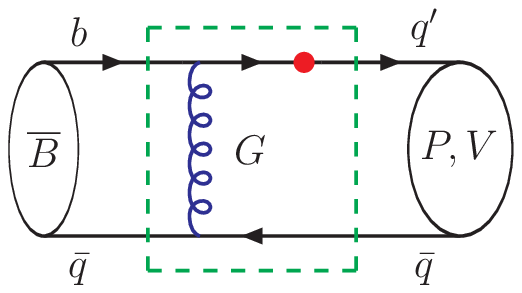} \quad
   \includegraphics[width=0.25\textwidth,bb=180 520 320 605]{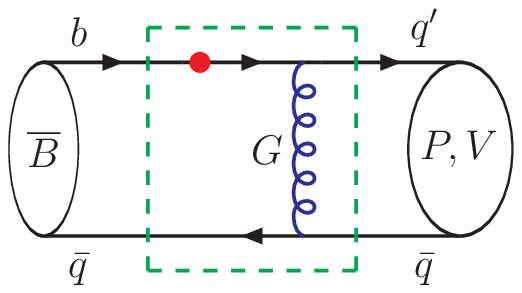}
   \\  \vspace{-3mm} (a) \hspace{0.25\textwidth} (b)
   \caption{Diagrams contributing to the $\overline{B}$ ${\to}$
   $P$, $V$ transition with the PQCD approach, where the dots
   denote an appropriate diquark current interaction and the
   dashed boxes represent the scattering amplitudes.}
   \label{fig:formfactor}
   \end{figure}

   As far as we know, the implications of hadronic WFs on
   transition form factors have been carefully studied with the PQCD
   approach in Refs. \cite{PhysRevD.71.034018,PhysRevD.103.056006,
   PhysRevD.74.014027,epjc.28.515,npb.625.239,npb.642.263,
   PhysRevD.89.094004,front.phys.16.24201},
   where HMEs for the transition form factors
   are expressed as the convolution integral of the scattering
   amplitudes and WFs of the initial and final mesons, and
   the lowest order approximation of the scattering amplitudes
   is illustrated with the one-gluon-exchange diagrams in
   Fig. \ref{fig:formfactor}.

   \begin{figure}[ht]
   \includegraphics[width=0.22\textwidth]{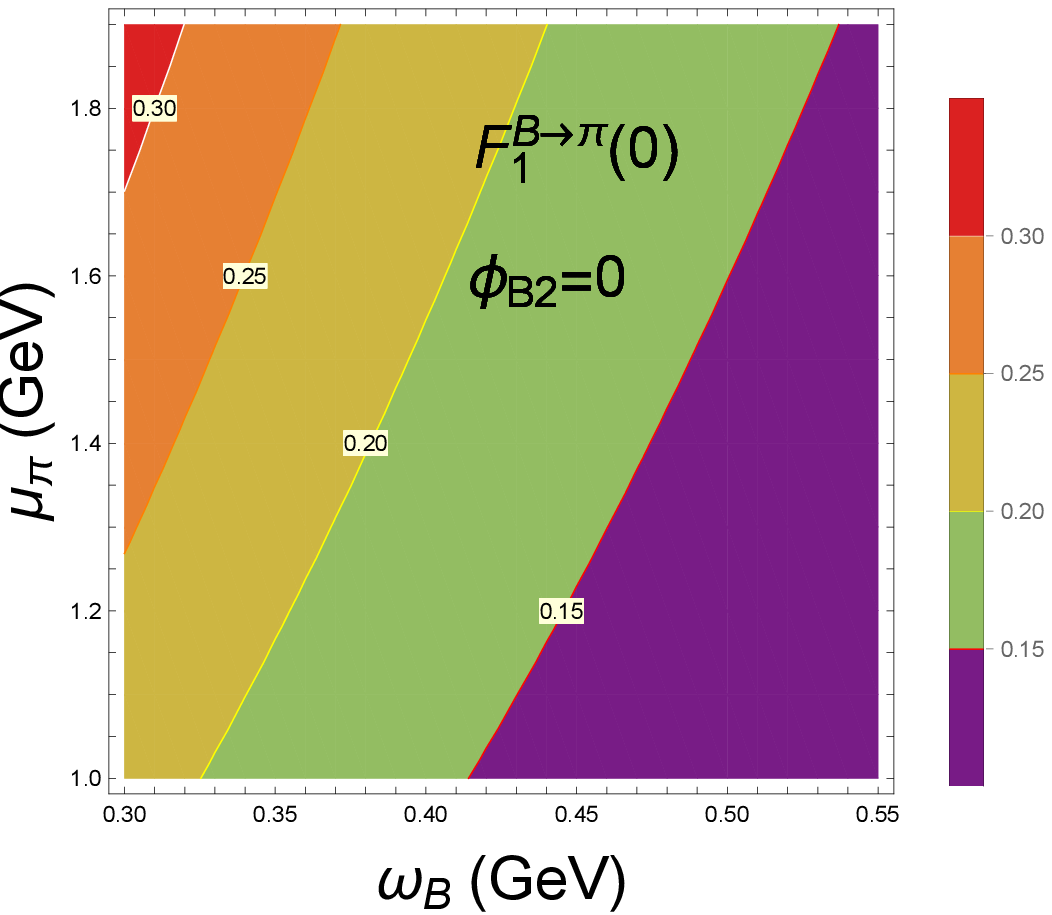} \quad
   \includegraphics[width=0.22\textwidth]{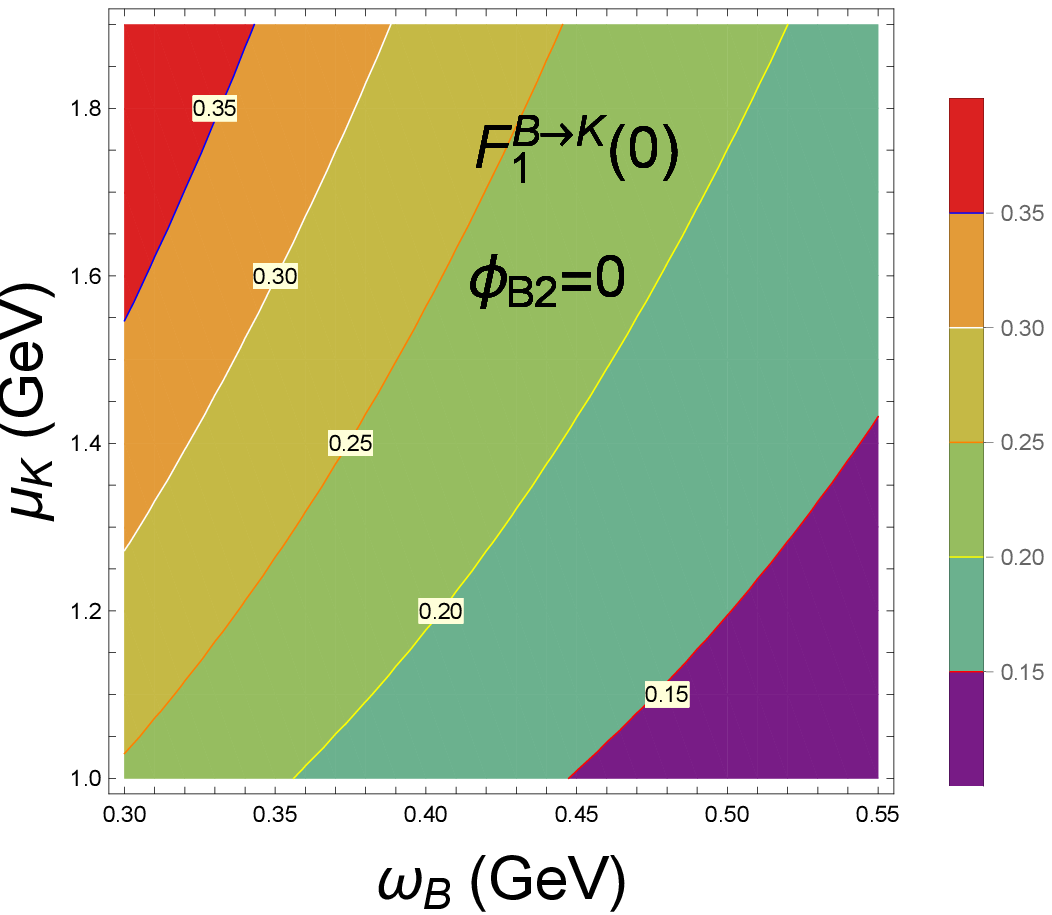} \quad
   \includegraphics[width=0.23\textwidth]{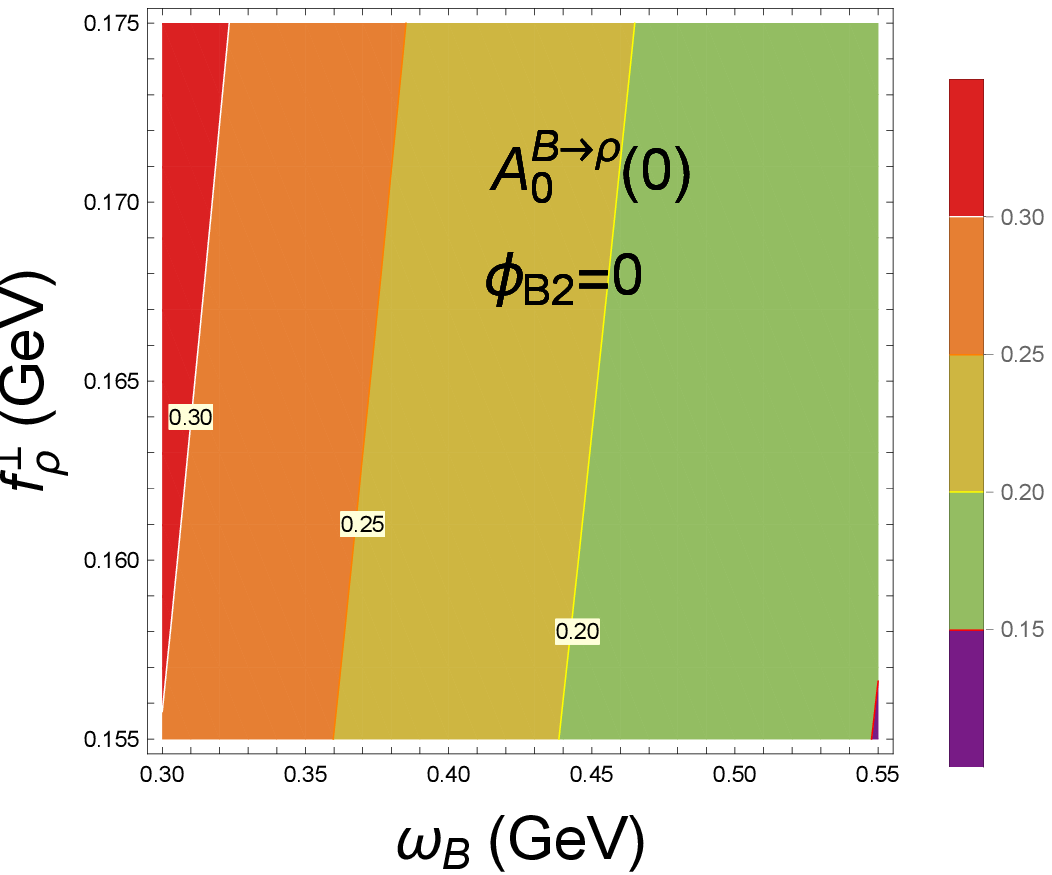}  \quad
   \includegraphics[width=0.23\textwidth]{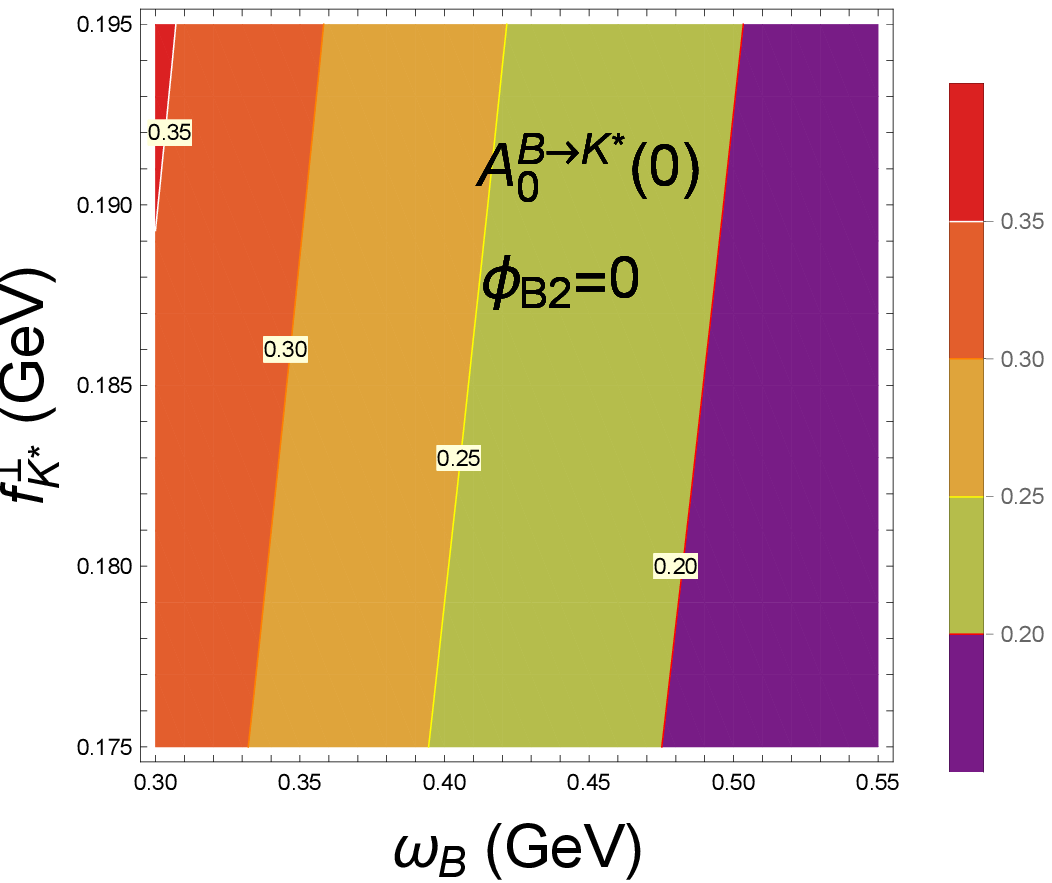}
   \\  \vspace{-3mm} (a)
       \hspace{0.22\textwidth} (b)
       \hspace{0.22\textwidth} (c)
       \hspace{0.22\textwidth} (d) \\ \vspace{4mm}
   \includegraphics[width=0.22\textwidth]{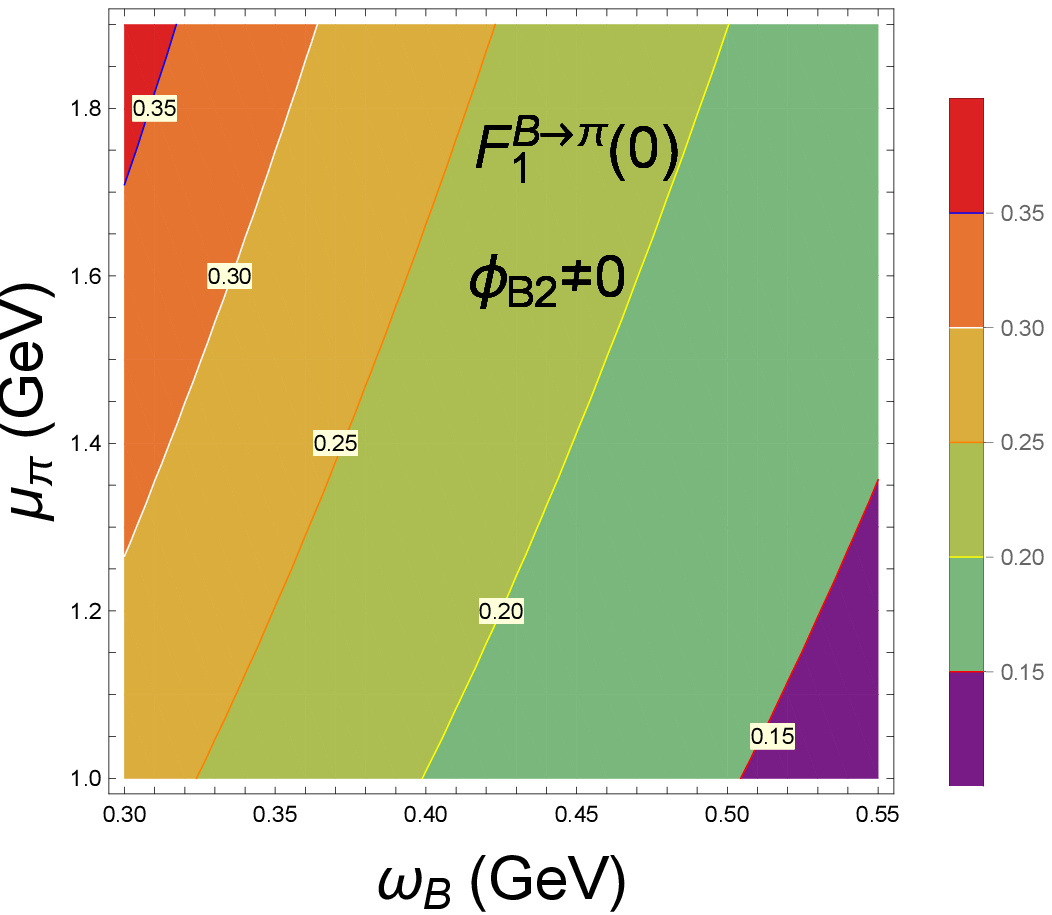} \quad
   \includegraphics[width=0.22\textwidth]{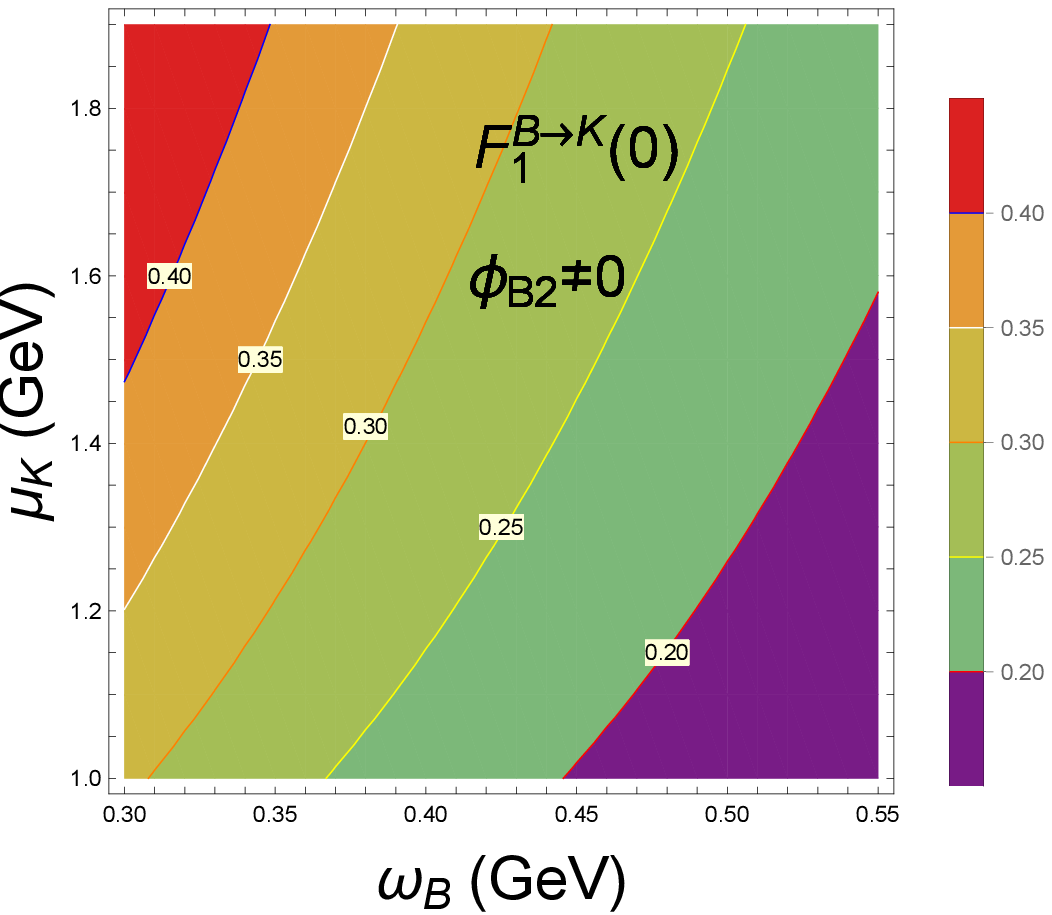} \quad
   \includegraphics[width=0.23\textwidth]{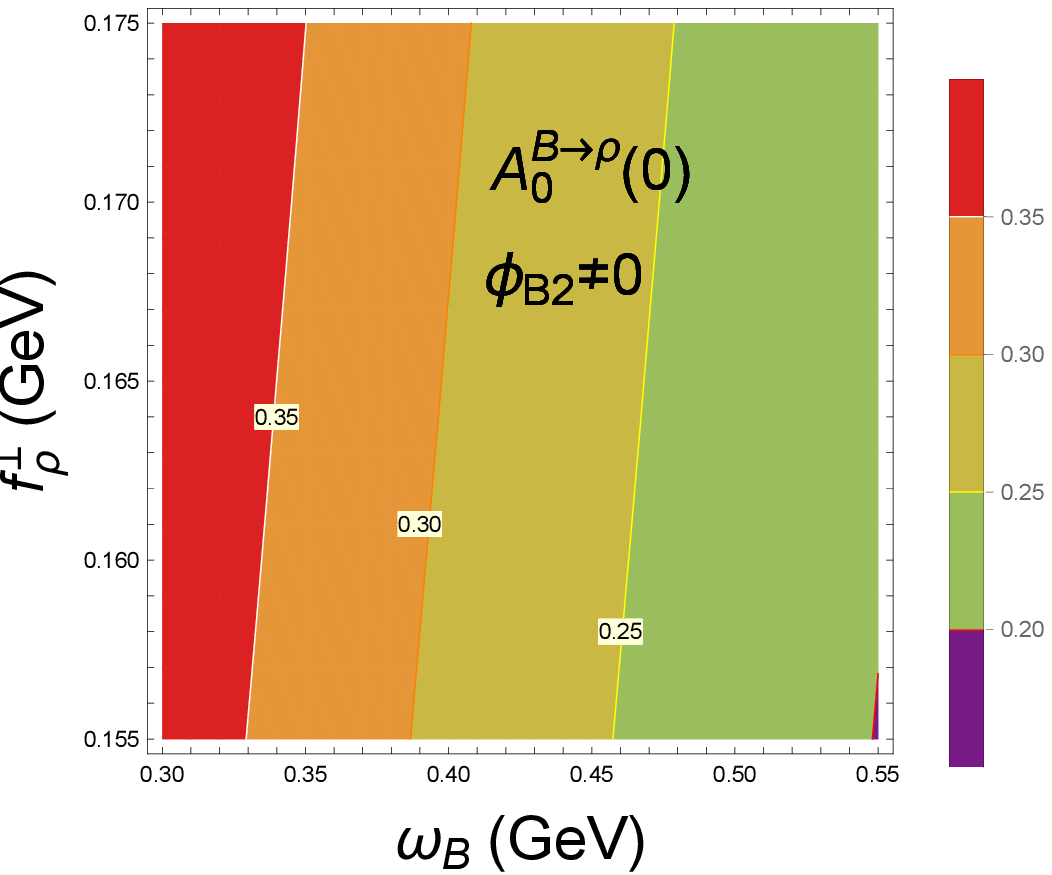}  \quad
   \includegraphics[width=0.23\textwidth]{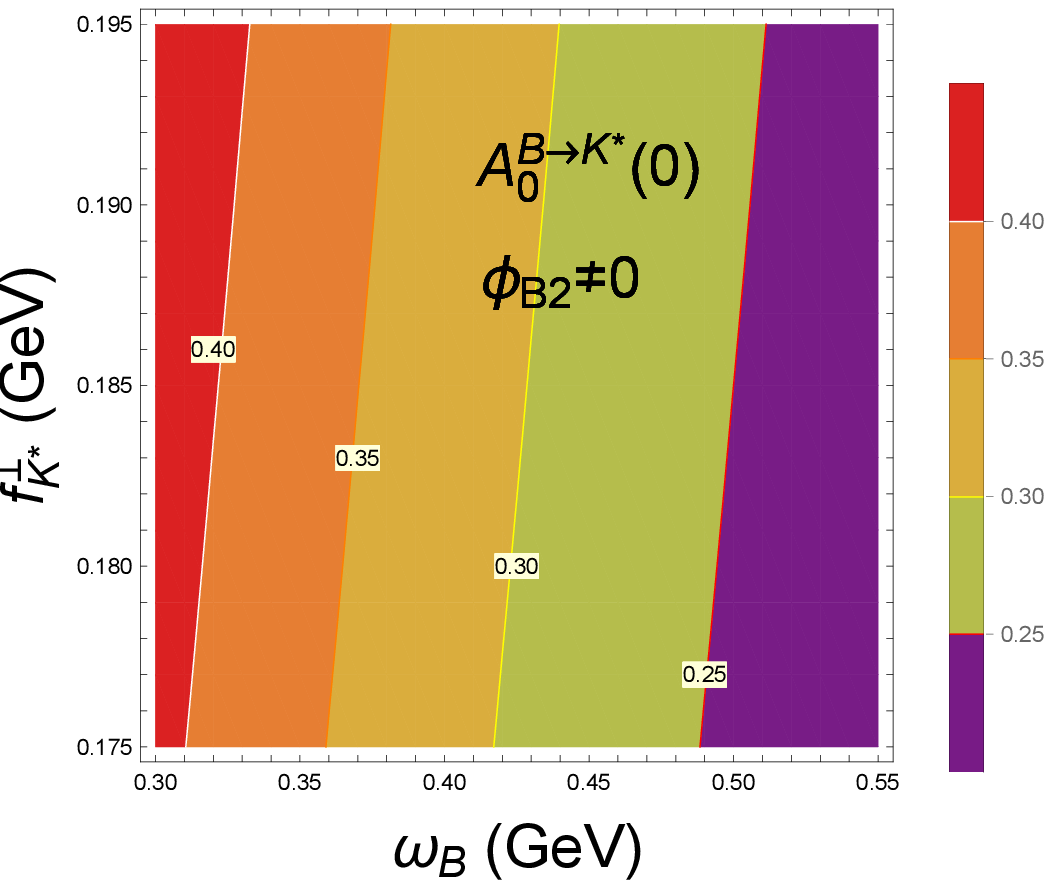}
   \\  \vspace{-3mm} (e)
       \hspace{0.22\textwidth} (f)
       \hspace{0.22\textwidth} (g)
       \hspace{0.22\textwidth} (h)
   \caption{Contour plot of the formfactors $F_{1}(q^{2})$
   and $A_{0}(q^{2})$ at $q^{2}$ $=$ $0$, where the values
   in (a,b,c,d) and (e,f,g,h) are calculated without and with
   contributions from ${\phi}_{B2}$.}
   \label{contour:formfactor}
   \end{figure}
   \begin{table}[ht]
   \caption{Contributions from different twist hadronic DAs to
   the formfactors $F_{1}(q^{2})$ and $A_{0}(q^{2})$ at $q^{2}$ $=$
   $0$ using the PQCD approach, where ${\omega}_{B}$ $=$ $0.4$ GeV,
   ${\mu}_{P}$ $=$ $1.4$ GeV, ${\Sigma}_{P}$ $=$ ${\phi}_{P}^{a}$
   $+$ ${\phi}_{P}^{p}$ $+$ ${\phi}_{P}^{t}$, ${\Sigma}_{V}$ $=$
   ${\phi}_{V}^{v}$ $+$ ${\phi}_{V}^{t}$ $+$ ${\phi}_{V}^{s}$,
   and ${\Sigma}_{B}$ $=$ ${\phi}_{B1}$ $+$ ${\phi}_{B2}$.
   The ratio ${\phi}_{i}/{\Sigma}_{j}$ is expressed
   as a percentage.}
   \label{tab:formfactor}
   \begin{ruledtabular}
   \begin{tabular}{c|ccccccc}
     $F_{1}^{B{\to}{\pi}}(0)$
   & ${\phi}_{\pi}^{a}$ & ${\phi}_{\pi}^{p}$ & ${\phi}_{\pi}^{t}$
   & ${\Sigma}_{\pi}$
   & ${\phi}_{\pi}^{a}/{\Sigma}_{\pi}$
   & ${\phi}_{\pi}^{p}/{\Sigma}_{\pi}$
   & ${\phi}_{\pi}^{t}/{\Sigma}_{\pi}$  \\ \hline
     ${\phi}_{B1}$
   & $ 0.064$ & $ 0.106$ & $ 0.019$ & $ 0.188$ & $  34.0$& $  56.0$& $   9.9$ \\
     ${\phi}_{B2}$
   & $ 0.045$ & $-0.003$ & $-0.000$ & $ 0.042$ & $ 107.5$& $  -6.8$& $  -0.7$ \\
     ${\Sigma}_{B}$
   & $ 0.109$ & $ 0.103$ & $ 0.018$ & $ 0.230$ & $  47.4$& $  44.7$& $   8.0$ \\
     ${\phi}_{B2}/{\Sigma}_{B}$
   & $  41.1$ & $  -2.8$ & $  -1.6$ & $  18.1$  \\ \hline \hline
     $F_{1}^{B{\to}K}(0)$
   & ${\phi}_{K}^{a}$ & ${\phi}_{K}^{p}$ & ${\phi}_{K}^{t}$
   & ${\Sigma}_{K}$
   & ${\phi}_{K}^{a}/{\Sigma}_{K}$
   & ${\phi}_{K}^{p}/{\Sigma}_{K}$
   & ${\phi}_{K}^{t}/{\Sigma}_{K}$ \\ \hline
     ${\phi}_{B1}$
   & $ 0.081$ & $ 0.131$ & $ 0.018$ & $ 0.230$ & $  35.3$ & $  56.9$ & $   7.8$  \\
     ${\phi}_{B2}$
   & $ 0.056$ & $-0.004$ & $-0.000$ & $ 0.053$ & $ 107.3$ & $  -6.9$ & $  -0.5$  \\
     ${\Sigma}_{B}$
   & $ 0.138$ & $ 0.127$ & $ 0.018$ & $ 0.282$ & $  48.7$ & $  45.0$ & $   6.3$  \\
     ${\phi}_{B2}/{\Sigma}_{B}$
   & $  41.0$ & $  -2.8$ & $  -1.4$ & $  18.6$  \\ \hline \hline
     $A_{0}^{B{\to}{\rho}}(0)$
   & ${\phi}_{\rho}^{v}$ & ${\phi}_{\rho}^{t}$ & ${\phi}_{\rho}^{s}$
   & ${\Sigma}_{\rho}$
   & ${\phi}_{\rho}^{v}/{\Sigma}_{\rho}$
   & ${\phi}_{\rho}^{t}/{\Sigma}_{\rho}$
   & ${\phi}_{\rho}^{s}/{\Sigma}_{\rho}$ \\ \hline
     ${\phi}_{B1}$
   & $ 0.097$ & $ 0.090$ & $ 0.044$ & $ 0.231$ & $  41.8$ & $  39.1$ & $  19.1$  \\
     ${\phi}_{B2}$
   & $ 0.069$ & $-0.002$ & $-0.001$ & $ 0.067$ & $ 103.6$ & $  -2.7$ & $  -0.9$  \\
     ${\Sigma}_{B}$
   & $ 0.166$ & $ 0.088$ & $ 0.044$ & $ 0.298$ & $  55.7$ & $  29.7$ & $  14.6$  \\
     ${\phi}_{B2}/{\Sigma}_{B}$
   & $  41.8$ & $  -2.0$ & $  -1.4$ & $  22.4$  \\ \hline \hline
     $A_{0}^{B{\to}K^{\ast}}(0)$
   & ${\phi}_{K^{\ast}}^{v}$ & ${\phi}_{K^{\ast}}^{t}$
   & ${\phi}_{K^{\ast}}^{s}$ & ${\Sigma}_{K^{\ast}}$
   & ${\phi}_{K^{\ast}}^{v}/{\Sigma}_{K^{\ast}}$
   & ${\phi}_{K^{\ast}}^{t}/{\Sigma}_{K^{\ast}}$
   & ${\phi}_{K^{\ast}}^{s}/{\Sigma}_{K^{\ast}}$ \\ \hline
     ${\phi}_{B1}$
   & $ 0.098$ & $ 0.106$ & $ 0.052$ & $ 0.256$ & $  38.1$ & $  41.4$ & $  20.5$  \\
     ${\phi}_{B2}$
   & $ 0.070$ & $-0.003$ & $-0.001$ & $ 0.067$ & $ 104.5$ & $  -3.7$ & $  -0.8$  \\
     ${\Sigma}_{B}$
   & $ 0.168$ & $ 0.104$ & $ 0.052$ & $ 0.323$ & $  52.0$ & $  32.0$ & $  16.0$  \\
     ${\phi}_{B2}/{\Sigma}_{B}$
   & $  42.0$ & $  -2.4$ & $  -1.1$ & $  20.9$
   \end{tabular}
   \end{ruledtabular}
   \end{table}

   It is well known that two form factors, $F_{1}(q^{2})$
   and $A_{0}(q^{2})$ corresponding to the vector and axial-vector
   currents of the weak interactions, respectively,
   are directly related to $B$ ${\to}$ $PV$ decays.
   The detailed definitions and explicit expressions of form factors
   can be found in Ref. \cite{epjc.28.515}, where the contributions
   from the higher twist DAs are considered properly.
   The dependences of form factors on certain input parameters
   are shown in Fig. \ref{contour:formfactor}.
   It is easily seen from Fig. \ref{contour:formfactor} that
   (1)
   the form factors $F_{1}$ and $A_{0}$ are highly sensitive to
   the shape parameter ${\omega}_{B}$ of $B$ mesonic WFs and
   the contributions from ${\phi}_{B2}$.
   In general, the values of the form factors $F_{1}$ and $A_{0}$
   decrease with increasing ${\omega}_{B}$.
   This type of regular phenomenon has also been also found in
   previous studies \cite{PhysRevD.71.034018,PhysRevD.103.056006,
   PhysRevD.74.014027,npb.625.239,npb.642.263,PhysRevD.64.112002}.
   (2)
   In addition, the form factors $F_{1}$ are also dependent on
   the value of the chiral parameter ${\mu}_{P}$.
   For a more comprehensive analysis, the numerical results
   of the form factors with specific inputs are listed in
   Table \ref{tab:formfactor}.
   It is clear from Table \ref{tab:formfactor} that
   (1)
   when the contributions from ${\phi}_{B2}$ are not considered,
   the total shares of the formfactors $F_{1}$ from the twist-3 DAs
   ${\phi}_{P}^{p,t}$ of the recoiled light pseudoscalar meson far
   outweigh those from the leading twist DAs ${\phi}_{P}^{a}$,
   and account for more than $60\%$.
   The total shares of the formfactors $A_{0}$ from the twist-3 DAs
   ${\phi}_{V}^{t,s}$ of the recoiled vector meson, which is
   approximately $60\%$, far exceeds those from the twist-2 DAs
   ${\phi}_{V}^{v}$.
   (2)
   When only the contributions from the twist-2 DAs ${\phi}_{P}^{a}$
   and ${\phi}_{V}^{v}$ are considered, the shares of the formfactors
   $F_{1}$ and $A_{0}$ from the $B$ mesonic WFs ${\phi}_{B2}$
   are approximately $40\%$.
   (3) When the contributions from both the twist-2 and twist-3 DAs
   ${\phi}_{P}^{a,p,t}$ and  ${\phi}_{V}^{v,t,s}$ are considered,
   the shares of the formfactors $F_{1}$ and $A_{0}$ from the $B$
   mesonic WFs ${\phi}_{B2}$ are about $20\%$.
   The contributions from ${\phi}_{B2}$ to the formfactors have
   been investigated in previous studies \cite{PhysRevD.71.034018,
   PhysRevD.103.056006, PhysRevD.74.014027,epjc.28.515,npb.625.239,
   npb.642.263,PhysRevD.89.094004}.
   The general consensus seems to be that the
   unnegligible contributions
   from ${\phi}_{B2}$ to the formfactors should be given due
   attention.
   Here, we would like to point out that for the arguments on
   the reliability of the perturbative calculation of the form
   factors using the PQCD approach, which is not the focus
   of this study, one can refer to detailed analyses,
   for example, in Refs.
   \cite{epjc.28.515,npb.625.239,npb.642.263}.

   \section{branching ratios and $CP$ violating asymmetries}
   \label{sec:branch}
   According to the above analysis, it is clear that the
   contributions from higher twist DAs are important to HMEs
   for nonleptonic $B$ decays using the PQCD approach.
   In most phenomenological studies of the $B$ ${\to}$ $PV$
   decays with the PQCD approach, the shares of both the twist-2
   DAs (${\phi}_{P}^{a}$ and ${\phi}_{V}^{v}$) and twist-3 DAs
   (${\phi}_{P}^{p,t}$ and ${\phi}_{V}^{t,s}$) for the final mesons
   have been carefully and commonly considered, such as in Refs.
   \cite{PhysRevD.64.112002,epjc.23.275,epjc.72.1923,PhysRevD.90.074018,   PhysRevD.75.014019,epjc.59.49,PhysRevD.104.016025,PhysRevD.74.094020}.
   In contrast, the possible influence of the $B$ mesonic WFs
   ${\phi}_{B}^{-}$ or ${\phi}_{B2}$ on nonleptonic $B$ decays
   garners significantly less attention.
   In this paper, our main purpose is to investigate the effects
   of $B$ mesonic WFs ${\phi}_{B2}$ on the $B$ ${\to}$ $PV$ decays
   using the PQCD approach.
   \begin{figure}[h]
   \includegraphics[width=0.21\textwidth,bb=180 520 320 660]{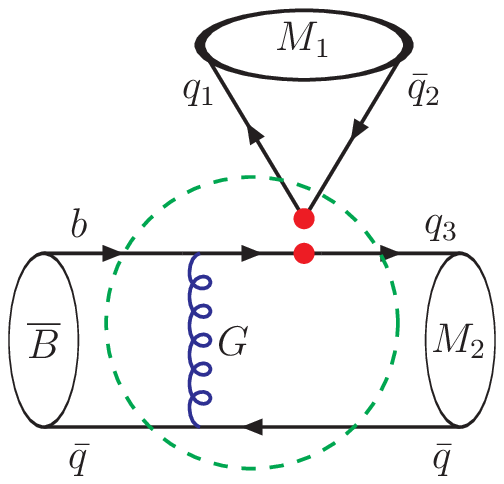}\quad
   \includegraphics[width=0.21\textwidth,bb=180 520 320 660]{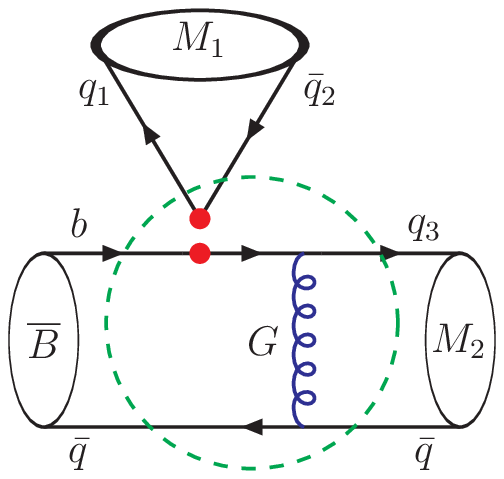}\quad
   \includegraphics[width=0.21\textwidth,bb=180 520 320 660]{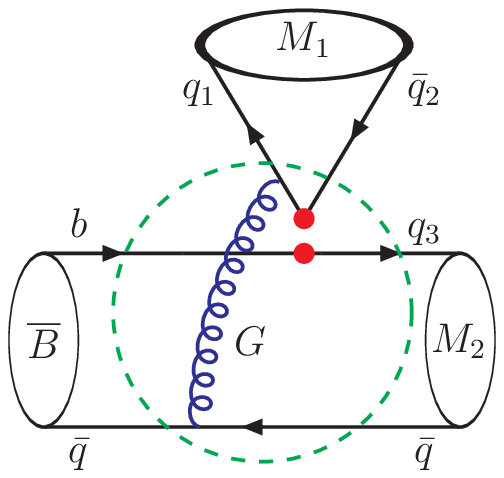}\quad
   \includegraphics[width=0.21\textwidth,bb=180 520 320 660]{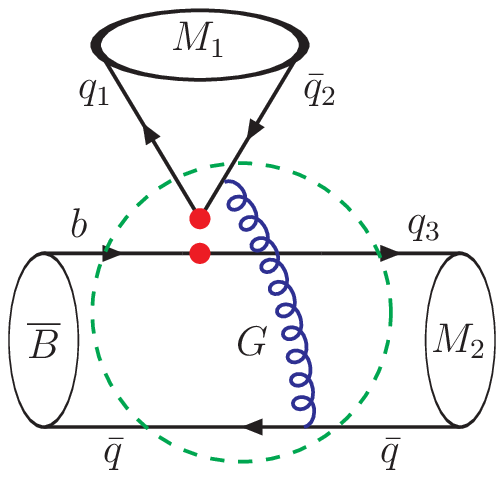}
   \\  \vspace{-3mm} (a)
       \hspace{0.19\textwidth} (b)
       \hspace{0.19\textwidth} (c)
       \hspace{0.19\textwidth} (d)
   \\ \vspace{5mm}
   \includegraphics[width=0.21\textwidth,bb=200 525 340 640]{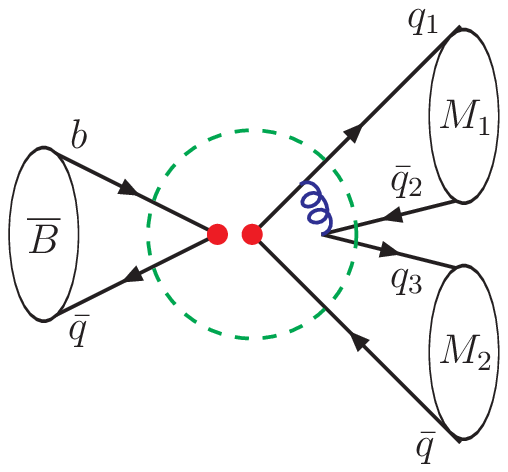}\quad
   \includegraphics[width=0.21\textwidth,bb=200 525 340 640]{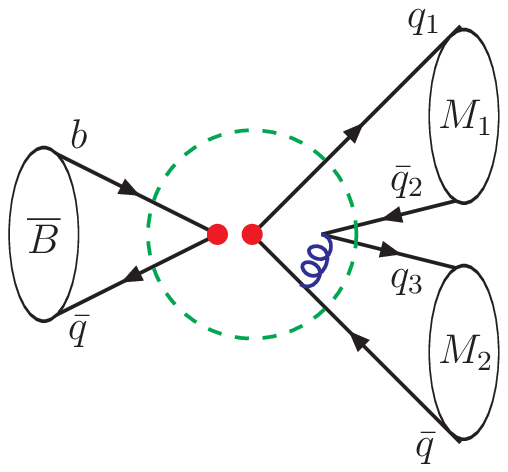}\quad
   \includegraphics[width=0.21\textwidth,bb=200 525 340 640]{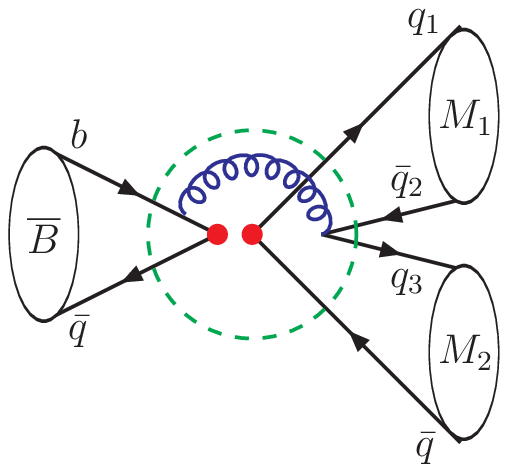}\quad
   \includegraphics[width=0.21\textwidth,bb=200 525 340 640]{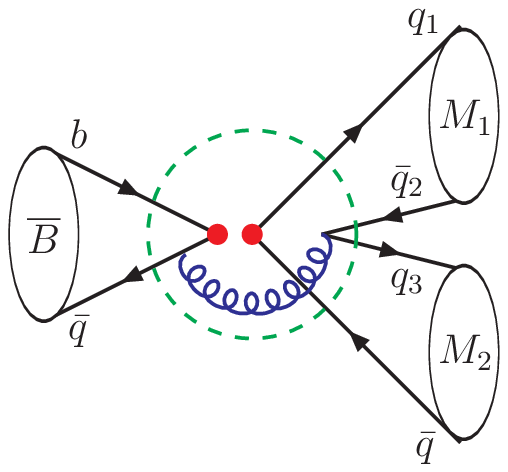}
   \\  \vspace{-3mm} (e)
       \hspace{0.19\textwidth} (f)
       \hspace{0.19\textwidth} (g)
       \hspace{0.19\textwidth} (h)
   \caption{Feynman diagrams contributing to the $\overline{B}$
    ${\to}$ $PV$ decays with the PQCD approach,
    where $M_{1,2}$ $=$ $P$ and $V$,
    the dots denote appropriate interactions and the dashed
    circles represent the scattering amplitudes.
    (a) and (b) are factorizable emission diagrams.
    (c) and (d) are nonfactorizable emission diagrams.
    (e) and (f) are factorizable annihilation diagrams.
    (g) and (h) are nonfactorizable annihilation diagrams.}
   \label{fig:feynman}
   \end{figure}

   The leading order Feynman diagrams are shown in
   Fig. \ref{fig:feynman}.
   The analytical expressions of each subdiagram amplitude are
   listed in Appendix \ref{sec:block}.
   It is clearly seen that
   (1)
   for the factorizable annihilation diagrams (e) and (f),
   the initial $B$ meson is completely disconnected from the final
   state $PV$ system, where the disconnected $B$ meson corresponds
   to its decay constant and should have nothing to do with its
   WFs ${\phi}_{B2}$. These arguments are fully verified by
   Eqs.(\ref{amp-epv-left}-\ref{amp-fvp-sp}).
   (2)
   For the emission diagrams (a-d) and the nonfactorizable
   annihilation diagrams (g-h), the $B$ meson always connects
   with either one or two of the final states via the
   one-gluon-exchange interactions.
   Therefore, these corresponding amplitudes would generally
   be affected by the $B$ mesonic WFs ${\phi}_{B2}$ and should be
   updated accordingly.

   The decay amplitudes for $B$ ${\to}$ $PV$ decays with the
   PQCD approach are expressed as the sum of a series of
   multidimensional convolutions,
   \begin{eqnarray} & &
  {\cal A}(B{\to}PV)\, =\,
  {\langle}PV{\vert}{\cal H}_{\rm eff}{\vert}B{\rangle}
   \nonumber \\ &=&
   \frac{G_{F}}{\sqrt{2}}\,\sum\limits_{i}\,{\cal F}_{i}
  {\int}\,dx_{1}\,dx_{2}\,dx_{3}\,db_{1}\,db_{2}\,db_{3}\,
  {\cal T}_{i}(t_{i},x_{1},b_{1},x_{2},b_{2},x_{3},b_{3})
   \nonumber \\ & & \qquad
   C_{i}(t_{i})\,
  {\Phi}_{B}(x_{1},b_{1})\,e^{-S_{B}}\,
  {\Phi}_{P}(x_{2},b_{2})\,e^{-S_{P}}\,
  {\Phi}_{V}(x_{3},b_{3})\,e^{-S_{V}}
   \label{eq:amplitude},
   \end{eqnarray}
   where ${\cal F}_{i}$ is the CKM factor, and
   the rescattering functions ${\cal T}_{i}$ are represented
   by the dashed circles in Fig. \ref{fig:feynman}.
   The calculation expressions for the $B$ ${\to}$ $PV$ decays
   are listed in detail in Appendix \ref{sec:mode}.

   In the rest frame of the $B$ meson, the $CP$-averaged
   branching ratios are defined as,
   \begin{equation}
  {\cal B}\, =\,
   \frac{{\tau}_{B}}{16{\pi}}\,
   \frac{p_{\rm cm}}{m_{B}^{2}}\, \big\{
  {\vert}{\cal A}(B{\to}f){\vert}^{2}+
  {\vert}{\cal A}(\overline{B}{\to}\bar{f}){\vert}^{2} \big\}
   \label{branching-ratio-definition},
   \end{equation}
  where ${\tau}_{B}$ is the lifetime of the $B$ meson,
  ${\tau}_{B_{u}}$ $=$ $1.638(4)$ ps, and
  ${\tau}_{B_{d}}$ $=$ $1.519(4)$ ps \cite{pdg2020}.
  $p_{\rm cm}$ is the common center-of-mass momentum of
  final states.

  For the charged $B_{u}$ meson decays, the direct
  $CP$ violating asymmetry arising from interferences
  among different amplitudes is defined as,
   \begin{equation}
  {\cal A}_{CP}\, =\,
   \frac{{\Gamma}(B^{-}{\to}f)-{\Gamma}(B^{+}{\to}\bar{f})}
        {{\Gamma}(B^{-}{\to}f)+{\Gamma}(B^{+}{\to}\bar{f})}
   \, =\,
   \frac{{\vert}\mathcal{A}(B^{-}{\to}f){\vert}^{2}
        -{\vert}\mathcal{A}(B^{+}{\to}\bar{f}){\vert}^{2}}
        {{\vert}\mathcal{A}(B^{-}{\to}f){\vert}^{2}
        +{\vert}\mathcal{A}(B^{+}{\to}\bar{f}){\vert}^{2}}
   \label{direct-CP-definition-01}.
   \end{equation}

  For the neutral $B_{d}$ meson decays, the effects of the
  $B^{0}$-$\overline{B}^{0}$ mixing should be considered.
  The time-dependent $CP$ violating asymmetry is
  defined as:
   \begin{equation}
  {\cal A}_{CP}(t)\, =\,
   \frac{ {\Gamma}(\overline{B}^{0}(t){\to}f)
        - {\Gamma}(B^{0}(t){\to}\bar{f})}
        { {\Gamma}(\overline{B}^{0}(t){\to}f)
        + {\Gamma}(B^{0}(t){\to}\bar{f})}
   \label{time-dependent-CP-asymmetry-01}.
   \end{equation}
  The $CP$ violating asymmetries can, in principle,
  be divided into three cases according to the final states
  \cite{pdg2020,PhysRevD.59.014005,PhysRevD.65.094025}.
  For the sake of simplification, the following
  conventional symbols will be defined and used,
  {\em i.e.},
   \begin{equation}
   A_{f}\, =\, {\cal A}(B^{0}(0){\to}f),
   \qquad
   \bar{A}_{f}\, =\, {\cal A}(\overline{B}^{0}(0){\to}f)
   \label{pdr59.014005-eq.36-a},
   \end{equation}
   \begin{equation}
   A_{\bar{f}}\, =\, {\cal A}(B^{0}(0){\to}\bar{f}),
   \qquad
   \bar{A}_{\bar{f}}\, =\, {\cal A}(\overline{B}^{0}(0){\to}\bar{f})
   \label{pdr59.014005-eq.36-b}.
   \end{equation}

  \begin{itemize}
  \item case 1:
  The final states come from either $B^{0}$ decays or
  $\overline{B}^{0}$ decays, but not both, {\em i.e.},
  $\overline{B}^{0}$ ${\to}$ $f$ and $B^{0}$ ${\to}$ $\bar{f}$
  with $f$ ${\ne}$ $\bar{f}$, for example, the
  $\overline{B}^{0}$ ${\to}$ ${\pi}^{+}K^{{\ast}-}$ decay.
  The $CP$ asymmetries are immune to the
  $B^{0}$-$\overline{B}^{0}$ mixing,
  and have a similar definition as to the direct $CP$ asymmetry
  in Eq.(\ref{direct-CP-definition-01}).
  \item case 2:
  The final states are the eigenstates of the $CP$ transformation,
  {\em i.e.}, $f^{CP}$ $=$ ${\eta}_{f}\,\bar{f}$ with the
  eigenvalue ${\vert}{\eta}_{f}{\vert}$ $=$ $1$.
  The final states can come from both $B^{0}$ decays and
  $\overline{B}^{0}$ decays, {\em i.e.}, $\overline{B}^{0}$
  ${\to}$ $f$ ${\gets}$ $B^{0}$, for example,
  the $\overline{B}^{0}$ ${\to}$ ${\pi}^{0}{\rho}^{0}$ decay.

  For the $B^{0}$-$\overline{B}^{0}$ mixing, SM predicts that
  the ratio of the decay width difference ${\Delta}{\Gamma}$
  of mass eigenstates to the total decay width ${\Gamma}$ is small,
  that is, ${\Delta}{\Gamma}/{\Gamma}$ $=$ $0.001{\pm}0.010$
  from data \cite{pdg2020}.
  In the most general calculation, it is usually assumed that
  ${\Delta}{\Gamma}$ $=$ $0$; thus the $CP$ asymmetries
  can be expressed as \cite{pdg2020}:
   \begin{equation}
  {\cal A}_{CP}(t)\, =\,
   S_{f}\,{\sin}({\Delta}m\,t)
  -C_{f}\,{\cos}({\Delta}m\,t)
   \label{mixing-CP-asymmetry-06},
   \end{equation}
   \begin{equation}
   S_{f}\, =\,
   \frac{ 2\,{\cal I}m({\lambda}_{f}) }
        { 1+{\vert}{\lambda}_{f}{\vert}^{2} },
   \qquad
   C_{f}\, =\,
   \frac{ 1-{\vert}{\lambda}_{f}{\vert}^{2} }
        { 1+{\vert}{\lambda}_{f}{\vert}^{2} },
   \qquad
   {\lambda}_{f}\, =\,
   \frac{ q }{ p }\, \frac{ \bar{A}_{f} }{ A_{f} }
   \label{mixing-CP-asymmetry-definition-02},
   \end{equation}
  where $q/p$ $=$ $V_{tb}^{\ast}\,V_{td}/V_{tb}\,V_{td}^{\ast}$
  describes the $B^{0}$-$\overline{B}^{0}$ mixing.
  Sometimes, the time-integrated $CP$ asymmetries are written as:
   \begin{eqnarray}
  {\cal A}_{CP} &=&
   \frac{  \displaystyle
          {\int}_{0}^{\infty}dt\,{\Gamma}(\overline{B}^{0}(t){\to}f)
        - {\int}_{0}^{\infty}dt\,{\Gamma}(B^{0}(t){\to}\bar{f})  }
        { \displaystyle
          {\int}_{0}^{\infty}dt\,{\Gamma}(\overline{B}^{0}(t){\to}f)
        + {\int}_{0}^{\infty}dt\,{\Gamma}(B^{0}(t){\to}\bar{f})  }
   \label{time-dependent-CP-asymmetry-02}
   \\ &=&
   \frac{ x }{ 1+x^{2} }\,S_{f}
  -\frac{ 1 }{ 1+x^{2} }\,C_{f}
   \label{time-dependent-CP-asymmetry-03},
   \end{eqnarray}
  with $x$ $=$ ${\Delta}m/{\Gamma}$ $=$ $0.769(4)$ \cite{pdg2020}
  for the $B^{0}$-$\overline{B}^{0}$ system,
  where ${\Delta}m$ $=$ $0.5065(19)\,{\rm ps}^{-1}$ \cite{pdg2020}
  is the mass difference of the mass eigenstates.
  \item case 3:
  The final states are not the eigenstates of the $CP$ transformation,
  however, both $f$ and $\bar{f}$ are the common
  final states of $\overline{B}^{0}$ and $B^{0}$, {\em i.e.},
  $\overline{B}^{0}$ ${\to}$ ($f$ \& $\bar{f}$) ${\gets}$ $B^{0}$,
  for example, the $\overline{B}^{0}$ ${\to}$ ${\pi}^{+}{\rho}^{-}$,
  ${\pi}^{-}{\rho}^{+}$ decays.

  The four time-dependent partial decay widths can be expressed
  as \cite{PhysRevD.59.014005,PhysRevD.65.094025}:
   \begin{equation}
  {\Gamma}(B^{0}(t){\to}f)\, =\,
   \frac{1}{2}\, e^{-{\Gamma}\,t}\,
   \big( {\vert}A_{f}{\vert}^{2}
        +{\vert}\bar{A}_{f}{\vert}^{2} \big)\, \Big\{ 1
  +a_{{\epsilon}^{\prime}}\,{\cos}({\Delta}m\,t)
  +a_{{\epsilon}+{\epsilon}^{\prime}}\, {\sin}({\Delta}m\,t) \Big\}
   \label{pdr59.014005-eq.39-a},
   \end{equation}
   \begin{equation}
  {\Gamma}(B^{0}(t){\to}\bar{f})\, =\,
   \frac{1}{2}\, e^{-{\Gamma}\,t}\,
   \big( {\vert}\bar{A}_{\bar{f}}{\vert}^{2}
        +{\vert}A_{\bar{f}}{\vert}^{2} \big)\, \Big\{ 1
   +\bar{a}_{{\epsilon}^{\prime}}\,{\cos}({\Delta}m\,t)
   +\bar{a}_{{\epsilon}+{\epsilon}^{\prime}}\,
  {\sin}({\Delta}m\,t) \Big\}
   \label{pdr59.014005-eq.39-c},
   \end{equation}
   \begin{equation}
  {\Gamma}(\overline{B}^{0}(t){\to}f)\, =\,
   \frac{1}{2}\, e^{-{\Gamma}\,t}\,
   \big( {\vert}A_{f}{\vert}^{2}
        +{\vert}\bar{A}_{f}{\vert}^{2} \big)\, \Big\{ 1
  -a_{{\epsilon}^{\prime}}\,{\cos}({\Delta}m\,t)
  -a_{{\epsilon}+{\epsilon}^{\prime}}\, {\sin}({\Delta}m\,t) \Big\}
   \label{pdr59.014005-eq.39-d},
   \end{equation}
   \begin{equation}
  {\Gamma}(\overline{B}^{0}(t){\to}\bar{f}) \, =\,
   \frac{1}{2}\, e^{-{\Gamma}\,t}\,
   \big( {\vert}\bar{A}_{\bar{f}}{\vert}^{2}
        +{\vert}A_{\bar{f}}{\vert}^{2} \big)\, \Big\{ 1
   -\bar{a}_{{\epsilon}^{\prime}}\,{\cos}({\Delta}m\,t)
   -\bar{a}_{{\epsilon}+{\epsilon}^{\prime}}\,
  {\sin}({\Delta}m\,t) \Big\}
   \label{pdr59.014005-eq.39-b},
   \end{equation}
   with the following definitions,
   \begin{equation}
   a_{{\epsilon}^{\prime}}\, =\,
    \frac{ 1-{\vert}{\lambda}_{f}{\vert}^{2} }
         { 1+{\vert}{\lambda}_{f}{\vert}^{2} },
   \qquad
   a_{{\epsilon}+{\epsilon}^{\prime}}\, =\,
   \frac{ -2\,{\cal I}m({\lambda}_{f}) }
        { 1+{\vert}{\lambda}_{f}{\vert}^{2} },
   \qquad
   {\lambda}_{f}\, =\,
    \frac{ V_{tb}^{\ast}\,V_{td} }{ V_{tb}\,V_{td}^{\ast} }\,
    \frac{ \bar{A}_{f} }{ A_{f} }
   \label{pdr59.014005-eq.40-a},
   \end{equation}
   \begin{equation}
   \bar{a}_{{\epsilon}^{\prime}}\, =\,
    \frac{ 1-{\vert}\bar{\lambda}_{f}{\vert}^{2} }
         { 1+{\vert}\bar{\lambda}_{f}{\vert}^{2} },
   \qquad
   \bar{a}_{{\epsilon}+{\epsilon}^{\prime}}\, =\,
   \frac{ -2\,{\cal I}m(\bar{\lambda}_{f}) }
        { 1+{\vert}\bar{\lambda}_{f}{\vert}^{2} },
   \qquad
   \bar{\lambda}_{f}\, =\,
    \frac{ V_{tb}^{\ast}\,V_{td} }{ V_{tb}\,V_{td}^{\ast} }\,
    \frac{ \bar{A}_{\bar{f}} }{ A_{\bar{f}} }
   \label{pdr59.014005-eq.40-b}.
   \end{equation}
   Besides ${\cal A}_{CP}$ in Eq.(\ref{time-dependent-CP-asymmetry-03}),
   the $CP$ asymmetries can also be
   expressed by the physical quantities $a_{{\epsilon}^{\prime}}$,
   $a_{{\epsilon}+{\epsilon}^{\prime}}$,
   $\bar{a}_{{\epsilon}^{\prime}}$ and
   $\bar{a}_{{\epsilon}+{\epsilon}^{\prime}}$.
   \end{itemize}
   \begin{figure}[ht]
   \includegraphics[width=0.3\textwidth]{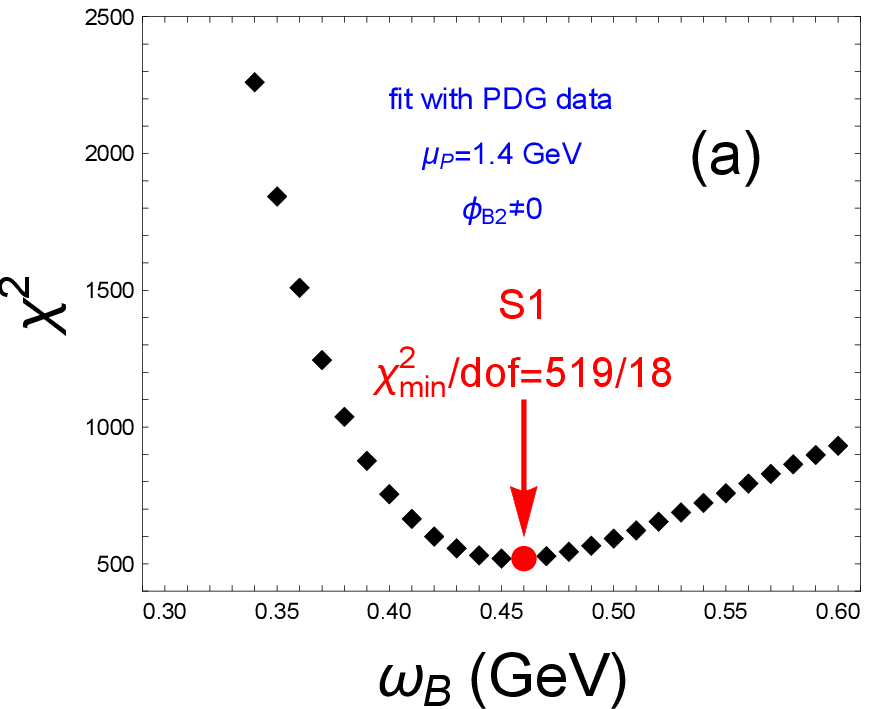}   \quad
   \includegraphics[width=0.3\textwidth]{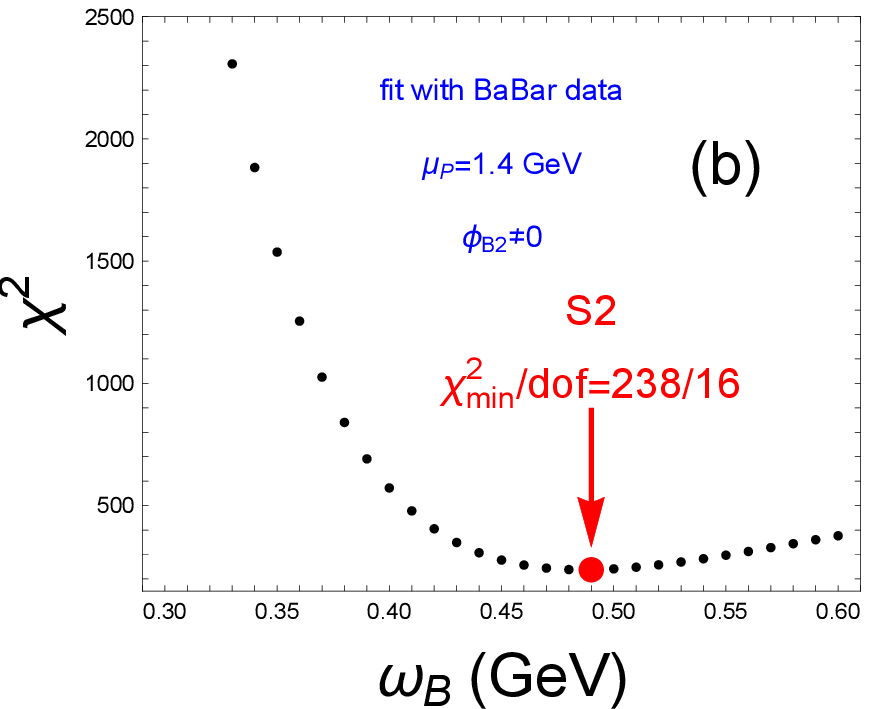} \quad
   \includegraphics[width=0.3\textwidth]{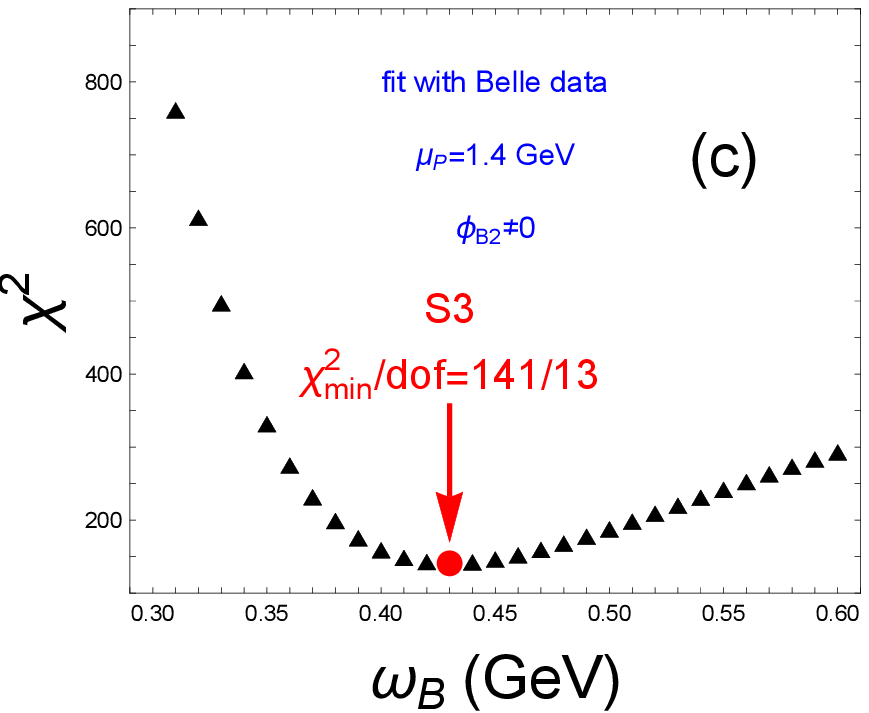}
   \caption{Distribution of ${\chi}^{2}$ vs the shape parameter
   ${\omega}_{B}$, where the red points at the arrowheads
   correspond to the optimal values.}
   \label{fig:chi-w}
   \end{figure}

   According to the previous analysis of the formfactors in Fig.
   \ref{contour:formfactor}, it is natural to suppose that the
   theoretical results of the branching ratios would be strongly
   dependent on the shape parameter ${\omega}_{B}$.
   In this paper, we optimize the parameter ${\omega}_{B}$
   using the minimum ${\chi}^{2}$ method,
   \begin{equation}
  {\chi}^{2}\, =\, \sum\limits_{i}{\chi}^{2}_{i}\, =\, \sum\limits_{i}
   \frac{ ({\cal B}_{i}^{\rm th.}-{\cal B}_{i}^{\rm exp.})^{2} }
        { {\sigma}_{i}^{2} }
   \label{eq:chi2},
   \end{equation}
   where ${\cal B}_{i}^{\rm th.}$ and ${\cal B}_{i}^{\rm exp.}$
   denote the theoretical results and experimental data on the branching
   ratio, respectively. ${\sigma}_{i}$ denotes the errors of
   experimental measurements.
   The distribution of ${\chi}^{2}$ vs the shape parameter
   ${\omega}_{B}$ is shown in Fig. \ref{fig:chi-w},
   where the contributions from the $B$ mesonic WFs ${\phi}_{B2}$
   are considered.
   Three optimal scenarios of the shape parameter ${\omega}_{B}$
   corresponding to experimental data from the PDG, BaBar
   and Belle groups
   are obtained with the chiral mass ${\mu}_{P}$ $=$ $1.4$ GeV,
   {\em i.e.},
   \begin{itemize}
   \item scenario 1 (S1): ${\omega}_{B}$ $=$ $0.46$ GeV from PDG data
             with ${\chi}^{2}_{\rm min.}$/dof ${\approx}$
             $519/18$ ${\approx}$ $29$,
   \item scenario 2 (S2): ${\omega}_{B}$ $=$ $0.49$ GeV from BaBar data
             with ${\chi}^{2}_{\rm min.}$/dof ${\approx}$
             $238/16$ ${\approx}$ $15$,
   \item scenario 3 (S3): ${\omega}_{B}$ $=$ $0.43$ GeV from Belle data
             with ${\chi}^{2}_{\rm min.}$/dof ${\approx}$
             $141/13$ ${\approx}$ $11$.
   \end{itemize}

   As is well known, the errors of the PDG group from a weighted
   average of selected data are generally smaller than those of any
   independent experimental groups. Therefore,
   it is clear from Eq.(\ref{eq:chi2}) that the relatively
   smaller (larger) errors of the PDG (Belle) data result in the
   relatively larger (smaller) value of ${\chi}^{2}_{\rm min.}$/dof.

  {\renewcommand{\baselinestretch}{1.25}
   \begin{table}[ht]
   \caption{The $CP$-averaged branching ratios (in the
   unit of $10^{-6}$) for the $B_{u}$ ${\to}$ $PV$ decays.
   The central theoretical values are calculated
   with three scenario parameters of ${\omega}_{B}$ to compare
   with data from the PDG, BaBar and  Belle groups \cite{pdg2020}.
   The first theoretical uncertainties arise
   from the variations of ${\omega}_{B}$ $=$
   $0.46{\pm}0.01$ GeV for scenario S1,
   ${\omega}_{B}$ $=$ $0.49{\pm}0.01$ GeV for scenario S2 and
   ${\omega}_{B}$ $=$ $0.43{\pm}0.01$ GeV for scenario S3, respectively.
   The second theoretical uncertainties come from the
   variations of ${\mu}_{P}$ $=$ $1.4{\pm}0.1$ GeV. }
   \label{tab:branch-bu}
   \begin{ruledtabular}
   \begin{tabular}{cccccc}
     \multicolumn{2}{c}{mode}
   & $B^{-}$ ${\to}$ ${\pi}^{-}{\rho}^{0}$
   & $B^{-}$ ${\to}$ ${\pi}^{0}{\rho}^{-}$
   & $B^{-}$ ${\to}$ ${\pi}^{-}{\omega}$
   & $B^{-}$ ${\to}$ ${\pi}^{-}{\phi}$  \\ \hline
     data & PDG
   & $8.3{\pm}1.2$
   & $10.9{\pm}1.4$
   & $6.9{\pm}0.5$
   & $(3.2{\pm}1.5){\times}10^{-2}$ \\
     \multirow{2}{*}{S1}
   & ${\phi}_{B1}$ + ${\phi}_{B2}$
   & $   4.25^{+  0.22 +  0.01 }_{-  0.21 -  0.01 }$
   & $   6.91^{+  0.39 +  0.46 }_{-  0.37 -  0.45 }$
   & $   3.83^{+  0.20 +  0.01 }_{-  0.19 -  0.01 }$
   & $    4.9^{+   0.4 +   0.6 }_{-   0.4 -   0.5 }{\times}10^{-2}$ \\
   & ${\phi}_{B1}$
   & $   2.66^{+  0.15 +  0.01 }_{-  0.14 -  0.01 }$
   & $   4.56^{+  0.28 +  0.38 }_{-  0.26 -  0.36 }$
   & $   2.47^{+  0.14 +  0.01 }_{-  0.13 -  0.01 }$
   & $    4.2^{+   0.3 +   0.5 }_{-   0.3 -   0.5 }{\times}10^{-2}$ \\ \hline
     data & BaBar
   & $8.1{\pm}1.7$
   & $10.2{\pm}1.7$
   & $6.7{\pm}0.6$
   & \\
     \multirow{2}{*}{S2}
   & ${\phi}_{B1}$ + ${\phi}_{B2}$
   & $   3.66^{+  0.19 +  0.01 }_{-  0.18 -  0.00 }$
   & $   5.87^{+  0.32 +  0.38 }_{-  0.30 -  0.37 }$
   & $   3.28^{+  0.17 +  0.01 }_{-  0.16 -  0.01 }$
   & $    3.9^{+   0.3 +   0.4 }_{-   0.3 -   0.4 }{\times}10^{-2}$ \\
   & ${\phi}_{B1}$
   & $   2.26^{+  0.12 +  0.01 }_{-  0.12 -  0.01 }$
   & $   3.83^{+  0.22 +  0.31 }_{-  0.21 -  0.30 }$
   & $   2.10^{+  0.12 +  0.01 }_{-  0.11 -  0.01 }$
   & $    3.3^{+   0.3 +   0.4 }_{-   0.3 -   0.4 }{\times}10^{-2}$ \\ \hline
     data & Belle
   & $8.0{\pm}2.4$
   & $13.2{\pm}3.0$
   & $6.9{\pm}0.8$
   & \\
     \multirow{2}{*}{S3}
   & ${\phi}_{B1}$ + ${\phi}_{B2}$
   & $   4.96^{+  0.26 +  0.01 }_{-  0.25 -  0.00 }$
   & $   8.17^{+  0.48 +  0.56 }_{-  0.45 -  0.54 }$
   & $   4.47^{+  0.24 +  0.01 }_{-  0.23 -  0.01 }$
   & $    6.2^{+   0.5 +   0.7 }_{-   0.5 -   0.7 }{\times}10^{-2}$ \\
   & ${\phi}_{B1}$
   & $   3.13^{+  0.18 +  0.01 }_{-  0.17 -  0.01 }$
   & $   5.45^{+  0.34 +  0.46 }_{-  0.32 -  0.44 }$
   & $   2.91^{+  0.17 +  0.01 }_{-  0.16 -  0.01 }$
   & $    5.3^{+   0.4 +   0.6 }_{-   0.4 -   0.6 }{\times}10^{-2}$ \\ \hline \hline
     \multicolumn{2}{c}{mode}
   & $B^{-}$ ${\to}$ $K^{-}K^{{\ast}0}$
   & $B^{-}$ ${\to}$ $K^{0}K^{{\ast}-}$
   & $B^{-}$ ${\to}$ ${\pi}^{0}K^{{\ast}-}$
   & $B^{-}$ ${\to}$ $K^{-}{\rho}^{0}$ \\ \hline
     data & PDG
   & $0.59{\pm}0.08$
   &
   & $6.8{\pm}0.9$
   & $3.7{\pm}0.5$  \\
     \multirow{2}{*}{S1}
   & ${\phi}_{B1}$ + ${\phi}_{B2}$
   & $   0.35^{+  0.02 +  0.05 }_{-  0.02 -  0.05 }$
   & $    4.7^{+   0.1 +   1.3 }_{-   0.1 -   0.8 }{\times}10^{-2}$
   & $   2.90^{+  0.19 +  0.22 }_{-  0.18 -  0.22 }$
   & $   1.04^{+  0.01 +  0.03 }_{-  0.01 -  0.01 }$ \\
   & ${\phi}_{B1}$
   & $   0.24^{+  0.02 +  0.04 }_{-  0.01 -  0.04 }$
   & $    4.8^{+   0.1 +   1.3 }_{-   0.1 -   1.0 }{\times}10^{-2}$
   & $   1.96^{+  0.14 +  0.19 }_{-  0.13 -  0.18 }$
   & $   0.88^{+  0.01 +  0.03 }_{-  0.01 -  0.01 }$ \\ \hline
     data & BaBar
   &
   &
   & $6.4{\pm}1.0$
   & $3.56{\pm}0.73$  \\
     \multirow{2}{*}{S2}
   & ${\phi}_{B1}$ + ${\phi}_{B2}$
   & $   0.29^{+  0.02 +  0.04 }_{-  0.02 -  0.04 }$
   & $    4.4^{+   0.1 +   1.0 }_{-   0.1 -   0.7 }{\times}10^{-2}$
   & $   2.40^{+  0.15 +  0.18 }_{-  0.14 -  0.18 }$
   & $   1.01^{+  0.01 +  0.02 }_{-  0.01 -  0.01 }$ \\
   & ${\phi}_{B1}$
   & $   0.19^{+  0.01 +  0.03 }_{-  0.01 -  0.03 }$
   & $    4.5^{+   0.1 +   1.1 }_{-   0.1 -   0.8 }{\times}10^{-2}$
   & $   1.61^{+  0.11 +  0.15 }_{-  0.10 -  0.15 }$
   & $   0.86^{+  0.01 +  0.02 }_{-  0.01 -  0.01 }$ \\ \hline
     data & Belle
   &
   &
   &
   & $3.89{\pm}0.64$  \\
     \multirow{2}{*}{S3}
   & ${\phi}_{B1}$ + ${\phi}_{B2}$
   & $   0.42^{+  0.03 +  0.06 }_{-  0.03 -  0.06 }$
   & $    5.1^{+   0.2 +   1.6 }_{-   0.1 -   1.0 }{\times}10^{-2}$
   & $   3.52^{+  0.24 +  0.27 }_{-  0.22 -  0.26 }$
   & $   1.08^{+  0.02 +  0.03 }_{-  0.01 -  0.01 }$ \\
   & ${\phi}_{B1}$
   & $   0.29^{+  0.02 +  0.05 }_{-  0.02 -  0.05 }$
   & $    5.3^{+   0.2 +   1.6 }_{-   0.2 -   1.2 }{\times}10^{-2}$
   & $   2.41^{+  0.18 +  0.23 }_{-  0.16 -  0.22 }$
   & $   0.91^{+  0.01 +  0.03 }_{-  0.01 -  0.01 }$ \\ \hline \hline
     \multicolumn{2}{c}{mode}
   & $B^{-}$ ${\to}$ $K^{-}{\omega}$
   & $B^{-}$ ${\to}$ ${\pi}^{-}\overline{K}^{{\ast}0}$
   & $B^{-}$ ${\to}$ $\overline{K}^{0}{\rho}^{-}$
   & $B^{-}$ ${\to}$ $K^{-}{\phi}$ \\ \hline
     data & PDG
   & $6.5{\pm}0.4$
   & $10.1{\pm}0.8$
   & $7.3{\pm}1.2$
   & $8.8{\pm}0.7$  \\
     \multirow{2}{*}{S1}
   & ${\phi}_{B1}$ + ${\phi}_{B2}$
   & $   3.43^{+  0.17 +  0.26 }_{-  0.16 -  0.30 }$
   & $   4.44^{+  0.31 +  0.44 }_{-  0.28 -  0.42 }$
   & $   1.36^{+  0.00 +  0.12 }_{-  0.00 -  0.05 }$
   & $  13.48^{+  0.96 +  2.32 }_{-  0.89 -  2.27 }$ \\
   & ${\phi}_{B1}$
   & $   2.76^{+  0.13 +  0.28 }_{-  0.12 -  0.30 }$
   & $   3.04^{+  0.22 +  0.36 }_{-  0.20 -  0.34 }$
   & $   1.36^{+  0.01 +  0.15 }_{-  0.01 -  0.09 }$
   & $   9.50^{+  0.73 +  1.94 }_{-  0.67 -  1.87 }$ \\ \hline
     data & BaBar
   & $6.3{\pm}0.6$
   & $10.1{\pm}2.0$
   & $6.5{\pm}2.2$
   & $9.2{\pm}0.8$  \\
     \multirow{2}{*}{S2}
   & ${\phi}_{B1}$ + ${\phi}_{B2}$
   & $   2.99^{+  0.14 +  0.22 }_{-  0.13 -  0.25 }$
   & $   3.65^{+  0.24 +  0.36 }_{-  0.23 -  0.34 }$
   & $   1.36^{+  0.00 +  0.10 }_{-  0.00 -  0.04 }$
   & $  11.00^{+  0.77 +  1.87 }_{-  0.71 -  1.83 }$ \\
   & ${\phi}_{B1}$
   & $   2.41^{+  0.11 +  0.23 }_{-  0.10 -  0.25 }$
   & $   2.48^{+  0.17 +  0.29 }_{-  0.16 -  0.28 }$
   & $   1.34^{+  0.01 +  0.12 }_{-  0.01 -  0.07 }$
   & $   7.64^{+  0.57 +  1.55 }_{-  0.53 -  1.49 }$ \\ \hline
     data & Belle
   & $6.8{\pm}0.6$
   & $9.67{\pm}1.10$
   &
   & $9.60{\pm}1.40$  \\
     \multirow{2}{*}{S3}
   & ${\phi}_{B1}$ + ${\phi}_{B2}$
   & $   3.98^{+  0.21 +  0.32 }_{-  0.19 -  0.36 }$
   & $   5.44^{+  0.39 +  0.53 }_{-  0.36 -  0.51 }$
   & $   1.38^{+  0.01 +  0.16 }_{-  0.01 -  0.07 }$
   & $  16.60^{+  1.21 +  2.89 }_{-  1.12 -  2.83 }$ \\
   & ${\phi}_{B1}$
   & $   3.19^{+  0.16 +  0.33 }_{-  0.15 -  0.36 }$
   & $   3.76^{+  0.29 +  0.44 }_{-  0.26 -  0.42 }$
   & $   1.38^{+  0.01 +  0.18 }_{-  0.01 -  0.11 }$
   & $  11.88^{+  0.93 +  2.43 }_{-  0.86 -  2.35 }$
   \end{tabular}
   \end{ruledtabular}
   \end{table}
   \renewcommand{\baselinestretch}{1.41}
   \begin{table}[ht]
   \caption{The $CP$-averaged branching ratios (in the unit
   of $10^{-6}$) for the $B_{d}$ ${\to}$ $PV$ decays. Other
   legends are the same as those of Table \ref{tab:branch-bu}.}
   \label{tab:branch-bd}
   \begin{ruledtabular}
   \begin{tabular}{ccccccc}
     \multicolumn{2}{c}{mode}
   & $\overline{B}^{0}$ ${\to}$ ${\pi}^{+}{\rho}^{-}$
   & $\overline{B}^{0}$ ${\to}$ ${\pi}^{-}{\rho}^{+}$
   & $\overline{B}^{0}$ ${\to}$ ${\pi}^{0}{\omega}$
   & $\overline{B}^{0}$ ${\to}$ ${\pi}^{0}{\rho}^{0}$
   & $\overline{B}^{0}$ ${\to}$ ${\pi}^{0}{\phi}$
    \\ \hline
     data & PDG
   & \multicolumn{2}{c}{ $23.0{\pm}2.3$ }
   &  $<0.5$
   &  $2.0{\pm}0.5$
   &  $<0.15$ \\
     \multirow{2}{*}{S1}
   & ${\phi}_{B1}$ + ${\phi}_{B2}$
   & $  10.11^{+  0.59 +  0.60 }_{-  0.55 -  0.59 }$
   & $   7.52^{+  0.41 +  0.14 }_{-  0.38 -  0.13 }$
   & $   0.16^{+  0.01 +  0.01 }_{-  0.01 -  0.01 }$
   & $    7.0^{+   0.3 +   0.4 }_{-   0.3 -   0.3 }{\times}10^{-2}$
   & $    2.3^{+   0.2 +   0.3 }_{-   0.2 -   0.2 }{\times}10^{-2}$
    \\
   & ${\phi}_{B1}$
   & $   6.35^{+  0.40 +  0.49 }_{-  0.37 -  0.47 }$
   & $   4.58^{+  0.26 +  0.12 }_{-  0.25 -  0.11 }$
   & $   0.12^{+  0.01 +  0.01 }_{-  0.01 -  0.01 }$
   & $    5.8^{+   0.3 +   0.2 }_{-   0.3 -   0.1 }{\times}10^{-2}$
   & $    2.0^{+   0.2 +   0.2 }_{-   0.2 -   0.2 }{\times}10^{-2}$
    \\ \hline
     data & BaBar
   &  \multicolumn{2}{c}{ $22.6{\pm}2.8$ }
   &  $<0.5$
   &  $1.4{\pm}0.7$
   &  $<0.28$ \\
     \multirow{2}{*}{S2}
   & ${\phi}_{B1}$ + ${\phi}_{B2}$
   & $   8.56^{+  0.48 +  0.50 }_{-  0.45 -  0.49 }$
   & $   6.44^{+  0.34 +  0.12 }_{-  0.32 -  0.11 }$
   & $   0.13^{+  0.01 +  0.01 }_{-  0.01 -  0.01 }$
   & $    6.2^{+   0.3 +   0.3 }_{-   0.2 -   0.3 }{\times}10^{-2}$
   & $    1.8^{+   0.1 +   0.2 }_{-   0.1 -   0.2 }{\times}10^{-2}$
    \\
   & ${\phi}_{B1}$
   & $   5.32^{+  0.32 +  0.41 }_{-  0.30 -  0.39 }$
   & $   3.88^{+  0.22 +  0.10 }_{-  0.20 -  0.09 }$
   & $   0.10^{+  0.01 +  0.01 }_{-  0.01 -  0.01 }$
   & $    5.1^{+   0.2 +   0.1 }_{-   0.2 -   0.1 }{\times}10^{-2}$
   & $    1.5^{+   0.1 +   0.2 }_{-   0.1 -   0.2 }{\times}10^{-2}$
    \\ \hline
     data & Belle
   &  \multicolumn{2}{c}{ $22.6{\pm}4.5$ }
   &  $<2.0$
   &  $3.0{\pm}0.9$
   &  $<0.15$ \\
     \multirow{2}{*}{S3}
   & ${\phi}_{B1}$ + ${\phi}_{B2}$
   & $  11.99^{+  0.72 +  0.73 }_{-  0.67 -  0.71 }$
   & $   8.82^{+  0.49 +  0.17 }_{-  0.46 -  0.16 }$
   & $   0.19^{+  0.01 +  0.01 }_{-  0.01 -  0.01 }$
   & $    8.0^{+   0.4 +   0.4 }_{-   0.3 -   0.4 }{\times}10^{-2}$
   & $    2.9^{+   0.2 +   0.3 }_{-   0.2 -   0.3 }{\times}10^{-2}$
    \\
   & ${\phi}_{B1}$
   & $   7.64^{+  0.49 +  0.60 }_{-  0.46 -  0.58 }$
   & $   5.43^{+  0.32 +  0.14 }_{-  0.30 -  0.14 }$
   & $   0.14^{+  0.01 +  0.01 }_{-  0.01 -  0.01 }$
   & $    6.7^{+   0.3 +   0.2 }_{-   0.3 -   0.2 }{\times}10^{-2}$
   & $    2.5^{+   0.2 +   0.3 }_{-   0.2 -   0.3 }{\times}10^{-2}$
    \\ \hline \hline
     \multicolumn{2}{c}{mode}
   & $\overline{B}^{0}$ ${\to}$ $\overline{K}^{0}K^{{\ast}0}$
   & $\overline{B}^{0}$ ${\to}$ $K^{0}\overline{K}^{{\ast}0}$
   & $\overline{B}^{0}$ ${\to}$ ${\pi}^{+}K^{{\ast}-}$
   & $\overline{B}^{0}$ ${\to}$ $K^{-}{\rho}^{+}$
   & $\overline{B}^{0}$ ${\to}$ ${\pi}^{0}\overline{K}^{{\ast}0}$
    \\ \hline
     data & PDG
   &  \multicolumn{2}{c}{ $<0.96$}
   &  $7.5{\pm}0.4$
   &  $7.0{\pm}0.9$
   &  $3.3{\pm}0.6$ \\
     \multirow{2}{*}{S1}
   & ${\phi}_{B1}$ + ${\phi}_{B2}$
   & $   0.23^{+  0.02 +  0.04 }_{-  0.02 -  0.03 }$
   & $   0.16^{+  0.01 +  0.02 }_{-  0.01 -  0.02 }$
   & $   3.61^{+  0.24 +  0.34 }_{-  0.22 -  0.33 }$
   & $   1.80^{+  0.04 +  0.19 }_{-  0.03 -  0.13 }$
   & $   1.30^{+  0.09 +  0.16 }_{-  0.08 -  0.15 }$
    \\
   & ${\phi}_{B1}$
   & $   0.15^{+  0.01 +  0.03 }_{-  0.01 -  0.03 }$
   & $   0.12^{+  0.01 +  0.02 }_{-  0.00 -  0.02 }$
   & $   2.53^{+  0.17 +  0.29 }_{-  0.16 -  0.27 }$
   & $   1.58^{+  0.03 +  0.19 }_{-  0.03 -  0.14 }$
   & $   0.94^{+  0.07 +  0.13 }_{-  0.06 -  0.13 }$
    \\ \hline
     data & BaBar
   &  \multicolumn{2}{c}{ $<1.9$}
   &  $8.0{\pm}1.4$
   &  $6.6{\pm}0.9$
   &  $3.3{\pm}0.6$ \\
     \multirow{2}{*}{S2}
   & ${\phi}_{B1}$ + ${\phi}_{B2}$
   & $   0.19^{+  0.01 +  0.03 }_{-  0.01 -  0.03 }$
   & $   0.14^{+  0.01 +  0.02 }_{-  0.01 -  0.01 }$
   & $   2.99^{+  0.19 +  0.29 }_{-  0.18 -  0.27 }$
   & $   1.71^{+  0.03 +  0.16 }_{-  0.03 -  0.10 }$
   & $   1.07^{+  0.07 +  0.13 }_{-  0.06 -  0.13 }$
    \\
   & ${\phi}_{B1}$
   & $   0.12^{+  0.01 +  0.02 }_{-  0.01 -  0.02 }$
   & $   0.11^{+  0.01 +  0.02 }_{-  0.01 -  0.01 }$
   & $   2.09^{+  0.13 +  0.24 }_{-  0.12 -  0.23 }$
   & $   1.51^{+  0.02 +  0.15 }_{-  0.02 -  0.11 }$
   & $   0.77^{+  0.05 +  0.11 }_{-  0.05 -  0.10 }$
    \\ \hline
     data & Belle
   &
   &
   &  $8.4{\pm}1.5$
   &  $15.1{\pm}4.2$
   & \\
     \multirow{2}{*}{S3}
   & ${\phi}_{B1}$ + ${\phi}_{B2}$
   & $   0.28^{+  0.02 +  0.04 }_{-  0.02 -  0.04 }$
   & $   0.18^{+  0.01 +  0.02 }_{-  0.01 -  0.02 }$
   & $   4.40^{+  0.30 +  0.42 }_{-  0.28 -  0.40 }$
   & $   1.93^{+  0.05 +  0.24 }_{-  0.05 -  0.16 }$
   & $   1.60^{+  0.12 +  0.20 }_{-  0.11 -  0.19 }$
    \\
   & ${\phi}_{B1}$
   & $   0.19^{+  0.01 +  0.04 }_{-  0.01 -  0.03 }$
   & $   0.14^{+  0.01 +  0.02 }_{-  0.01 -  0.02 }$
   & $   3.10^{+  0.23 +  0.35 }_{-  0.21 -  0.34 }$
   & $   1.68^{+  0.04 +  0.23 }_{-  0.04 -  0.17 }$
   & $   1.15^{+  0.09 +  0.17 }_{-  0.08 -  0.16 }$ \\ \hline \hline
     \multicolumn{2}{c}{mode}
   & $\overline{B}^{0}$ ${\to}$ $\overline{K}^{0}{\rho}^{0}$
   & $\overline{B}^{0}$ ${\to}$ $\overline{K}^{0}{\omega}$
   & $\overline{B}^{0}$ ${\to}$ $\overline{K}^{0}{\phi}$
   & $\overline{B}^{0}$ ${\to}$ $K^{+}K^{{\ast}-}$
   & $\overline{B}^{0}$ ${\to}$ $K^{-}K^{{\ast}+}$
    \\ \hline
     data & PDG
   &  $3.4{\pm}1.1$
   &  $4.8{\pm}0.4$
   &  $7.3{\pm}0.7$
   &  \multicolumn{2}{c}{ $<0.4$}
     \\
     \multirow{2}{*}{S1}
   & ${\phi}_{B1}$ + ${\phi}_{B2}$
   & $   0.81^{+  0.02 +  0.18 }_{-  0.02 -  0.14 }$
   & $   3.54^{+  0.17 +  0.27 }_{-  0.16 -  0.31 }$
   & $  12.31^{+  0.88 +  2.15 }_{-  0.81 -  2.11 }$
   & $    5.1^{+   0.1 +   0.1 }_{-   0.1 -   0.1 }{\times}10^{-2}$
   & $    2.6^{+   0.1 +   0.1 }_{-   0.1 -   0.1 }{\times}10^{-2}$
    \\
   & ${\phi}_{B1}$
   & $   0.77^{+  0.02 +  0.16 }_{-  0.02 -  0.12 }$
   & $   2.81^{+  0.13 +  0.28 }_{-  0.12 -  0.31 }$
   & $   8.64^{+  0.66 +  1.79 }_{-  0.61 -  1.73 }$
   & $    2.9^{+   0.1 +   0.1 }_{-   0.1 -   0.1 }{\times}10^{-2}$
   & $    2.2^{+   0.1 +   0.1 }_{-   0.0 -   0.1 }{\times}10^{-2}$
    \\ \hline
     data & BaBar
   &  $4.4{\pm}0.8$
   &  $5.4{\pm}0.9$
   &  $7.1{\pm}0.7$
   &
   &
     \\
     \multirow{2}{*}{S2}
   & ${\phi}_{B1}$ + ${\phi}_{B2}$
   & $   0.76^{+  0.02 +  0.15 }_{-  0.02 -  0.11 }$
   & $   3.09^{+  0.14 +  0.22 }_{-  0.13 -  0.26 }$
   & $  10.04^{+  0.70 +  1.73 }_{-  0.65 -  1.70 }$
   & $    4.7^{+   0.1 +   0.0 }_{-   0.1 -   0.1 }{\times}10^{-2}$
   & $    2.4^{+   0.1 +   0.1 }_{-   0.1 -   0.1 }{\times}10^{-2}$
    \\
   & ${\phi}_{B1}$
   & $   0.72^{+  0.02 +  0.13 }_{-  0.01 -  0.10 }$
   & $   2.47^{+  0.11 +  0.23 }_{-  0.10 -  0.26 }$
   & $   6.95^{+  0.52 +  1.43 }_{-  0.48 -  1.38 }$
   & $    2.7^{+   0.1 +   0.1 }_{-   0.1 -   0.1 }{\times}10^{-2}$
   & $    2.0^{+   0.1 +   0.1 }_{-   0.0 -   0.1 }{\times}10^{-2}$
    \\ \hline
     data & Belle
   &  $6.1{\pm}1.6$
   &  $4.5{\pm}0.5$
   &  $9.0{\pm}2.3$
   &
   &
     \\
     \multirow{2}{*}{S3}
   & ${\phi}_{B1}$ + ${\phi}_{B2}$
   & $   0.89^{+  0.03 +  0.23 }_{-  0.03 -  0.17 }$
   & $   4.09^{+  0.21 +  0.33 }_{-  0.20 -  0.37 }$
   & $  15.15^{+  1.10 +  2.68 }_{-  1.02 -  2.62 }$
   & $    5.4^{+   0.1 +   0.0 }_{-   0.1 -   0.1 }{\times}10^{-2}$
   & $    2.8^{+   0.1 +   0.1 }_{-   0.1 -   0.1 }{\times}10^{-2}$
    \\
   & ${\phi}_{B1}$
   & $   0.84^{+  0.03 +  0.20 }_{-  0.03 -  0.15 }$
   & $   3.24^{+  0.17 +  0.34 }_{-  0.15 -  0.37 }$
   & $  10.80^{+  0.85 +  2.25 }_{-  0.78 -  2.17 }$
   & $    3.2^{+   0.1 +   0.1 }_{-   0.1 -   0.1 }{\times}10^{-2}$
   & $    2.3^{+   0.1 +   0.1 }_{-   0.1 -   0.1 }{\times}10^{-2}$
   \end{tabular}
   \end{ruledtabular}
   \end{table} }
   \begin{table}[ht]
   \caption{Previous results of branching ratios (in the unit
   of $10^{-6}$) for the $B$ ${\to}$ $PV$ decays including the LO
   and NLO contributions using the PQCD approach, where NLO and NLOG
   represent those without and with the Glauber effects, and the contributions
   from the $B$ mesonic WFs ${\phi}_{B2}$ are not considered.
   If there are many theoretical uncertainties, the total
   uncertainties are given by the square roots of the sums of all
   quadratic errors. The details and meanings of the uncertainties
   can be found in their respective references.}
   \label{tab:previous-branching-ratio}
   \begin{ruledtabular}
   \begin{tabular}{lcccccc}
      \multicolumn{1}{c}{mode}
   &  \multicolumn{3}{c}{LO}
   &  \multicolumn{2}{c}{NLO}
   &  NLOG \\ \cline{1-1} \cline{2-4} \cline{5-6} \cline{7-7}
      $B^{-}$ ${\to}$ ${\pi}^{-}{\rho}^{0}$
   &  $10.4^{+3.9}_{-4.0}$ \cite{epjc.23.275}
   &  $9.0$ \cite{epjc.72.1923}
   &  $4.61{\pm}0.36$ \cite{PhysRevD.104.016025} 
   &  $5.4^{+1.6}_{-1.2}$ \cite{epjc.72.1923} 
   &  $6.5$ \cite{PhysRevD.90.074018} 
   &  $7.2$ \cite{PhysRevD.90.074018} 
      \\
      $B^{-}$ ${\to}$ ${\pi}^{0}{\rho}^{-}$
   &
   &  $14.1$ \cite{epjc.72.1923} 
   &  $8.73{\pm}0.25$ \cite{PhysRevD.104.016025} 
   &  $9.6^{+2.8}_{-2.6}$ \cite{epjc.72.1923} 
   &  $13.3$ \cite{PhysRevD.90.074018} 
   &  $9.3$ \cite{PhysRevD.90.074018} 
      \\
      $B^{-}$ ${\to}$ ${\pi}^{-}{\omega}$
   &  $11.3^{+3.6}_{-3.2}$ \cite{epjc.23.275} 
   &  $8.4$ \cite{epjc.72.1923} 
   &
   &  $4.6^{+1.4}_{-1.1}$ \cite{epjc.72.1923} 
   &  $5.4$ \cite{PhysRevD.90.074018} 
   &  $6.1$ \cite{PhysRevD.90.074018} 
      \\
      $B^{-}$ ${\to}$ $K^{-}K^{{\ast}0}$
   &  $0.31^{+0.12}_{-0.08}$ \cite{PhysRevD.75.014019} 
   &  $0.42$ \cite{epjc.59.49} 
   &  $0.48{\pm}0.02$ \cite{PhysRevD.104.016025} 
   &  $0.32^{+0.12}_{-0.08}$ \cite{epjc.59.49} 
   &
      \\
      $B^{-}$ ${\to}$ $K^{0}K^{{\ast}-}$
   &  $1.83^{+0.68}_{-0.47}$ \cite{PhysRevD.75.014019} 
   &  $0.20$ \cite{epjc.59.49} 
   &
   &  $0.21^{+0.14}_{-0.12}$ \cite{epjc.59.49} 
      \\
      $B^{-}$ ${\to}$ ${\pi}^{0}K^{{\ast}-}$
   &  $4.0$ \cite{PhysRevD.74.094020} 
   &
   &  $3.51{\pm}0.19$ \cite{PhysRevD.104.016025} 
   &  $4.3^{+5.0}_{-2.2}$ \cite{PhysRevD.74.094020} 
      \\
      $B^{-}$ ${\to}$ $K^{-}{\rho}^{0}$
   &  $2.5$ \cite{PhysRevD.74.094020} 
   &
   &  $2.24{\pm}0.41$ \cite{PhysRevD.104.016025} 
   &  $5.1^{+4.1}_{-2.8}$ \cite{PhysRevD.74.094020} 
      \\
      $B^{-}$ ${\to}$ $K^{-}{\omega}$
   &  $2.1$ \cite{PhysRevD.74.094020} 
   &
   &
   &  $10.6^{+10.4}_{-~5.8}$ \cite{PhysRevD.74.094020} 
      \\
      $B^{-}$ ${\to}$ ${\pi}^{-}\overline{K}^{{\ast}0}$
   &  $5.5$ \cite{PhysRevD.74.094020} 
   &
   &  $5.17{\pm}0.23$ \cite{PhysRevD.104.016025} 
   &  $6.0^{+2.8}_{-1.5}$ \cite{PhysRevD.74.094020} 
      \\
      $B^{-}$ ${\to}$ $\overline{K}^{0}{\rho}^{-}$
   &  $3.6$ \cite{PhysRevD.74.094020} 
   &
   &  $3.39{\pm}0.55$ \cite{PhysRevD.104.016025} 
   &  $8.7^{+6.8}_{-4.4}$ \cite{PhysRevD.74.094020} 
      \\
      $B^{-}$ ${\to}$ $K^{-}{\phi}$
   &  $13.8$ \cite{PhysRevD.74.094020} 
   &  $10.2$ \cite{PhysRevD.64.112002} 
   &
   &  $7.8^{+5.9}_{-1.8}$ \cite{PhysRevD.74.094020} 
      \\
      $\overline{B}^{0}$ ${\to}$ ${\pi}^{\pm}{\rho}^{\mp}$
   &
   &  $41.3$ \cite{epjc.72.1923}
   &  $23.3{\pm}0.8$ \cite{PhysRevD.104.016025} 
   &  $25.7^{+7.7}_{-6.4}$ \cite{epjc.72.1923} 
   &  $27.8$ \cite{PhysRevD.90.074018} 
   &  $30.8$ \cite{PhysRevD.90.074018} 
      \\
      $\overline{B}^{0}$ ${\to}$ ${\pi}^{0}{\omega}$
   &
   &  $0.22$ \cite{epjc.72.1923} 
   &
   &  $0.32^{+0.08}_{-0.10}$ \cite{epjc.72.1923} 
   &  $0.04$ \cite{PhysRevD.90.074018} 
   &  $0.85$ \cite{PhysRevD.90.074018} 
      \\
      $\overline{B}^{0}$ ${\to}$ ${\pi}^{0}{\rho}^{0}$
   &
   &  $0.15$ \cite{epjc.72.1923} 
   &  $0.026{\pm}0.002$ \cite{PhysRevD.104.016025} 
   &  $0.37^{+0.13}_{-0.10}$ \cite{epjc.72.1923} 
   &  $0.7$ \cite{PhysRevD.90.074018} 
   &  $1.1$ \cite{PhysRevD.90.074018} 
      \\
      $\overline{B}^{0}$ ${\to}$ ${\pi}^{+}K^{{\ast}-}$
   &  $5.1$ \cite{PhysRevD.74.094020} 
   &
   &  $4.93{\pm}0.28$ \cite{PhysRevD.104.016025} 
   &  $6.0^{+6.8}_{-2.6}$ \cite{PhysRevD.74.094020} 
      \\
      $\overline{B}^{0}$ ${\to}$ $K^{-}{\rho}^{+}$
   &  $4.7$ \cite{PhysRevD.74.094020} 
   &
   &  $4.4{\pm}0.6$ \cite{PhysRevD.104.016025} 
   &  $8.8^{+6.8}_{-4.5}$ \cite{PhysRevD.74.094020} 
      \\
      $\overline{B}^{0}$ ${\to}$ ${\pi}^{0}\overline{K}^{{\ast}0}$
   &  $1.5$ \cite{PhysRevD.74.094020} 
   &
   &  $1.73{\pm}0.10$ \cite{PhysRevD.104.016025} 
   &  $2.0^{+1.2}_{-0.6}$ \cite{PhysRevD.74.094020} 
      \\
      $\overline{B}^{0}$ ${\to}$ $\big\{$ \hspace{-5mm}
      {\renewcommand{\arraystretch}{0.1} $ \begin{array}{l}
       \overline{K}^{0}K^{{\ast}0} \\ K^{0}\overline{K}^{{\ast}0} \end{array}$}
   &  $1.96^{+0.79}_{-0.54}$ \cite{PhysRevD.75.014019} 
   &  $1.37$ \cite{epjc.59.49} 
   &
   &  $0.85^{+0.26}_{-0.21}$ \cite{epjc.59.49} 
      \\
      $\overline{B}^{0}$ ${\to}$ $\overline{K}^{0}{\rho}^{0}$
   &  $2.5$ \cite{PhysRevD.74.094020} 
   &
   &  $3.06{\pm}0.37$ \cite{PhysRevD.104.016025} 
   &  $4.8^{+4.3}_{-2.3}$ \cite{PhysRevD.74.094020} 
      \\
      $\overline{B}^{0}$ ${\to}$ $\overline{K}^{0}{\omega}$
   &  $1.9$ \cite{PhysRevD.74.094020} 
   &
   &
   &  $9.8^{+8.6}_{-4.9}$ \cite{PhysRevD.74.094020} 
      \\
      $\overline{B}^{0}$ ${\to}$ $\overline{K}^{0}{\phi}$
   &  $12.9$ \cite{PhysRevD.74.094020} 
   &
   &
   &  $7.3^{+5.4}_{-1.6}$ \cite{PhysRevD.74.094020} 
      \\
      $\overline{B}^{0}$ ${\to}$ $K^{\pm}K^{{\ast}{\mp}}$
   &  $0.07{\pm}0.01$ \cite{PhysRevD.75.014019} 
   &  $0.27$ \cite{epjc.59.49} 
   &
   &  $0.13^{+0.05}_{-0.07}$ \cite{epjc.59.49} 
   \end{tabular}
   \end{ruledtabular}
   \end{table}
  {\renewcommand{\baselinestretch}{1.41}
   \begin{table}[ht]
   \caption{The direct $CP$-violating asymmetries (${\cal A}_{CP}$,
   in the unit of percentage) for the $B_{u}$ ${\to}$ $PV$ decays.
   Other legends are the same as those of Table \ref{tab:branch-bu}.}
   \label{tab:cp-bu}
   \begin{ruledtabular}
   \begin{tabular}{cccccc}
     \multicolumn{2}{c}{mode}
   & $B^{-}$ ${\to}$ ${\pi}^{-}{\rho}^{0}$
   & $B^{-}$ ${\to}$ ${\pi}^{0}{\rho}^{-}$
   & $B^{-}$ ${\to}$ ${\pi}^{-}{\omega}$
   & $B^{-}$ ${\to}$ ${\pi}^{-}{\phi}$  \\ \hline
     data & PDG
   &  $0.9{\pm}1.9$
   &  $2{\pm}11$
   &  $-4{\pm}5$
   &  $9.8{\pm}51.1$
     \\
     \multirow{2}{*}{S1}
   & ${\phi}_{B1}$ + ${\phi}_{B2}$
   & $ -24.00^{+  0.61 +  1.16 }_{-  0.62 -  1.15 }$
   & $  18.35^{+  0.54 +  0.23 }_{-  0.53 -  0.24 }$
   & $  -1.67^{+  0.01 +  0.26 }_{-  0.01 -  0.26 }$
   & $0$
    \\
   & ${\phi}_{B1}$
   & $ -29.49^{+  0.81 +  1.25 }_{-  0.83 -  1.22 }$
   & $  22.40^{+  0.70 +  0.10 }_{-  0.68 -  0.11 }$
   & $  -4.36^{+  0.02 +  0.18 }_{-  0.02 -  0.18 }$
   & $0$
     \\ \hline
     data & BaBar
   &  $18{\pm}17$
   &  $-1{\pm}13$
   &  $-2{\pm}8$
     \\
     \multirow{2}{*}{S2}
   & ${\phi}_{B1}$ + ${\phi}_{B2}$
   & $ -25.90^{+  0.65 +  1.25 }_{-  0.66 -  1.23 }$
   & $  19.98^{+  0.56 +  0.26 }_{-  0.55 -  0.28 }$
   & $  -1.66^{+  0.01 +  0.26 }_{- -0.00 -  0.26 }$
   & $0$
     \\
   & ${\phi}_{B1}$
   & $ -32.02^{+  0.86 +  1.35 }_{-  0.88 -  1.32 }$
   & $  24.52^{+  0.73 +  0.13 }_{-  0.72 -  0.14 }$
   & $  -4.43^{+  0.03 +  0.18 }_{-  0.02 -  0.17 }$
   & $0$
     \\ \hline
     data & Belle
   &
   &  $6{\pm}18$
   &  $-2{\pm}9$
     \\
     \multirow{2}{*}{S3}
   & ${\phi}_{B1}$ + ${\phi}_{B2}$
   & $ -22.20^{+  0.58 +  1.08 }_{-  0.59 -  1.07 }$
   & $  16.80^{+  0.51 +  0.20 }_{-  0.50 -  0.21 }$
   & $  -1.68^{+  0.00 +  0.26 }_{-  0.00 -  0.26 }$
   & $0$
     \\
   & ${\phi}_{B1}$
   & $ -27.09^{+  0.77 +  1.15 }_{-  0.78 -  1.12 }$
   & $  20.38^{+  0.66 +  0.08 }_{-  0.65 -  0.09 }$
   & $  -4.28^{+  0.03 +  0.18 }_{-  0.03 -  0.18 }$
   & $0$
     \\ \hline \hline
     \multicolumn{2}{c}{mode}
   & $B^{-}$ ${\to}$ $K^{-}K^{{\ast}0}$
   & $B^{-}$ ${\to}$ $K^{0}K^{{\ast}-}$
   & $B^{-}$ ${\to}$ ${\pi}^{0}K^{{\ast}-}$
   & $B^{-}$ ${\to}$ $K^{-}{\rho}^{0}$ \\ \hline
     data & PDG
   &  $12.3{\pm}9.8$
   &
   &  $-39{\pm}21$
   &  $37{\pm}10$
     \\
     \multirow{2}{*}{S1}
   & ${\phi}_{B1}$ + ${\phi}_{B2}$
   & $  15.16^{+  0.34 +  0.70 }_{-  0.34 -  0.58 }$
   & $ -25.13^{+  2.03 + 13.90 }_{-  1.99 -  7.02 }$
   & $ -21.17^{+  0.82 +  0.16 }_{-  0.84 -  0.15 }$
   & $  82.79^{+  1.30 +  3.37 }_{-  1.32 -  5.16 }$
     \\
   & ${\phi}_{B1}$
   & $  19.60^{+  0.50 +  1.29 }_{-  0.51 -  1.03 }$
   & $ -11.19^{+  1.82 +  7.91 }_{-  1.77 -  4.25 }$
   & $ -24.32^{+  1.00 +  0.00 }_{-  1.02 -  0.01 }$
   & $  78.05^{+  1.52 +  0.57 }_{-  1.53 -  2.24 }$
     \\ \hline
     data & BaBar
   &
   &
   &  $-52{\pm}15$
   &  $44{\pm}17$
     \\
     \multirow{2}{*}{S2}
   & ${\phi}_{B1}$ + ${\phi}_{B2}$
   & $  16.19^{+  0.37 +  0.74 }_{-  0.35 -  0.60 }$
   & $ -19.01^{+  2.03 + 13.91 }_{-  2.05 -  8.18 }$
   & $ -23.76^{+  0.89 +  0.15 }_{-  0.91 -  0.14 }$
   & $  78.79^{+  1.35 +  3.20 }_{-  1.36 -  4.65 }$
     \\
   & ${\phi}_{B1}$
   & $  21.14^{+  0.54 +  1.39 }_{-  0.52 -  1.09 }$
   & $  -5.72^{+  1.82 +  8.30 }_{-  1.83 -  5.14 }$
   & $ -27.46^{+  1.07 +  0.04 }_{-  1.09 -  0.05 }$
   & $  73.45^{+  1.54 +  0.72 }_{-  1.54 -  2.04 }$
     \\ \hline
     data & Belle
   &
   &
   &
   &  $30{\pm}16$
     \\
     \multirow{2}{*}{S3}
   & ${\phi}_{B1}$ + ${\phi}_{B2}$
   & $  14.18^{+  0.32 +  0.68 }_{-  0.31 -  0.55 }$
   & $ -31.01^{+  1.93 + 13.31 }_{-  1.86 -  5.46 }$
   & $ -18.78^{+  0.75 +  0.16 }_{-  0.77 -  0.15 }$
   & $  86.60^{+  1.20 +  3.42 }_{-  1.24 -  5.63 }$
    \\
   & ${\phi}_{B1}$
   & $  18.12^{+  0.48 +  1.22 }_{-  0.47 -  0.96 }$
   & $ -16.43^{+  1.73 +  7.11 }_{-  1.66 -  3.10 }$
   & $ -21.41^{+  0.92 +  0.04 }_{-  0.95 -  0.04 }$
   & $  82.55^{+  1.45 +  0.28 }_{-  1.48 -  2.37 }$
     \\ \hline \hline
     \multicolumn{2}{c}{mode}
   & $B^{-}$ ${\to}$ $K^{-}{\omega}$
   & $B^{-}$ ${\to}$ ${\pi}^{-}\overline{K}^{{\ast}0}$
   & $B^{-}$ ${\to}$ $\overline{K}^{0}{\rho}^{-}$
   & $B^{-}$ ${\to}$ $K^{-}{\phi}$ \\ \hline
     data & PDG
   &  $-2{\pm}4$
   &  $-4{\pm}9$
   &  $-3{\pm}15$
   &  $2.4{\pm}2.8$
     \\
     \multirow{2}{*}{S1}
   & ${\phi}_{B1}$ + ${\phi}_{B2}$
   & $  20.64^{+  0.37 +  1.48 }_{-  0.38 -  1.06 }$
   & $  -1.13^{+  0.03 +  0.03 }_{-  0.03 -  0.03 }$
   & $   0.03^{+  0.09 +  0.55 }_{-  0.09 -  0.64 }$
   & $  -0.55^{+  0.01 +  0.02 }_{-  0.01 -  0.03 }$
     \\
   & ${\phi}_{B1}$
   & $  22.70^{+  0.32 +  2.07 }_{-  0.33 -  1.50 }$
   & $  -1.48^{+  0.05 +  0.05 }_{-  0.05 -  0.06 }$
   & $  -0.12^{+  0.09 +  0.36 }_{-  0.08 -  0.42 }$
   & $  -0.69^{+  0.02 +  0.04 }_{-  0.02 -  0.06 }$
     \\ \hline
     data & BaBar
   &  $-1{\pm}7$
   &  $-12{\pm}25$
   &  $21{\pm}31$
   &  $12.8{\pm}4.6$
     \\
     \multirow{2}{*}{S2}
   & ${\phi}_{B1}$ + ${\phi}_{B2}$
   & $  21.75^{+  0.35 +  1.42 }_{-  0.36 -  1.01 }$
   & $  -1.23^{+  0.03 +  0.03 }_{-  0.03 -  0.03 }$
   & $  -0.22^{+  0.08 +  0.51 }_{-  0.07 -  0.56 }$
   & $  -0.59^{+  0.01 +  0.02 }_{-  0.02 -  0.03 }$
     \\
   & ${\phi}_{B1}$
   & $  23.61^{+  0.27 +  1.99 }_{-  0.29 -  1.44 }$
   & $  -1.63^{+  0.05 +  0.06 }_{-  0.05 -  0.06 }$
   & $  -0.36^{+  0.08 +  0.34 }_{-  0.07 -  0.38 }$
   & $  -0.76^{+  0.02 +  0.04 }_{-  0.02 -  0.06 }$
     \\ \hline
     data & Belle
   &  $-3{\pm}4$
   &  $-14.9{\pm}6.8$
   &
   &  $1{\pm}13$
     \\
     \multirow{2}{*}{S3}
   & ${\phi}_{B1}$ + ${\phi}_{B2}$
   & $  19.49^{+  0.39 +  1.53 }_{-  0.40 -  1.09 }$
   & $  -1.04^{+  0.03 +  0.02 }_{-  0.03 -  0.03 }$
   & $   0.30^{+  0.10 +  0.58 }_{-  0.10 -  0.71 }$
   & $  -0.51^{+  0.01 +  0.02 }_{-  0.01 -  0.03 }$
     \\
   & ${\phi}_{B1}$
   & $  21.66^{+  0.36 +  2.13 }_{-  0.38 -  1.53 }$
   & $  -1.34^{+  0.04 +  0.05 }_{-  0.04 -  0.05 }$
   & $   0.14^{+  0.09 +  0.37 }_{-  0.09 -  0.47 }$
   & $  -0.64^{+  0.02 +  0.04 }_{-  0.02 -  0.05 }$
   \end{tabular}
   \end{ruledtabular}
   \end{table}
   \renewcommand{\baselinestretch}{1.37}
   \begin{table}[ht]
   \caption{The $CP$-violating asymmetries (in the unit of
   percentage) for the $B_{d}$ decays. Other legends
   are the same as those of Table \ref{tab:branch-bu}.}
   \label{tab:cp-bd-01}
   \begin{ruledtabular}
   \begin{tabular}{cccccc}
     ${\cal A}_{CP}$ & mode
   & $\overline{B}^{0}$ ${\to}$ ${\pi}^{+}{\rho}^{-}$
   & $\overline{B}^{0}$ ${\to}$ ${\pi}^{-}{\rho}^{+}$
   & $\overline{B}^{0}$ ${\to}$ ${\pi}^{+}K^{{\ast}-}$
   & $\overline{B}^{0}$ ${\to}$ $K^{-}{\rho}^{+}$
    \\ \hline
     data & PDG
   &  $13{\pm}6$
   &  $-8{\pm}8$
   &  $-27{\pm}4$
   &  $20{\pm}11$\\
     \multirow{2}{*}{S1}
   & ${\phi}_{B1}$ + ${\phi}_{B2}$
   & $   4.04^{+  0.27 +  0.77 }_{-  0.26 -  0.80 }$
   & $ -19.80^{+  0.35 +  0.18 }_{-  0.36 -  0.13 }$
   & $ -31.04^{+  1.13 +  0.29 }_{-  1.15 -  0.33 }$
   & $  86.31^{+  0.61 +  2.50 }_{-  0.71 -  4.95 }$
    \\
   & ${\phi}_{B1}$
   & $   7.20^{+  0.37 +  0.76 }_{-  0.36 -  0.80 }$
   & $ -22.88^{+  0.48 +  0.26 }_{-  0.48 -  0.18 }$
   & $ -35.44^{+  1.28 +  0.64 }_{-  1.29 -  0.74 }$
   & $  74.34^{+  0.83 +  3.15 }_{-  0.92 -  4.75 }$
     \\ \hline
     data & BaBar
   &  $9{\pm}7$
   &  $-12{\pm}9$
   &  $-29{\pm}11$
   &  $20{\pm}12$
     \\
     \multirow{2}{*}{S2}
   & ${\phi}_{B1}$ + ${\phi}_{B2}$
   & $   4.87^{+  0.29 +  0.77 }_{-  0.28 -  0.81 }$
   & $ -20.91^{+  0.38 +  0.12 }_{-  0.38 -  0.06 }$
   & $ -34.56^{+  1.19 +  0.36 }_{-  1.21 -  0.40 }$
   & $  83.91^{+  0.89 +  1.48 }_{-  0.97 -  3.71 }$
     \\
   & ${\phi}_{B1}$
   & $   8.33^{+  0.39 +  0.76 }_{-  0.39 -  0.80 }$
   & $ -24.36^{+  0.51 +  0.18 }_{-  0.51 -  0.10 }$
   & $ -39.34^{+  1.30 +  0.75 }_{-  1.30 -  0.85 }$
   & $  71.37^{+  1.06 +  2.15 }_{-  1.13 -  3.61 }$
     \\ \hline
     data & Belle
   &  $21{\pm}9$
   &  $8{\pm}19$
   &  $-21{\pm}13$
   &  $22{\pm}24$
     \\
     \multirow{2}{*}{S3}
   & ${\phi}_{B1}$ + ${\phi}_{B2}$
   & $   3.27^{+  0.25 +  0.76 }_{-  0.24 -  0.79 }$
   & $ -18.77^{+  0.33 +  0.23 }_{-  0.34 -  0.19 }$
   & $ -27.72^{+  1.06 +  0.24 }_{-  1.09 -  0.27 }$
   & $  87.84^{+  0.29 +  3.77 }_{-  0.40 -  6.31 }$
     \\
   & ${\phi}_{B1}$
   & $   6.14^{+  0.34 +  0.75 }_{-  0.33 -  0.80 }$
   & $ -21.48^{+  0.45 +  0.33 }_{-  0.46 -  0.26 }$
   & $ -31.64^{+  1.23 +  0.55 }_{-  1.25 -  0.63 }$
   & $  76.57^{+  0.55 +  4.41 }_{-  0.65 -  6.03 }$
     \\ \hline \hline
     ${\cal A}_{CP}$ & mode
   & $\overline{B}^{0}$ ${\to}$ ${\pi}^{0}\overline{K}^{{\ast}0}$
   & $\overline{B}^{0}$ ${\to}$ $\overline{K}^{0}{\rho}^{0}$
   & $\overline{B}^{0}$ ${\to}$ $\overline{K}^{0}{\omega}$
   & $\overline{B}^{0}$ ${\to}$ $\overline{K}^{0}{\phi}$
     \\ \hline
     data & PDG
   &  $-15{\pm}13$
   &  $-4{\pm}20$
   &  $0{\pm}40$
   &  $1{\pm}14$
   \\
     \multirow{2}{*}{S1}
   & ${\phi}_{B1}$ + ${\phi}_{B2}$
   & $  -0.47^{+  0.09 +  0.01 }_{-  0.08 -  0.01 }$
   & $  -5.47^{+  0.16 +  1.72 }_{-  0.15 -  2.25 }$
   & $   3.12^{+  0.04 +  0.35 }_{-  0.04 -  0.29 }$
   & $0$ \\
   & ${\phi}_{B1}$
   & $  -1.33^{+  0.12 +  0.03 }_{-  0.10 -  0.02 }$
   & $  -6.57^{+  0.20 +  1.80 }_{-  0.19 -  2.27 }$
   & $   4.58^{+  0.05 +  0.45 }_{-  0.05 -  0.35 }$
   & $0$ \\ \hline
     data & BaBar
   &  $-15{\pm}13$
   &  $-5{\pm}28$
   &  $-52{\pm}22$
   &  $5{\pm}19$
   \\
     \multirow{2}{*}{S2}
   & ${\phi}_{B1}$ + ${\phi}_{B2}$
   & $  -0.20^{+  0.10 +  0.01 }_{-  0.10 -  0.01 }$
   & $  -5.88^{+  0.12 +  1.65 }_{-  0.11 -  2.03 }$
   & $   3.23^{+  0.03 +  0.35 }_{-  0.03 -  0.29 }$
   & $0$ \\
   & ${\phi}_{B1}$
   & $  -0.96^{+  0.14 +  0.02 }_{-  0.13 -  0.01 }$
   & $  -7.08^{+  0.15 +  1.69 }_{-  0.14 -  2.02 }$
   & $   4.71^{+  0.04 +  0.44 }_{-  0.04 -  0.34 }$
   & $0$ \\ \hline
     data & Belle
   &
   &  $-3{\pm}28$
   &  $36{\pm}20$
   &  $-4{\pm}22$
     \\
     \multirow{2}{*}{S3}
   & ${\phi}_{B1}$ + ${\phi}_{B2}$
   & $  -0.69^{+  0.07 +  0.03 }_{-  0.06 -  0.03 }$
   & $  -4.97^{+  0.19 +  1.75 }_{-  0.18 -  2.42 }$
   & $   3.00^{+  0.04 +  0.36 }_{-  0.04 -  0.29 }$
   & $0$ \\
   & ${\phi}_{B1}$
   & $  -1.62^{+  0.09 +  0.06 }_{-  0.08 -  0.06 }$
   & $  -5.93^{+  0.24 +  1.86 }_{-  0.23 -  2.48 }$
   & $   4.41^{+  0.06 +  0.46 }_{-  0.06 -  0.36 }$
   & $0$ \\ \hline \hline
     \multicolumn{2}{c}{\multirow{2}{*}{mode}}
   & \multicolumn{2}{c}{$\overline{B}^{0}$ ${\to}$ ${\pi}^{0}{\rho}^{0}$}
   & \multicolumn{2}{c}{$\overline{B}^{0}$ ${\to}$ ${\pi}^{0}{\omega}$}
     \\ \cline{3-4} \cline{5-6}
     & & $C_{f}$ & $S_{f}$ & $C_{f}$ & $S_{f}$ \\ \hline
     data & PDG
   &  $27{\pm}24$
   &  $-23{\pm}34$
   &  \\
     \multirow{2}{*}{S1}
   & ${\phi}_{B1}$ + ${\phi}_{B2}$
   & $  -0.04^{+  0.14 +  0.31 }_{-  0.21 -  0.50 }$
   & $ -89.75^{+  0.46 +  2.11 }_{-  0.45 -  1.82 }$
   & $ -67.36^{+  1.06 +  0.75 }_{-  1.05 -  0.73 }$
   & $  24.00^{+  0.40 +  1.86 }_{-  0.40 -  1.78 }$
     \\
   & ${\phi}_{B1}$
   & $ -22.82^{+  0.03 +  0.11 }_{-  0.06 -  0.22 }$
   & $ -74.39^{+  0.50 +  3.22 }_{-  0.49 -  2.90 }$
   & $ -66.52^{+  1.05 +  0.33 }_{-  1.04 -  0.32 }$
   & $  39.32^{+  0.41 +  0.97 }_{-  0.42 -  0.89 }$
     \\ \hline
     data & BaBar
   &  $19{\pm}27$
   &  $-37{\pm}39$
   &  \\
     \multirow{2}{*}{S2}
   & ${\phi}_{B1}$ + ${\phi}_{B2}$
   & $   0.49^{+  0.21 +  0.26 }_{-  0.18 -  0.42 }$
   & $ -88.39^{+  0.46 +  2.27 }_{-  0.45 -  1.96 }$
   & $ -70.49^{+  1.05 +  0.82 }_{-  1.03 -  0.79 }$
   & $  25.15^{+  0.32 +  1.79 }_{-  0.38 -  1.74 }$
     \\
   & ${\phi}_{B1}$
   & $ -22.65^{+  0.09 +  0.10 }_{-  0.08 -  0.20 }$
   & $ -73.00^{+  0.42 +  3.28 }_{-  0.43 -  2.98 }$
   & $ -69.62^{+  1.03 +  0.36 }_{-  1.03 -  0.36 }$
   & $  40.43^{+  0.28 +  0.85 }_{-  0.35 -  0.79 }$
     \\ \hline
     data & Belle
   &  $49{\pm}46$
   &  $17{\pm}67$
   &  \\
     \multirow{2}{*}{S3}
   & ${\phi}_{B1}$ + ${\phi}_{B2}$
   & $  -0.57^{+  0.13 +  0.43 }_{-  0.11 -  0.57 }$
   & $ -91.10^{+  0.46 +  1.94 }_{-  0.45 -  1.64 }$
   & $ -64.23^{+  1.05 +  0.67 }_{-  1.03 -  0.65 }$
   & $  22.69^{+  0.46 +  1.91 }_{-  0.46 -  1.84 }$
     \\
   & ${\phi}_{B1}$
   & $ -22.90^{+  0.05 +  0.15 }_{-  0.01 -  0.23 }$
   & $ -76.00^{+  0.58 +  3.11 }_{-  0.60 -  2.78 }$
   & $ -63.47^{+  1.00 +  0.27 }_{-  1.00 -  0.27 }$
   & $  37.90^{+  0.53 +  1.07 }_{-  0.55 -  1.00 }$
   \end{tabular}
   \end{ruledtabular}
   \end{table}
   \renewcommand{\baselinestretch}{1.41}
   \begin{table}[ht]
   \caption{The $CP$-violating asymmetries (in the unit of
   percentage) for case 3 $B_{d}$ decays. Other legends
   are the same as those of Table \ref{tab:branch-bu}.}
   \label{tab:cp-bd-03}
   \begin{ruledtabular}
   \begin{tabular}{cccccc}
     \multicolumn{2}{c}{${\pi}^{+}{\rho}^{-}$,
     ${\pi}^{-}{\rho}^{+}$}
   & $a_{{\epsilon}^{\prime}}$
   & $a_{{\epsilon}+{\epsilon}^{\prime}}$
   & $\bar{a}_{{\epsilon}^{\prime}}$
   & $\bar{a}_{{\epsilon}+{\epsilon}^{\prime}}$ \\ \hline
     data & PDG
   &  $-3{\pm}7$
   &  $5{\pm}7$
   &
   &
     \\
     \multirow{2}{*}{S1}
   & ${\phi}_{B1}$ + ${\phi}_{B2}$
   & $ -19.23^{+  0.34 +  3.24 }_{-  0.34 -  3.06 }$
   & $   9.05^{+  0.03 +  1.00 }_{-  0.03 -  1.00 }$
   & $  27.32^{+  0.22 +  3.26 }_{-  0.22 -  3.46 }$
   & $  12.71^{+  0.09 +  1.36 }_{-  0.09 -  1.36 }$
    \\
   & ${\phi}_{B1}$
   & $ -22.66^{+  0.38 +  4.00 }_{-  0.38 -  3.72 }$
   & $   9.78^{+  0.03 +  1.14 }_{-  0.03 -  1.15 }$
   & $  32.52^{+  0.21 +  4.04 }_{-  0.20 -  4.37 }$
   & $   9.60^{+  0.04 +  1.60 }_{-  0.04 -  1.59 }$
     \\ \hline
     data & BaBar
   &  $1.6{\pm}6.9$
   &  $5.3{\pm}8.8$
   &
     \\
     \multirow{2}{*}{S2}
   & ${\phi}_{B1}$ + ${\phi}_{B2}$
   & $ -18.25^{+  0.31 +  3.18 }_{-  0.32 -  3.00 }$
   & $   8.97^{+  0.02 +  0.96 }_{-  0.03 -  0.96 }$
   & $  26.71^{+  0.19 +  3.24 }_{-  0.19 -  3.43 }$
   & $  12.99^{+  0.10 +  1.33 }_{-  0.10 -  1.34 }$
     \\
   & ${\phi}_{B1}$
   & $ -21.56^{+  0.34 +  3.95 }_{-  0.36 -  3.68 }$
   & $   9.88^{+  0.03 +  1.09 }_{-  0.04 -  1.09 }$
   & $  31.97^{+  0.17 +  4.05 }_{-  0.16 -  4.37 }$
   & $   9.70^{+  0.04 +  1.57 }_{-  0.04 -  1.57 }$
     \\ \hline
     data & Belle
   &  $-13{\pm}10$
   &  $6{\pm}14$
   &
   &
     \\
     \multirow{2}{*}{S3}
   & ${\phi}_{B1}$ + ${\phi}_{B2}$
   & $ -20.29^{+  0.36 +  3.30 }_{-  0.38 -  3.11 }$
   & $   9.13^{+  0.03 +  1.04 }_{-  0.03 -  1.04 }$
   & $  28.03^{+  0.26 +  3.28 }_{-  0.25 -  3.49 }$
   & $  12.44^{+  0.09 +  1.38 }_{-  0.09 -  1.38 }$
     \\
   & ${\phi}_{B1}$
   & $ -23.84^{+  0.41 +  4.05 }_{-  0.42 -  3.75 }$
   & $   9.68^{+  0.04 +  1.20 }_{-  0.03 -  1.19 }$
   & $  33.19^{+  0.25 +  4.02 }_{-  0.23 -  4.36 }$
   & $   9.50^{+  0.03 +  1.62 }_{-  0.04 -  1.61 }$
     \\ \hline \hline
     \multicolumn{2}{c}{$\overline{K}^{0}K^{{\ast}0}$,
     $K^{0}\overline{K}^{{\ast}0}$}
   & $a_{{\epsilon}^{\prime}}$
   & $a_{{\epsilon}+{\epsilon}^{\prime}}$
   & $\bar{a}_{{\epsilon}^{\prime}}$
   & $\bar{a}_{{\epsilon}+{\epsilon}^{\prime}}$ \\ \hline
     data & PDG
   & \\
     \multirow{2}{*}{S1}
   & ${\phi}_{B1}$ + ${\phi}_{B2}$
   & $ -76.49^{+  1.10 +  1.14 }_{-  1.05 -  0.53 }$
   & $  58.96^{+  1.43 +  4.66 }_{-  1.42 -  4.44 }$
   & $  76.49^{+  1.05 +  0.53 }_{-  1.10 -  1.14 }$
   & $ -58.96^{+  1.42 +  4.44 }_{-  1.43 -  4.66 }$
     \\
   & ${\phi}_{B1}$
   & $ -64.09^{+  1.43 +  1.69 }_{-  1.37 -  0.24 }$
   & $  61.42^{+  1.34 +  6.32 }_{-  1.34 -  5.61 }$
   & $  64.09^{+  1.37 +  0.24 }_{-  1.43 -  1.69 }$
   & $ -61.42^{+  1.34 +  5.61 }_{-  1.34 -  6.32 }$
     \\ \hline
     data & BaBar
   & \\
     \multirow{2}{*}{S2}
   & ${\phi}_{B1}$ + ${\phi}_{B2}$
   & $ -73.04^{+  1.25 +  1.01 }_{-  1.20 -  0.69 }$
   & $  63.22^{+  1.40 +  4.78 }_{-  1.41 -  4.61 }$
   & $  73.04^{+  1.20 +  0.69 }_{-  1.25 -  1.01 }$
   & $ -63.22^{+  1.41 +  4.61 }_{-  1.40 -  4.78 }$
     \\
   & ${\phi}_{B1}$
   & $ -59.64^{+  1.58 +  1.11 }_{-  1.53 -  0.43 }$
   & $  65.34^{+  1.24 +  6.41 }_{-  1.27 -  5.78 }$
   & $  59.64^{+  1.53 +  0.43 }_{-  1.58 -  1.11 }$
   & $ -65.34^{+  1.27 +  5.78 }_{-  1.24 -  6.41 }$
     \\ \hline
     data & Belle
   & \\
     \multirow{2}{*}{S3}
   & ${\phi}_{B1}$ + ${\phi}_{B2}$
   & $ -79.48^{+  0.95 +  1.56 }_{-  0.90 -  0.05 }$
   & $  54.71^{+  1.41 +  4.50 }_{-  1.40 -  4.25 }$
   & $  79.48^{+  0.90 +  0.05 }_{-  0.95 -  1.56 }$
   & $ -54.71^{+  1.40 +  4.25 }_{-  1.41 -  4.50 }$
    \\
   & ${\phi}_{B1}$
   & $ -68.05^{+  1.27 +  2.23 }_{-  1.22 -  0.89 }$
   & $  57.33^{+  1.37 +  6.13 }_{-  1.39 -  5.39 }$
   & $  68.05^{+  1.22 +  0.89 }_{-  1.27 -  2.23 }$
   & $ -57.33^{+  1.39 +  5.39 }_{-  1.37 -  6.13 }$
     \\ \hline \hline
     \multicolumn{2}{c}{$K^{+}K^{{\ast}-}$, $K^{-}K^{{\ast}+}$}
   & $a_{{\epsilon}^{\prime}}$
   & $a_{{\epsilon}+{\epsilon}^{\prime}}$
   & $\bar{a}_{{\epsilon}^{\prime}}$
   & $\bar{a}_{{\epsilon}+{\epsilon}^{\prime}}$ \\ \hline
     data & PDG
   &  \\
     \multirow{2}{*}{S1}
   & ${\phi}_{B1}$ + ${\phi}_{B2}$
   & $ -14.02^{+  0.79 +  0.86 }_{-  0.71 -  0.95 }$
   & $ -38.60^{+  0.01 +  0.60 }_{-  0.03 -  0.62 }$
   & $  67.26^{+  0.34 +  0.17 }_{-  0.32 -  0.10 }$
   & $ -11.47^{+  0.44 +  0.69 }_{-  0.47 -  0.69 }$
     \\
   & ${\phi}_{B1}$
   & $  11.39^{+  0.07 +  0.17 }_{-  0.05 -  0.23 }$
   & $ -41.60^{+  0.20 +  0.63 }_{-  0.17 -  0.56 }$
   & $  27.65^{+  0.89 +  0.42 }_{-  0.88 -  0.38 }$
   & $  -1.97^{+  0.05 +  1.45 }_{-  0.15 -  1.41 }$
     \\ \hline
     data & BaBar
   &  \\
     \multirow{2}{*}{S2}
   & ${\phi}_{B1}$ + ${\phi}_{B2}$
   & $ -16.09^{+  0.63 +  0.79 }_{-  0.72 -  0.87 }$
   & $ -38.54^{+  0.05 +  0.58 }_{-  0.04 -  0.63 }$
   & $  66.24^{+  0.40 +  0.15 }_{-  0.33 -  0.11 }$
   & $ -10.14^{+  0.46 +  0.73 }_{-  0.44 -  0.71 }$
     \\
   & ${\phi}_{B1}$
   & $  11.56^{+  0.01 +  0.07 }_{-  0.12 -  0.14 }$
   & $ -40.90^{+  0.25 +  0.63 }_{-  0.27 -  0.63 }$
   & $  25.04^{+  0.94 +  0.45 }_{-  0.86 -  0.46 }$
   & $  -1.80^{+  0.07 +  1.47 }_{-  0.04 -  1.38 }$
     \\ \hline
     data & Belle
   &  \\
     \multirow{2}{*}{S3}
   & ${\phi}_{B1}$ + ${\phi}_{B2}$
   & $ -11.66^{+  0.89 +  0.93 }_{-  0.81 -  1.03 }$
   & $ -38.60^{+  0.02 +  0.63 }_{-  0.04 -  0.66 }$
   & $  68.33^{+  0.31 +  0.13 }_{-  0.38 -  0.12 }$
   & $ -12.84^{+  0.48 +  0.69 }_{-  0.50 -  0.64 }$
     \\
   & ${\phi}_{B1}$
   & $  11.38^{+  0.05 +  0.25 }_{-  0.04 -  0.29 }$
   & $ -42.02^{+  0.09 +  0.67 }_{-  0.07 -  0.62 }$
   & $  30.49^{+  0.95 +  0.32 }_{-  1.03 -  0.34 }$
   & $  -2.33^{+  0.14 +  1.48 }_{-  0.19 -  1.35 }$
   \end{tabular}
   \end{ruledtabular}
   \end{table} }

\begin{figure}[ht]
   \includegraphics[width=0.93\textwidth,bb=10 60 290 210]{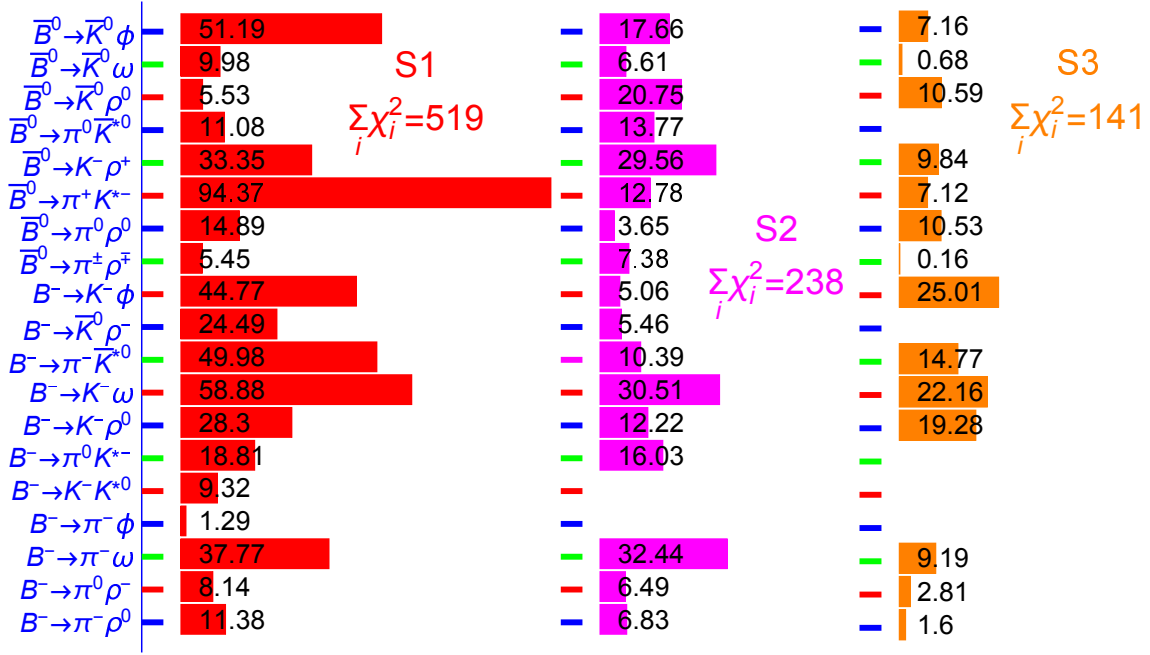}
   \caption{The ${\chi}_{i}^{2}$ distribution of branching
   ratios for three scenarios, where the numbers in the barcharts
   denote the values of ${\chi}_{i}^{2}$ for a specific process.}
   \label{fig:chi2c-b2}
   \end{figure}

   The numerical results of the $CP$-averaged branching ratios for
   three scenarios (S1, S2 and S3) using the PQCD approach together
   with experimental data are presented in Tables \ref{tab:branch-bu}
   and \ref{tab:branch-bd}, and the previous PQCD results without
   the contributions from the $B$ mesonic WFs ${\phi}_{B2}$ are
   listed in Table \ref{tab:previous-branching-ratio}.
   To obtain a clear and comprehensive impression of the agreement
   between the theoretical and experimental results, the ${\chi}_{i}^{2}$
   distributions are illustrated in Fig. \ref{fig:chi2c-b2}.
   The results of the $CP$ asymmetries are presented in Tables
   \ref{tab:cp-bu}, \ref{tab:cp-bd-01} and \ref{tab:cp-bd-03}.
   It should be pointed out that the uncertainties of our results
   only come from the parameters ${\omega}_{B}$ and ${\mu}_{P}$
   based on the previous analysis of form factors.
   The uncertainties from other factors, such as the Gegenbauer
   moments\footnotemark[1],
   \footnotetext[1]{Recently,
   the effects from the Gegenbauer moments to the branching ratios
   for $B_{u,d,s}$ ${\to}$ $PP$, $PV$ decays
   have been carefully studied in Ref. \cite{PhysRevD.104.016025}.
   The Gegenbauer moments of the twist-2 and twist-3 LCDAs
   for some pseudoscalar and vector mesons have been determined
   with a relatively higher precision through a global fit
   between the LO PQCD results and available data,
   where the shape parameter ${\omega}_{B}$ for the $B$ mesonic
   WFs is fixed, ${\omega}_{B}$ $=$ $0.4$ GeV.}
   different models of the mesonic WFs and etc,
   are not carefully scrutinized here, but deserve a more
   dedicated study.
   \begin{figure}[ht]
   \includegraphics[width=0.3\textwidth]{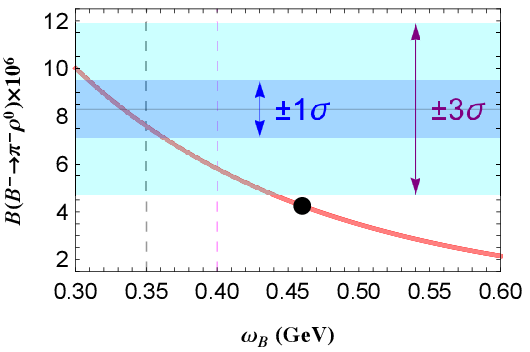} \quad
   \includegraphics[width=0.3\textwidth]{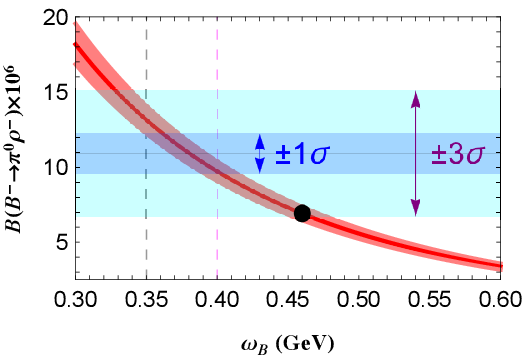} \quad
   \includegraphics[width=0.3\textwidth]{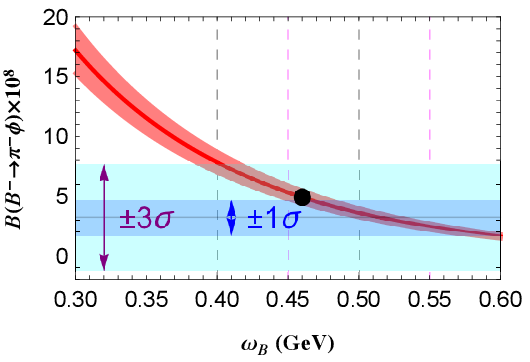}  \\
   \includegraphics[width=0.3\textwidth]{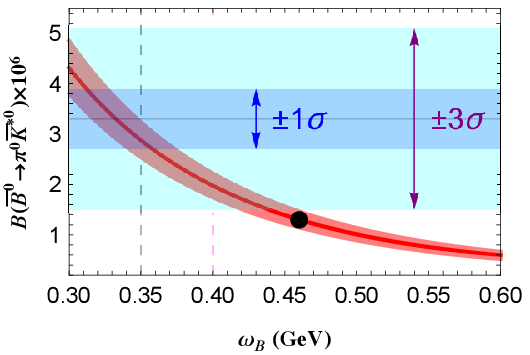}  \quad
   \includegraphics[width=0.3\textwidth]{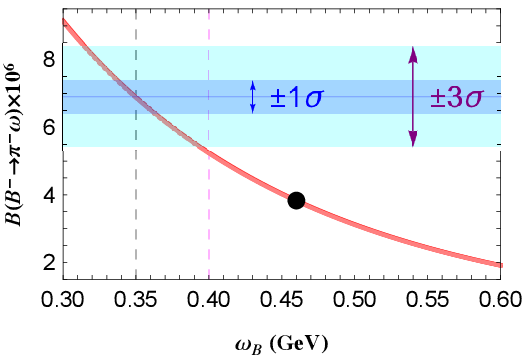}    \quad
   \includegraphics[width=0.3\textwidth]{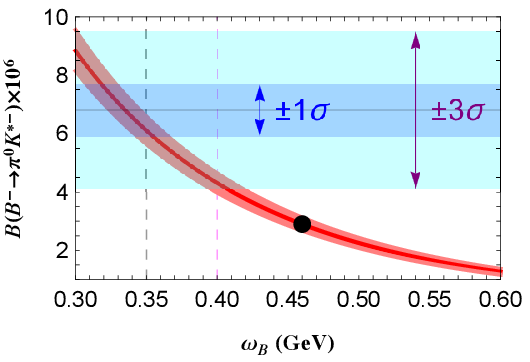}  \\
   \includegraphics[width=0.3\textwidth]{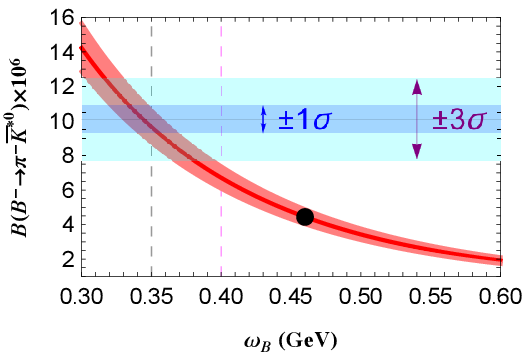}  \quad
   \includegraphics[width=0.3\textwidth]{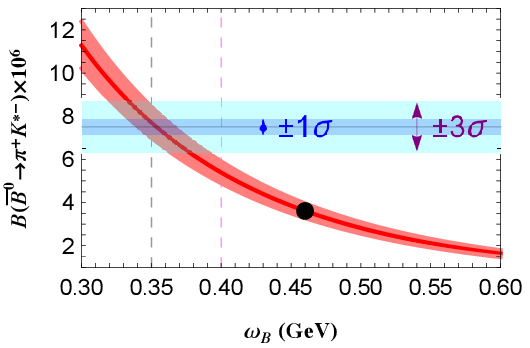}  \quad
   \includegraphics[width=0.3\textwidth]{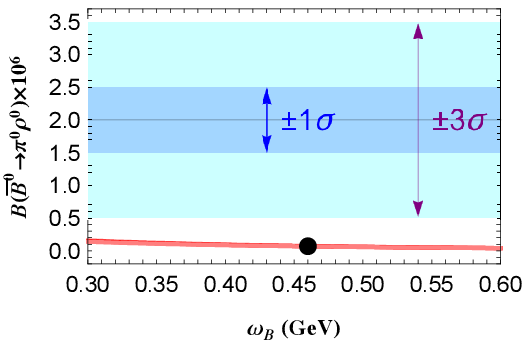} \\
   \includegraphics[width=0.3\textwidth]{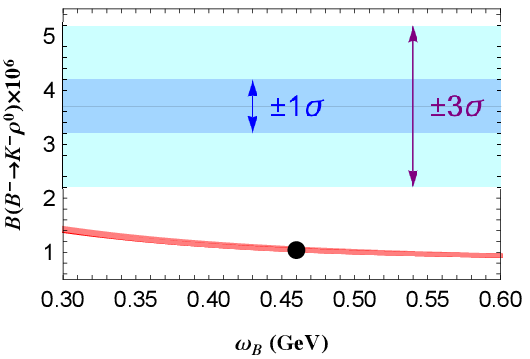}   \quad
   \includegraphics[width=0.3\textwidth]{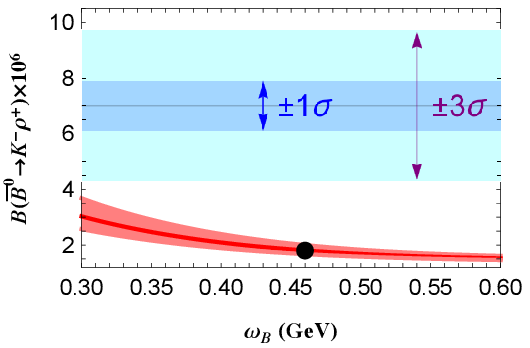}  \quad
   \includegraphics[width=0.3\textwidth]{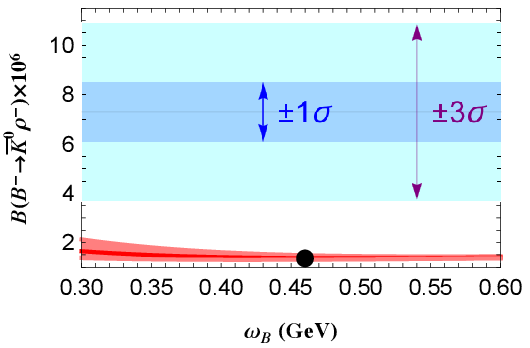}  \\
   \includegraphics[width=0.3\textwidth]{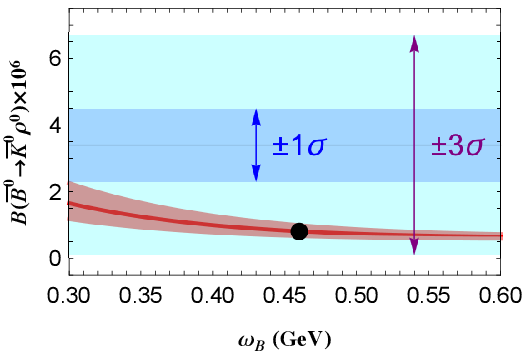}   \quad
   \includegraphics[width=0.3\textwidth]{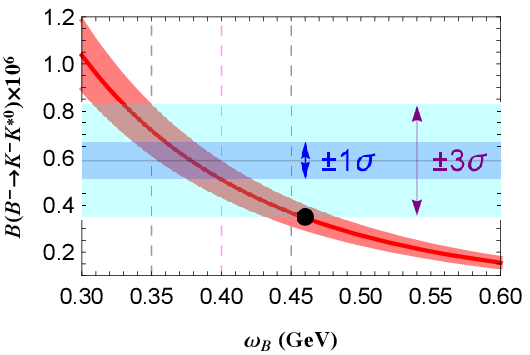}   \quad
   \includegraphics[width=0.3\textwidth]{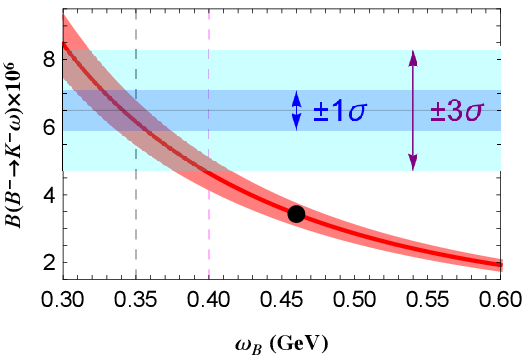}     \\
   \includegraphics[width=0.3\textwidth]{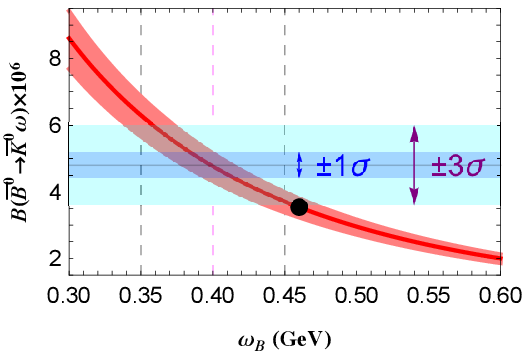}     \quad
   \includegraphics[width=0.3\textwidth]{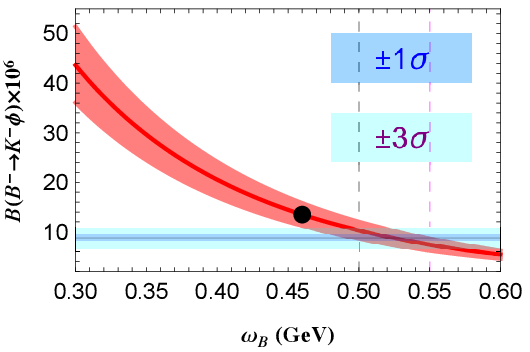}   \quad
   \includegraphics[width=0.3\textwidth]{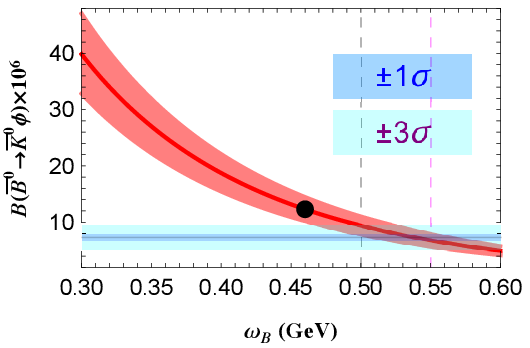}
   \caption{The branching ratios vs the shape parameter
   ${\omega}_{B}$, where the relatively narrower (wider)
   horizontal bands denote the PGD data within ${\pm}1\,{\sigma}$
   (${\pm}3\,{\sigma}$) regions, the curves in red are the PQCD results,
   including the contributions from ${\phi}_{B2}$, the curvy
   bands in pink denote the theoretical uncertainties from
   the chiral mass ${\mu}_{P}$ $=$ $1.4{\pm}0.1$ GeV, and the
   points in black denote the scenario S1 results.}
   \label{fig:br-w}
   \end{figure}

   (1)
   It is seen from Tables \ref{tab:branch-bu} and \ref{tab:branch-bd}
   that except for the $B^{-}$ ${\to}$ $\overline{K}^{0}{\rho}^{-}$
   and $K^{0}K^{{\ast}-}$ decays, the contributions from the $B$
   mesonic WFs ${\phi}_{B2}$ can enhance the branching ratios
   compared with those from ${\phi}_{B1}$.
   The contributions from ${\phi}_{B2}$ to the branching ratios are
   about $30\%$ and sometimes more, except for the $\overline{B}$
   ${\to}$ $K{\rho}$, $K{\omega}$, ${\pi}{\phi}$,
   ${\pi}^{0}{\rho}^{0}$, $K^{0}K^{{\ast}-}$ and
   $K^{-}K^{{\ast}+}$ decays.
   In addition, as shown in Tables \ref{tab:previous-branching-ratio},
   various results are obtained with the PQCD approach at
   the LO and NLO levels. The previous PQCD studies in Refs.
   \cite{epjc.72.1923,epjc.59.49,PhysRevD.74.094020}
   have shown that the NLO contributions can sometimes enhance and
   sometimes lessen the LO branching ratios.
   The shares from ${\phi}_{B2}$ to the branching ratios are
   comparable to the module of the shares from the NLO
   contributions.
   Taking the branching ratios for the $B^{-}$ ${\to}$
   ${\pi}^{-}{\rho}^{0}$ ($K^{-}K^{{\ast}0}$) decays as an example,
   the shares from ${\phi}_{B2}$ are about $37\%$ ($30\%$), and
   the shares from the NLO contributions are about $40\%$ \cite{epjc.72.1923}
   ($24\%$ \cite{epjc.59.49}).
   On the whole, considerably more effort is required to further improve
   the agreement between the theoretical results and data.

   (2)
   It is seen from Tables \ref{tab:cp-bu} and \ref{tab:cp-bd-01}
   that the participation of the WFs ${\phi}_{B2}$ results in a small
   reduction in the direct $CP$ asymmetries, except for the
   $\overline{B}$ ${\to}$ $K^{-}{\rho}$ and $K^{0}K^{{\ast}-}$
   decays.
   As is well known, the theoretical results of the $CP$
   asymmetries are highly sensitive to the strong phases.
   Therefore, it is essential to obtain the strong phases
   as accurately as possible.
   There are numerous sources of the strong phases, such as the higher
   order radiative corrections to HMEs, the final state interactions,
   and etcetera.
   In this paper, the $CP$ asymmetries are calculated at the LO
   order, and many factors that might affect these asymmetries
   are not carefully considered owing to our inadequate comprehension.
   For instance, there are still many theoretical and experimental
   discrepancies on the branching ratios.
   Therefore, our estimation of $CP$ asymmetries in
   Tables \ref{tab:cp-bu}, \ref{tab:cp-bd-01} and
   \ref{tab:cp-bd-03} cannot be taken too literally.
   Moreover, it is assumed that the current precision of most
   measurements of $CP$ asymmetries is too low to impose
   helpful constraints.
   Given the theoretical and experimental research status, the
   $CP$ asymmetries are not considered in the
   fit with Eq.(\ref{eq:chi2}).
   In addition, it is shown in the amplitudes Eq.(\ref{amp-pim-phi}),
   Eq.(\ref{piz-phi-amp}) and Eq.(\ref{kz-phi-amp}) that the tree
   amplitudes are absent and only the penguin contributions
   participate.
   These facts result in unavailable weak phase differences,
   which are an essential ingredient of the direct $CP$ asymmetries.
   So it is not surprising that the theoretical expectations of
   the direct $CP$ asymmetries for the $B^{-}$ ${\to}$
   ${\pi}^{-}{\phi}$ and $\overline{B}^{0}$ ${\to}$
   ${\pi}^{0}{\phi}$, $\overline{K}^{0}{\phi}$ decays are
   exactly zero.
   For the $\overline{B}^{0}$ ${\to}$ $\overline{K}^{0}K^{{\ast}0}$
   $+$ $K^{0}\overline{K}^{{\ast}0}$ decays which are induced
   by the pure penguin amplitudes in Eq.(\ref{kzb-kvz-amp}) and
   Eq.(\ref{kz-kvz-amp}), the parameters ${\lambda}_{f}$ of
   Eq.(\ref{pdr59.014005-eq.40-a}) and $\bar{\lambda}_{f}$ of
   Eq.(\ref{pdr59.014005-eq.40-b}) contain only the strong
   phase information. The measurements of the observables
   $a_{{\epsilon}^{\prime}}$ $=$ $-\bar{a}_{{\epsilon}^{\prime}}$
   and $a_{{\epsilon}+{\epsilon}^{\prime}}$ $=$
   $-\bar{a}_{{\epsilon}+{\epsilon}^{\prime}}$ would be helpful
   for testing our understanding on the strong interactions in
   nonleptonic $B$ meson weak decays.

   (3)
   It is clear in Fig. \ref{fig:chi2c-b2} that
   for the scenario S1, the goodness of fit between the PDG group
   data and PQCD results is still far from satisfactory.
   Among the 19 $B$ ${\to}$ $PV$ decays, there are only four decay
   modes with ${\chi}_{i}^{2}$ $<$ $9$, which indicates that the
   theoretical results on the branching ratios for the $B^{-}$
   ${\to}$ ${\pi}^{0}{\rho}^{-}$, ${\pi}^{-}{\phi}$,
   $\overline{B}^{0}$ ${\to}$ ${\pi}^{\pm}{\rho}^{\mp}$,
   $\overline{K}^{0}{\rho}^{0}$ decays agree with the PDG data
   within three standard experimental errors.
   The minimal ${\chi}_{i}^{2}$ ${\approx}$ $1.3$ is obtained for
   the $B^{-}$ ${\to}$ ${\pi}^{-}{\phi}$ decay, where the relative
   fitting error is significantly large and can reach up to about $47\%$.
   (Note: there is a general and conventional consensus in the
   elementary particle physics, {\em i.e.}, that a signal or event with
   a statistic significance of less than $3\,{\sigma}$, more than
   $3\,{\sigma}$, and more than $5\,{\sigma}$ are respectively
   known as a hint (or an indication), an evidence, and a discovery
   (or an observation or a confirmation), usually with a relative
   error greater than $33.3\%$, lower than $33.3\%$,
   and lower than $20\%$.)
   There are eight decay modes with ${\chi}_{i}^{2}$ $>$ $25$,
   which suggests that the discrepancies between the theoretical calculations
   and the data are larger than five standard experimental errors, and
   the theoretical results on the $B^{-}$ ${\to}$ ${\pi}^{-}{\omega}$,
   $K^{-}{\rho}^{0}$, $K^{-}{\omega}$, $K^{-}{\phi}$,
   ${\pi}^{-}\overline{K}^{{\ast}0}$ and $\overline{B}^{0}$ ${\to}$
   ${\pi}^{+}K^{{\ast}-}$, $K^{-}{\rho}^{+}$, $\overline{K}^{0}{\phi}$
   decays fail to provide a satisfactory explanation for the PDG data.
   The maximal ${\chi}_{i}^{2}$ ${\approx}$ $94$ is found for the
   $\overline{B}^{0}$ ${\to}$ ${\pi}^{+}K^{{\ast}-}$ decay, where
   the relative fitting error is significantly small at approximately $5\%$.
   It should be noted that the disagreement about the
   $\overline{B}^{0}$ ${\to}$ ${\pi}^{+}K^{{\ast}-}$ decay
   has been reported by previous PQCD studies in which only
   the ${\phi}_{B1}$ contributions were considered, for example,
   ${\chi}_{i}^{2}$ ${\approx}$ $36$ ($14$) with the LO (NLO)
   contributions \cite{PhysRevD.74.094020} and
   ${\chi}_{i}^{2}$ ${\approx}$ $41$ with the recent global analysis
   of $B$ decays \cite{PhysRevD.104.016025} at the LO level.
   For the scenario S2 (S3), there are seven (six) decay modes
   with ${\chi}_{i}^{2}$ $<$ $9$.
   $B^{-}$ ${\to}$ $K^{-}{\phi}$  is the only decay with
   ${\chi}_{i}^{2}$ just above $25$ in the scenario S3.
   In either scenario, both the $B^{-}$ ${\to}$ ${\pi}^{0}{\rho}^{-}$
   and $\overline{B}^{0}$ ${\to}$ ${\pi}^{\pm}{\rho}^{\mp}$ decays
   have ${\chi}_{i}^{2}$ $<$ $9$ at present.

   (4)
   It is seen from the Fig. \ref{fig:chi2c-b2} that there are
   four decays with ${\chi}_{i}^{2}$ ${\ge}$ $50$ in the
   scenario S1, including the $B^{-}$ ${\to}$ $K^{-}{\omega}$,
   ${\pi}^{-}\overline{K}^{{\ast}0}$ and $\overline{B}^{0}$
   ${\to}$ ${\pi}^{+}K^{{\ast}-}$, $\overline{K}^{0}{\phi}$
   decays.
   To further explore other possible underlying causes
   for the relatively larger ${\chi}_{\rm min}^{2}$ in the scenario
   S1, besides the relatively smaller errors of the PDG data,
   the relations of the branching ratios versus the shape
   parameter ${\omega}_{B}$ are shown in Fig. \ref{fig:br-w}.
   There are several clear and attractive phenomena evident in
   Fig. \ref{fig:br-w}.
   (i)
   Most of the branching ratios decrease with
   the increase of the parameter ${\omega}_{B}$.
   This situation is similar to that of form factors in
   Fig. \ref{contour:formfactor}.
   It is easy to understand this phenomenon, because
   the decay amplitudes are usually
   proportional to the formfactors.
   (ii)
   The current PDG data on the $B^{-}$ ${\to}$
   $\overline{K}^{0}{\rho}^{-}$, $K^{-}{\rho}^{0}$ and
   $\overline{B}^{0}$ ${\to}$ $K^{-}{\rho}^{+}$,
   ${\pi}^{0}{\rho}^{0}$ decays cannot be satisfactorily
   explained using the PQCD approach within $3\,{\sigma}$
   regions, no matter which value is taken for the
   parameter within $0.3$ GeV ${\le}$ ${\omega}_{B}$ ${\le}$
   $0.6$ GeV. These four decays contribute a large
   ${\chi}_{i}^{2}$ $>$ $9$.
   In addition, the branching ratios of these four decays are
   insensitive to the parameter ${\omega}_{B}$.
   (iii)
   When the pseudoscalar pion meson is one of the final states,
   branching ratios change significantly with the parameter
   ${\omega}_{B}$, except for the above mentioned
   $\overline{B}^{0}$ ${\to}$ ${\pi}^{0}{\rho}^{0}$ decay.
   A small ${\omega}_{B}$ $<$ $0.4$ GeV is commonly favored
   by most $B$ decays, except for the $B^{-}$ ${\to}$
   ${\pi}^{-}{\phi}$ decay.
   Although the PGD data opt for a large ${\omega}_{B}$ for the
   $B^{-}$ ${\to}$ ${\pi}^{-}{\phi}$ decay,
   the value of ${\omega}_{B}$ $=$ $0.4$ GeV can also marginally
   meet the experimental $3\,{\sigma}$ constraints.
   The scenario S1 parameter ${\omega}_{B}$ $=$ $0.46$ GeV is
   somewhat large and should be decreased for the $B$ ${\to}$
   ${\pi}V$ decays, which results in the extraordinarily
   large ${\chi}_{i}^{2}$ ${\approx}$ $38$, $50$, and $94$
   corresponding respectively to the
   $\overline{B}$ ${\to}$ ${\pi}^{-}{\omega}$,
   ${\pi}^{-}\overline{K}^{{\ast}0}$, and
   ${\pi}^{+}\overline{K}^{{\ast}0}$ decays.
   (iv)
   When the pseudoscalar kaon meson is one of the final states,
   the PDG data impose inconsistent inconsistent requirements on
   the parameter ${\omega}_{B}$, {\em i.e.}, ${\omega}_{B}$ $<$
   $0.45$ GeV for the $\overline{B}$ ${\to}$ $\overline{K}{\omega}$,
   $K^{-}K^{{\ast}0}$ decays, while ${\omega}_{B}$ $>$
   $0.5$ GeV for the $\overline{B}$ ${\to}$ $\overline{K}{\phi}$
   decays.
   The positive and negative deviations from ${\omega}_{B}$
   $=$ $0.46$ GeV facilitate ${\chi}_{i}^{2}$ for the
   $\overline{B}$ ${\to}$ $K^{-}{\omega}$, $K^{-}{\phi}$ and
   $\overline{K}^{0}{\phi}$ decays to $59$, $45$, and $51$,
   respectively.
   (v)
   The general conclusion about the scenario S1 is that
   on the one hand, a relatively small ${\omega}_{B}$ is
   favored by the $\overline{B}$ ${\to}$ $\overline{K}{\omega}$
   decays and most of the $\overline{B}$ ${\to}$ ${\pi}V$ decays.
   On the other hand, a relatively large ${\omega}_{B}$ is
   favored by the $\overline{B}$ ${\to}$ $P{\phi}$ decays.
   Furthermore, the $\overline{B}$ ${\to}$
   $\overline{K}{\rho}$ decays are insensitive to the
   changes of ${\omega}_{B}$.
   \begin{figure}[ht]
   \includegraphics[width=0.45\textwidth]{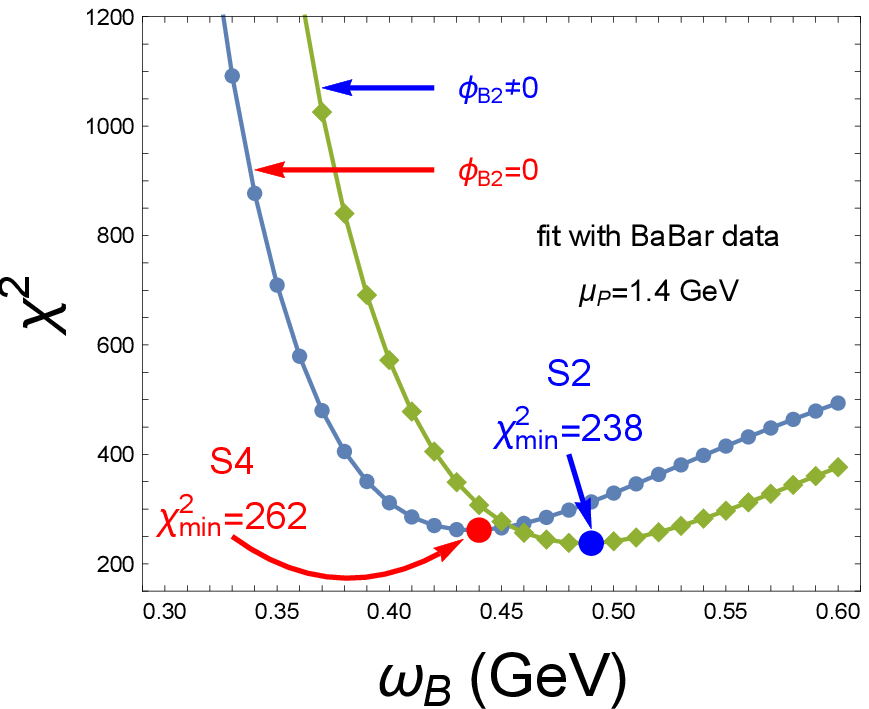} \qquad
   \includegraphics[width=0.45\textwidth]{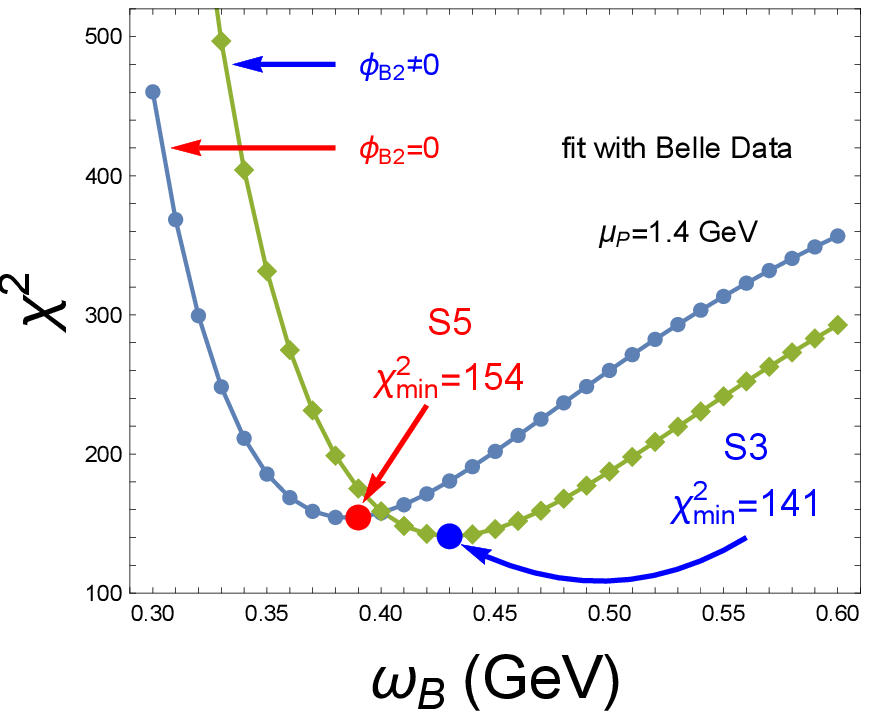}
   \caption{The distribution of ${\chi}^{2}$ versus the shape
   parameter ${\omega}_{B}$ with and without the contributions
   of ${\phi}_{B2}$.}
   \label{fig:chi2w-b1}
   \end{figure}
   \begin{figure}[ht]
   \includegraphics[width=0.45\textwidth,bb=10 25 270 200]{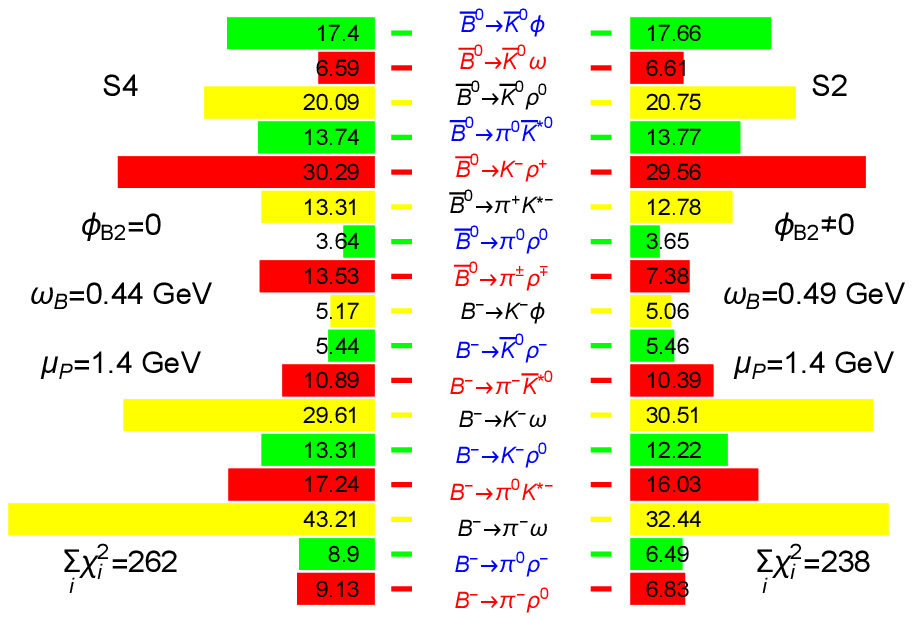}
   \includegraphics[width=0.45\textwidth,bb=10 30 270 200]{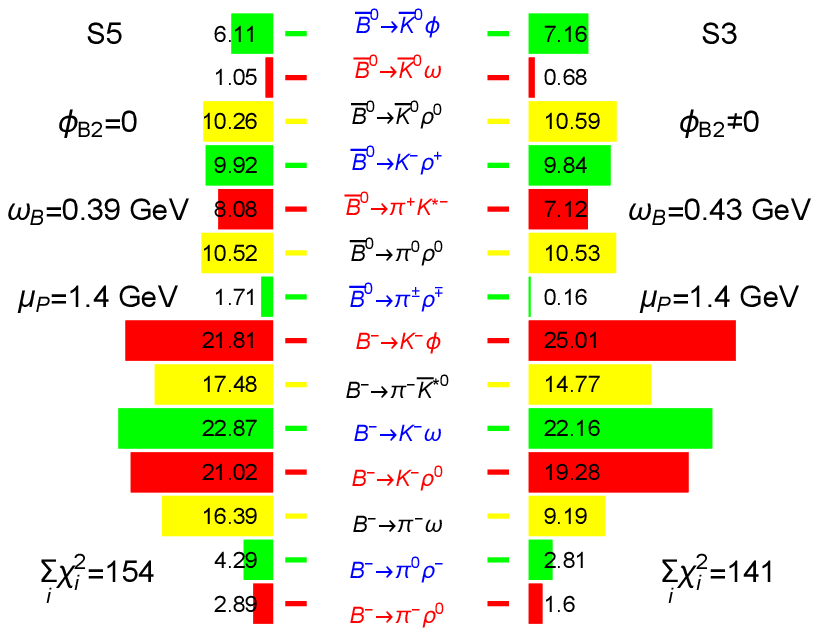}
   \caption{The ${\chi}_{i}^{2}$ distribution of the branching ratios
   with and without the contributions of ${\phi}_{B2}$,
   where the numbers in the barcharts denote the values of
   ${\chi}_{i}^{2}$ for a specific process.}
   \label{fig:chi2c-b1}
   \end{figure}

   (5)
   It is seen from Table \ref{tab:previous-branching-ratio} that
   previous PQCD results without the contributions from the $B$
   mesonic WFs ${\phi}_{B2}$ could also provide a satisfactory
   explanation for many of the $B$ ${\to}$ $PV$ decays by choosing
   appropriate parameters, such as ${\omega}_{B}$ and so on.
   In other words, to some extent, the effects of
   ${\phi}_{B2}$ on nonleptonic $B$ decays could be replaced
   by other scenarios of input parameters, which might be one
   reason why the contributions of ${\phi}_{B2}$ were often not
   seriously considered  in previous studies.
   To further illustrate the influences of ${\phi}_{B2}$ on the $B$
   ${\to}$ $PV$ decays, consistencies between the experimental
   data and the PQCD results with different ${\omega}_{B}$ are shown
   in Fig. \ref{fig:chi2w-b1} and \ref{fig:chi2c-b1}.
   It is clear from Fig. \ref{fig:chi2w-b1} that
   when contributions from the $B$ mesonic WFs ${\phi}_{B2}$
   are not taken into account, the optimal parameter ${\omega}_{B}$
   and the minimum ${\chi}^{2}$ corresponding to experimental data
   from the BaBar and Belle groups are, respectively,
   \begin{itemize}
   \item scenario 4 (S4): ${\omega}_{B}$ $=$ $0.44$ GeV from
      BaBar data with ${\chi}^{2}_{\rm min.}$ ${\approx}$ $262$,
   \item scenario 5 (S5): ${\omega}_{B}$ $=$ $0.39$ GeV from
      Belle data with ${\chi}^{2}_{\rm min.}$ ${\approx}$ $154$.
   \end{itemize}
   The value of ${\omega}_{B}$ for the scenario S4 (S5) is less
   than that for the scenario S2 (S3).
   The value of ${\chi}^{2}_{\rm min.}$ for the scenario S4 (S5)
   is larger than that for the scenario S2 (S3).
   It is seen from Fig. \ref{fig:chi2c-b1} that
   (i)
   there are five decay modes with ${\chi}^{2}_{i}$ $<$ $9$
   for the scenario S4, which is fewer than the seven decay modes
   with ${\chi}^{2}_{i}$ $<$ $9$ for the scenario S2.
   (ii)
   There are six same decay modes with ${\chi}^{2}_{i}$ $<$ $9$
   for both the scenario S3 and S5, including the
   $B^{-}$ ${\to}$ ${\pi}^{-}{\rho}^{0}$, ${\pi}^{0}{\rho}^{-}$
   and $\overline{B}^{0}$ ${\to}$ ${\pi}^{\pm}{\rho}^{\mp}$,
   ${\pi}^{+}\overline{K}^{{\ast}-}$, $\overline{K}^{0}{\omega}$,
   $\overline{K}^{0}{\phi}$ decays.
   Among these six same decays, except for the $\overline{B}^{0}$
   ${\to}$ $\overline{K}^{0}{\phi}$ decay, the ${\chi}^{2}_{i}$
   of the other decays for the scenario S5 are larger than
   that for the scenario S3.
   A conclusion from the comparative analysis of
   Fig. \ref{fig:chi2w-b1} and \ref{fig:chi2c-b1}
   is that a more comprehensive agreement of branching
   ratios between the PQCD calculations and experimental data
   can be improved by the participation of the $B$ mesonic WFs
   ${\phi}_{B2}$.

  \section{Summary}
  \label{sec:summary}
  In this paper, inspired by the experimental prospect of $B$
  meson physics, we reinvestigate the $B$ ${\to}$
  $PV$ decays ($P$ $=$ ${\pi}$ and $K$) at the LO order
  using the PQCD approach
  within the SM, by considering the $B$ mesonic WFs ${\phi}_{B2}$,
  which have been excluded in previous phenomenological studies.
  In the convolution integrals of HMEs of nonleptonic $B$
  decays, the WFs ${\phi}_{B2}$ are involved in the emission
  amplitudes and nonfactorizable annihilation amplitudes.
  The contributions from ${\phi}_{B2}$ can enhance the hadronic
  transition formfactors $F_{1}(0)$ and $A_{0}(0)$.
  The formfactors are highly sensitive to the shape parameter
  ${\omega}_{B}$ of $B$ mesonic WFs.
  By fitting the PQCD results on the branching ratios and
  experimental data using the minimum ${\chi}^{2}$ method,
  it is found that the participation of ${\phi}_{B2}$ is
  helpful for improving the comprehensive agreement between
  the PQCD calculations and experimental data.
  The shares of ${\phi}_{B2}$ should be given due attention and
  studied meticulously for nonleptonic $B$ decays.
  When the contributions from ${\phi}_{B2}$ are considered,
  three optimal scenarios for the parameter ${\omega}_{B}$ are found.
  The PQCD results on branching ratios and $CP$ asymmetries
  are updated with these three scenarios.
  It is found that with any one of these three scenarios,
  the ${\phi}_{B2}$ contributions can increase most branching
  ratios, except for the $B^{-}$ ${\to}$ $K^{0}K^{{\ast}-}$,
  $\overline{K}^{0}{\rho}^{-}$ decays, and on the other hand,
  these contributions
  can decrease most direct $CP$ asymmetries, except in the
  $\overline{B}$ ${\to}$ $K^{-}{\rho}$ and $K^{0}K^{{\ast}-}$ decays.
  At the same time, it should be noted that there are
  still several discrepancies between the PQCD results and
  available data to a greater or lesser extent.
  More worthwhile endeavors on nonleptonic $B$ decays
  are required, experimentally and theoretically.
  From the experimental perspective, an increasing number of accurate
  measurement results are being produced within existing and
  future experiments.
  From the theoretical perspective, at the quark level,
  other possible mechanisms of the interactions and higher
  order corrections to the scattering amplitudes are highly
  important; at the hadron level, some
  appropriate models for mesonic WFs including higher
  twist components are essential.

  \section*{Acknowledgements}
  The work is supported by the National Natural Science Foundation
  of China (Grant Nos. 11705047, U1632109, 11875122), the Natural
  Science Foundation of Henan Province (Grant No. 222300420479),
  the Excellent Youth Foundation
  of Henan Province (Grant No. 212300410010),  and the Youth Talent
  Support Program of Henan Province (Grant No. ZYQR201912178).

   \begin{widetext}
   \begin{appendix}

   \section{the amplitudes for the $B$ ${\to}$ $PV$ decays}
   \label{sec:mode}
   Using the $SU(3)$ flavor structure, a more concise and compact
   amplitude for the $B$ ${\to}$ $PP$,  $PV$ decays is given by
   Eq.(12) in Ref. \cite{PhysRevD.104.016025} with the PQCD approach.
   The analytical expressions are explicitly listed  below.
   \begin{eqnarray} & &
  {\cal A}(B_{u}^{-}{\to}{\pi}^{-}{\rho}^{0})
   \nonumber \\ &=&
   \frac{G_{F}}{2}\, V_{ub}\,V_{ud}^{\ast}\, \big\{
     a_{1}\, \big[
     {\cal A}_{ab}^{LL}({\pi},{\rho})
   + {\cal A}_{ef}^{LL}({\pi},{\rho})
   - {\cal A}_{ef}^{LL}({\rho},{\pi}) \big]
   + a_{2}\, {\cal A}_{ab}^{LL}({\rho},{\pi})
   \nonumber \\ & & \hspace{0.11\textwidth}
   + C_{2}\, \big[
     {\cal A}_{cd}^{LL}({\pi},{\rho})
   + {\cal A}_{gh}^{LL}({\pi},{\rho})
   - {\cal A}_{gh}^{LL}({\rho},{\pi}) \big]
   + C_{1}\, {\cal A}_{cd}^{LL}({\rho},{\pi}) \big\}
   \nonumber \\ &-&
   \frac{G_{F}}{2}\, V_{tb}\,V_{td}^{\ast}\, \big\{
     (a_{4}+a_{10}) \, \big[
     {\cal A}_{ab}^{LL}({\pi},{\rho})
   + {\cal A}_{ef}^{LL}({\pi},{\rho})
   - {\cal A}_{ef}^{LL}({\rho},{\pi}) \big]
   \nonumber \\ & & \hspace{0.12\textwidth}
   + (a_{6}+a_{8})\, \big[
     {\cal A}_{ab}^{SP}({\pi},{\rho})
   + {\cal A}_{ef}^{SP}({\pi},{\rho})
   - {\cal A}_{ef}^{SP}({\rho},{\pi}) \big]
   \nonumber \\ & & \hspace{0.12\textwidth}
   - (a_{4} -\frac{3}{2}\,a_{7}
   - \frac{3}{2}\,a_{9} -\frac{1}{2}\,a_{10} ) \,
     {\cal A}_{ab}^{LL}({\rho},{\pi})
   \nonumber \\ & & \hspace{0.12\textwidth}
   + (C_{3}+C_{9}) \, \big[
     {\cal A}_{cd}^{LL}({\pi},{\rho})
   + {\cal A}_{gh}^{LL}({\pi},{\rho})
   - {\cal A}_{gh}^{LL}({\rho},{\pi}) \big]
   \nonumber \\ & & \hspace{0.12\textwidth}
   + (C_{5}+C_{7})\, \big[
     {\cal A}_{cd}^{SP}({\pi},{\rho})
   + {\cal A}_{gh}^{SP}({\pi},{\rho})
   - {\cal A}_{gh}^{SP}({\rho},{\pi}) \big]
   \nonumber \\ & & \hspace{0.12\textwidth}
   - (C_{3} - \frac{3}{2}\,C_{10} -\frac{1}{2}\,C_{9} ) \,
     {\cal A}_{cd}^{LL}({\rho},{\pi})
   + \frac{3}{2}\,C_{8}\,
     {\cal A}_{cd}^{LR}({\rho},{\pi})
   \nonumber \\ & & \hspace{0.12\textwidth}
   - (C_{5}-\frac{1}{2}\,C_{7})\,
     {\cal A}_{cd}^{SP}({\rho},{\pi}) \big\}
   \label{amp-pim-rhoz},
   \end{eqnarray}
   \begin{eqnarray} & &
  {\cal A}(B_{u}^{-}{\to}{\rho}^{-}{\pi}^{0})
   \nonumber \\ &=&
   \frac{G_{F}}{2}\, V_{ub}\,V_{ud}^{\ast}\, \big\{
     a_{1}\, \big[
     {\cal A}_{ab}^{LL}({\rho},{\pi})
   + {\cal A}_{ef}^{LL}({\rho},{\pi})
   - {\cal A}_{ef}^{LL}({\pi},{\rho}) \big]
   + a_{2}\, {\cal A}_{ab}^{LL}({\pi},{\rho})
   \nonumber \\ & & \hspace{0.11\textwidth}
   + C_{2}\, \big[
     {\cal A}_{cd}^{LL}({\rho},{\pi})
   + {\cal A}_{gh}^{LL}({\rho},{\pi})
   - {\cal A}_{gh}^{LL}({\pi},{\rho}) \big]
   + C_{1}\, {\cal A}_{cd}^{LL}({\pi},{\rho}) \big\}
   \nonumber \\ &-&
   \frac{G_{F}}{2}\, V_{tb}\,V_{td}^{\ast}\, \big\{
     (a_{4}+a_{10}) \, \big[
     {\cal A}_{ab}^{LL}({\rho},{\pi})
   + {\cal A}_{ef}^{LL}({\rho},{\pi})
   - {\cal A}_{ef}^{LL}({\pi},{\rho}) \big]
   \nonumber \\ & & 
   + (a_{6}+a_{8})\, \big[
     {\cal A}_{ef}^{SP}({\rho},{\pi})
   - {\cal A}_{ef}^{SP}({\pi},{\rho}) \big]
   - (a_{6}-\frac{1}{2}\,a_{8} ) \,
     {\cal A}_{ab}^{SP}({\pi},{\rho})
   \nonumber \\ & &
   - (a_{4}+\frac{3}{2}\,a_{7}
   - \frac{3}{2}\,a_{9} -\frac{1}{2}\,a_{10} ) \,
     {\cal A}_{ab}^{LL}({\pi},{\rho})
   + (C_{5}+C_{7}) \, {\cal A}_{cd}^{SP}({\rho},{\pi})
   \nonumber \\ & & 
   + (C_{3}+C_{9}) \, \big[
     {\cal A}_{cd}^{LL}({\rho},{\pi})
   + {\cal A}_{gh}^{LL}({\rho},{\pi})
   - {\cal A}_{gh}^{LL}({\pi},{\rho}) \big]
   \nonumber \\ & & 
   + (C_{5}+C_{7})\, \big[
     {\cal A}_{gh}^{SP}({\rho},{\pi})
   - {\cal A}_{gh}^{SP}({\pi},{\rho}) \big]
   - (C_{5}-\frac{1}{2}\,C_{7})\,
     {\cal A}_{cd}^{SP}({\pi},{\rho})
   \nonumber \\ & & 
   - (C_{3} - \frac{3}{2}\,C_{10} -\frac{1}{2}\,C_{9} ) \,
     {\cal A}_{cd}^{LL}({\pi},{\rho})
   + \frac{3}{2}\,C_{8}\, {\cal A}_{cd}^{LR}({\pi},{\rho}) \big\}
   \label{amp-piz-rhom},
   \end{eqnarray}
   \begin{eqnarray} & &
  {\cal A}(B_{u}^{-}{\to}{\pi}^{-}{\omega})
   \nonumber \\ &=&
   \frac{G_{F}}{2}\, V_{ub}\,V_{ud}^{\ast}\, \big\{
     a_{1}\, \big[
     {\cal A}_{ab}^{LL}({\pi},{\omega})
   + {\cal A}_{ef}^{LL}({\pi},{\omega})
   + {\cal A}_{ef}^{LL}({\omega},{\pi}) \big]
   + a_{2}\, {\cal A}_{ab}^{LL}({\omega},{\pi})
   \nonumber \\ & & \hspace{0.11\textwidth}
   + C_{2}\, \big[
     {\cal A}_{cd}^{LL}({\pi},{\omega})
   + {\cal A}_{gh}^{LL}({\pi},{\omega})
   + {\cal A}_{gh}^{LL}({\omega},{\pi}) \big]
   + C_{1}\, {\cal A}_{cd}^{LL}({\omega},{\pi}) \big\}
   \nonumber \\ &-&
   \frac{G_{F}}{2}\, V_{tb}\,V_{td}^{\ast}\, \big\{
     (a_{4}+a_{10}) \, \big[
     {\cal A}_{ab}^{LL}({\pi},{\omega})
   + {\cal A}_{ef}^{LL}({\pi},{\omega})
   + {\cal A}_{ef}^{LL}({\omega},{\pi}) \big]
   \nonumber \\ & & \hspace{0.12\textwidth}
   + (a_{6}+a_{8})\, \big[
     {\cal A}_{ab}^{SP}({\pi},{\omega})
   + {\cal A}_{ef}^{SP}({\pi},{\omega})
   + {\cal A}_{ef}^{SP}({\omega},{\pi}) \big]
   \nonumber \\ & & \hspace{0.12\textwidth}
   + (2\,a_{3}+a_{4} +2\,a_{5}+ \frac{1}{2}\,a_{7}
   + \frac{1}{2}\,a_{9} -\frac{1}{2}\,a_{10} ) \,
     {\cal A}_{ab}^{LL}({\omega},{\pi})
   \nonumber \\ & & \hspace{0.12\textwidth}
   + (C_{3}+C_{9}) \, \big[
     {\cal A}_{cd}^{LL}({\pi},{\omega})
   + {\cal A}_{gh}^{LL}({\pi},{\omega})
   + {\cal A}_{gh}^{LL}({\omega},{\pi}) \big]
   \nonumber \\ & & \hspace{0.12\textwidth}
   + (C_{5}+C_{7})\, \big[
     {\cal A}_{cd}^{SP}({\pi},{\omega})
   + {\cal A}_{gh}^{SP}({\pi},{\omega})
   + {\cal A}_{gh}^{SP}({\omega},{\pi}) \big]
   \nonumber \\ & & \hspace{0.12\textwidth}
   + (C_{3} +2\,C_{4} - \frac{1}{2}\,C_{9} + \frac{1}{2}\,C_{10} ) \,
     {\cal A}_{cd}^{LL}({\omega},{\pi})
   \nonumber \\ & & \hspace{0.12\textwidth}
   + (2\,C_{6}+\frac{1}{2}\,C_{8} )\,
     {\cal A}_{cd}^{LR}({\omega},{\pi})
   + (C_{5}-\frac{1}{2}\,C_{7})\,
     {\cal A}_{cd}^{SP}({\omega},{\pi}) \big\}
   \label{amp-pim-omega},
   \end{eqnarray}
   \begin{eqnarray} & &
  {\cal A}(B_{u}^{-}{\to}{\pi}^{-}{\phi})
   \nonumber \\ &=&
   - \frac{G_{F}}{\sqrt{2}}\, V_{tb}\,V_{td}^{\ast}\, \big\{
     (a_{3}+a_{5}-\frac{1}{2}\,a_{7}-\frac{1}{2}\,a_{9}) \,
     {\cal A}_{ab}^{LL}({\phi},{\pi})
   \nonumber \\ & &
   + (C_{4}-\frac{1}{2}\,C_{10})\,
     {\cal A}_{cd}^{LL}({\phi},{\pi})
   + (C_{6}-\frac{1}{2}\,C_{8})\,
     {\cal A}_{cd}^{LR}({\phi},{\pi}) \big\}
   \label{amp-pim-phi},
   \end{eqnarray}
    \begin{eqnarray} & &
   {\cal A}(B_{u}^{-}{\to}K^{-}K^{{\ast}0})
    \nonumber \\ &=&
    \frac{G_{F}}{\sqrt{2}}\, V_{ub}\,V_{ud}^{\ast}\,
    \big\{ a_{1}\, {\cal A}_{ef}^{LL}(K^{\ast},\overline{K})
          +C_{2}\, {\cal A}_{gh}^{LL}(K^{\ast},\overline{K}) \big\}
    \nonumber \\ &-&
    \frac{G_{F}}{\sqrt{2}}\, V_{tb}\,V_{td}^{\ast}\, \big\{
    ( a_{4}-\frac{1}{2}\,a_{10} )\,
    {\cal A}_{ab}^{LL}(K^{\ast},\overline{K})
   +( C_{3}-\frac{1}{2}\,C_{9} )\,
    {\cal A}_{cd}^{LL}(K^{\ast},\overline{K})
    \nonumber \\ & & \hspace{0.11\textwidth}
   +( a_{4}+a_{10} )\,
    {\cal A}_{ef}^{LL}(K^{\ast},\overline{K})
   +( C_{3}+C_{9} )\,
    {\cal A}_{gh}^{LL}(K^{\ast},\overline{K})
    \nonumber \\ & & \hspace{0.11\textwidth}
   +( a_{6}+a_{8} )\,
    {\cal A}_{ef}^{SP}(K^{\ast},\overline{K})
   +( C_{5}+C_{7} )\,
    {\cal A}_{gh}^{SP}(K^{\ast},\overline{K})
    \nonumber \\ & & \hspace{0.11\textwidth}
   +( C_{5}-\frac{1}{2}\,C_{7} )\,
    {\cal A}_{cd}^{SP}(K^{\ast},\overline{K}) \big\}
   \label{amp-bu-km-kvz-amp},
   \end{eqnarray}
    \begin{eqnarray} & &
   {\cal A}(B_{u}^{-}{\to}K^{{\ast}-}K^{0})
    \nonumber \\ &=&
    \frac{G_{F}}{\sqrt{2}}\, V_{ub}\,V_{ud}^{\ast}\,
    \big\{ a_{1}\, {\cal A}_{ef}^{LL}(K,\overline{K}^{\ast})
          +C_{2}\, {\cal A}_{gh}^{LL}(K,\overline{K}^{\ast}) \big\}
    \nonumber \\ &-&
    \frac{G_{F}}{\sqrt{2}}\, V_{tb}\,V_{td}^{\ast}\, \big\{
    ( a_{4}-\frac{1}{2}\,a_{10} )\,
    {\cal A}_{ab}^{LL}(K,\overline{K}^{\ast})
   +( a_{4}+a_{10} )\,
    {\cal A}_{ef}^{LL}(K,\overline{K}^{\ast})
    \nonumber \\ & & \hspace{0.11\textwidth}
   +( a_{6}-\frac{1}{2}\,a_{8} )\,
    {\cal A}_{ab}^{SP}(K,\overline{K}^{\ast})
   +( a_{6}+a_{8} )\,
    {\cal A}_{ef}^{SP}(K,\overline{K}^{\ast})
     \nonumber \\ & & \hspace{0.11\textwidth}
   +( C_{3}-\frac{1}{2}\,C_{9} )\,
    {\cal A}_{cd}^{LL}(K,\overline{K}^{\ast})
   +( C_{3}+C_{9} )\,
    {\cal A}_{gh}^{LL}(K,\overline{K}^{\ast})
    \nonumber \\ & & \hspace{0.11\textwidth}
   +( C_{5}-\frac{1}{2}\,C_{7} )\,
    {\cal A}_{cd}^{SP}(K,\overline{K}^{\ast})
   +( C_{5}+C_{7} )\,
    {\cal A}_{gh}^{SP}(K,\overline{K}^{\ast}) \big\}
   \label{amp-bu-kz-kvm-amp},
   \end{eqnarray}
   \begin{eqnarray} & &
  {\cal A}(B_{u}^{-}{\to}K^{{\ast}-}{\pi}^{0})
   \nonumber \\ &=&
   \frac{G_{F}}{2}\, V_{ub}\,V_{us}^{\ast}\,
   \big\{ a_{1}\, \big[
     {\cal A}_{ab}^{LL}(\overline{K}^{\ast},{\pi})
   + {\cal A}_{ef}^{LL}(\overline{K}^{\ast},{\pi}) \big]
   + a_{2}\,{\cal A}_{ab}^{LL}({\pi},\overline{K}^{\ast})
   \nonumber \\ & & \hspace{0.11\textwidth}
   + C_{2} \big[
     {\cal A}_{cd}^{LL}(\overline{K}^{\ast},{\pi})
   + {\cal A}_{gh}^{LL}(\overline{K}^{\ast},{\pi}) \big]
   + C_{1}\,{\cal A}_{cd}^{LL}({\pi},\overline{K}^{\ast}) \big\}
   \nonumber \\ &-&
   \frac{G_{F}}{2}\, V_{tb}\,V_{ts}^{\ast}\, \big\{
     ( a_{4}+a_{10} )\, \big[
     {\cal A}_{ab}^{LL}(\overline{K}^{\ast},{\pi})
   + {\cal A}_{ef}^{LL}(\overline{K}^{\ast},{\pi}) \big]
   \nonumber \\ & & \hspace{0.11\textwidth}
   + ( a_{6}+a_{8} )\,
     {\cal A}_{ef}^{SP}(\overline{K}^{\ast},{\pi})
   - \frac{3}{2}\, (a_{7}-a_{9})\,
     {\cal A}_{ab}^{LL}({\pi},\overline{K}^{\ast})
   \nonumber \\ & & \hspace{0.11\textwidth}
   + ( C_{3}+C_{9} )\, \big[
     {\cal A}_{cd}^{LL}(\overline{K}^{\ast},{\pi})
   + {\cal A}_{gh}^{LL}(\overline{K}^{\ast},{\pi}) \big]
   \nonumber \\ & & \hspace{0.11\textwidth}
   + ( C_{5}+C_{7} )\, \big[
     {\cal A}_{cd}^{SP}(\overline{K}^{\ast},{\pi})
   + {\cal A}_{gh}^{SP}(\overline{K}^{\ast},{\pi}) \big]
   \nonumber \\ & & \hspace{0.11\textwidth}
   + \frac{3}{2}\, C_{8}\,
     {\cal A}_{cd}^{LR}({\pi},\overline{K}^{\ast})
   + \frac{3}{2}\, C_{10}\,
     {\cal A}_{cd}^{LL}({\pi},\overline{K}^{\ast}) \big\}
   \label{piz-kvm-amp},
   \end{eqnarray}
   \begin{eqnarray} & &
  {\cal A}(B_{u}^{-}{\to}K^{-}{\rho}^{0})
   \nonumber \\ &=&
   \frac{G_{F}}{2}\, V_{ub}\,V_{us}^{\ast}\,
   \big\{ a_{1}\, \big[
     {\cal A}_{ab}^{LL}(\overline{K},{\rho})
   + {\cal A}_{ef}^{LL}(\overline{K},{\rho}) \big]
   + a_{2}\,{\cal A}_{ab}^{LL}({\rho},\overline{K})
   \nonumber \\ & & \hspace{0.11\textwidth}
   + C_{2} \big[
     {\cal A}_{cd}^{LL}(\overline{K},{\rho})
   + {\cal A}_{gh}^{LL}(\overline{K},{\rho}) \big]
   + C_{1}\,{\cal A}_{cd}^{LL}({\rho},\overline{K}) \big\}
   \nonumber \\ &-&
   \frac{G_{F}}{2}\, V_{tb}\,V_{ts}^{\ast}\, \big\{
     ( a_{4}+a_{10} )\, \big[
     {\cal A}_{ab}^{LL}(\overline{K},{\rho})
   + {\cal A}_{ef}^{LL}(\overline{K},{\rho}) \big]
   \nonumber \\ & & \hspace{0.11\textwidth}
   + ( a_{6}+a_{8} )\, \big[
     {\cal A}_{ab}^{SP}(\overline{K},{\rho})
   + {\cal A}_{ef}^{SP}(\overline{K},{\rho}) \big]
   \nonumber \\ & & \hspace{0.11\textwidth}
   + \frac{3}{2}\, (a_{7}+a_{9})\,
     {\cal A}_{ab}^{LL}({\rho},\overline{K})
   \nonumber \\ & & \hspace{0.11\textwidth}
   + ( C_{3}+C_{9} )\, \big[
     {\cal A}_{cd}^{LL}(\overline{K},{\rho})
   + {\cal A}_{gh}^{LL}(\overline{K},{\rho}) \big]
   \nonumber \\ & & \hspace{0.11\textwidth}
   + ( C_{5}+C_{7} )\, \big[
     {\cal A}_{cd}^{SP}(\overline{K},{\rho})
   + {\cal A}_{gh}^{SP}(\overline{K},{\rho}) \big]
   \nonumber \\ & & \hspace{0.11\textwidth}
   + \frac{3}{2}\, C_{8}\,
     {\cal A}_{cd}^{LR}({\rho},\overline{K})
   + \frac{3}{2}\, C_{10}\,
     {\cal A}_{cd}^{LL}({\rho},\overline{K}) \big\}
   \label{km-rhoz-amp},
   \end{eqnarray}
   \begin{eqnarray} & &
  {\cal A}(B_{u}^{-}{\to}K^{-}{\omega})
   \nonumber \\ &=&
   \frac{G_{F}}{2}\, V_{ub}\,V_{us}^{\ast}\,
   \big\{ a_{1}\, \big[
     {\cal A}_{ab}^{LL}(\overline{K},{\omega})
   + {\cal A}_{ef}^{LL}(\overline{K},{\omega}) \big]
   + a_{2}\,{\cal A}_{ab}^{LL}({\omega},\overline{K})
   \nonumber \\ & & \hspace{0.11\textwidth}
   + C_{2} \big[
     {\cal A}_{cd}^{LL}(\overline{K},{\omega})
   + {\cal A}_{gh}^{LL}(\overline{K},{\omega}) \big]
   + C_{1}\,{\cal A}_{cd}^{LL}({\omega},\overline{K}) \big\}
   \nonumber \\ &-&
   \frac{G_{F}}{2}\, V_{tb}\,V_{ts}^{\ast}\, \big\{
     ( a_{4}+a_{10} )\, \big[
     {\cal A}_{ab}^{LL}(\overline{K},{\omega})
   + {\cal A}_{ef}^{LL}(\overline{K},{\omega}) \big]
   \nonumber \\ & & \hspace{0.11\textwidth}
   + ( a_{6}+a_{8} )\, \big[
     {\cal A}_{ab}^{SP}(\overline{K},{\omega})
   + {\cal A}_{ef}^{SP}(\overline{K},{\omega}) \big]
   \nonumber \\ & & \hspace{0.11\textwidth}
   + (2\,a_{3}+2\,a_{5}+\frac{1}{2}\,a_{7}+\frac{1}{2}\,a_{9})\,
     {\cal A}_{ab}^{LL}({\omega},\overline{K})
   \nonumber \\ & & \hspace{0.11\textwidth}
   + ( C_{3}+C_{9} )\, \big[
     {\cal A}_{cd}^{LL}(\overline{K},{\omega})
   + {\cal A}_{gh}^{LL}(\overline{K},{\omega}) \big]
   \nonumber \\ & & \hspace{0.11\textwidth}
   + ( C_{5}+C_{7} )\, \big[
     {\cal A}_{cd}^{SP}(\overline{K},{\omega})
   + {\cal A}_{gh}^{SP}(\overline{K},{\omega}) \big]
   \nonumber \\ & & \hspace{0.11\textwidth}
   + (2\,C_{6}+\frac{1}{2}\, C_{8})\,
     {\cal A}_{cd}^{LR}({\omega},\overline{K})
   + (2\,C_{4}+\frac{1}{2}\, C_{10})\,
     {\cal A}_{cd}^{LL}({\omega},\overline{K}) \big\}
   \label{km-w-amp},
   \end{eqnarray}
   \begin{eqnarray} & &
  {\cal A}(B_{u}^{-}{\to}{\pi}^{-}\overline{K}^{{\ast}0})
   \nonumber \\ &=&
   \frac{G_{F}}{\sqrt{2}}\, V_{ub}\,V_{us}^{\ast}\,
   \big\{ a_{1}\, {\cal A}_{ef}^{LL}(\overline{K}^{\ast},{\pi})
   + C_{2}\, {\cal A}_{gh}^{LL}(\overline{K}^{\ast},{\pi}) \big\}
   \nonumber \\ &-&
   \frac{G_{F}}{\sqrt{2}}\, V_{tb}\,V_{ts}^{\ast}\,
   \big\{ (a_{4}-\frac{1}{2}\,a_{10})\,
  {\cal A}_{ab}^{LL}(\overline{K}^{\ast},{\pi})
  + (C_{3}-\frac{1}{2}\,C_{9})\,
  {\cal A}_{cd}^{LL}(\overline{K}^{\ast},{\pi})
   \nonumber \\ & & \hspace{0.12\textwidth}
   + (a_{4}+a_{10})\,
  {\cal A}_{ef}^{LL}(\overline{K}^{\ast},{\pi})
  + (C_{3}+C_{9})\,
  {\cal A}_{gh}^{LL}(\overline{K}^{\ast},{\pi})
   \nonumber \\ & & \hspace{0.12\textwidth}
   + (a_{6}+a_{8})\,
  {\cal A}_{ef}^{SP}(\overline{K}^{\ast},{\pi})
   + (C_{5}+C_{7})\,
  {\cal A}_{gh}^{SP}(\overline{K}^{\ast},{\pi})
   \nonumber \\ & & \hspace{0.12\textwidth}
   + (C_{5}-\frac{1}{2}\,C_{7})\,
  {\cal A}_{cd}^{SP}(\overline{K}^{\ast},{\pi}) \big\}
   \label{pim-kvz-amp},
   \end{eqnarray}
   \begin{eqnarray} & &
  {\cal A}(B_{u}^{-}{\to}{\rho}^{-}\overline{K}^{0})
   \nonumber \\ &=&
   \frac{G_{F}}{\sqrt{2}}\, V_{ub}\,V_{us}^{\ast}\,
   \big\{ a_{1}\, {\cal A}_{ef}^{LL}(\overline{K},{\rho})
   + C_{2}\, {\cal A}_{gh}^{LL}(\overline{K},{\rho}) \big\}
   \nonumber \\ &-&
   \frac{G_{F}}{\sqrt{2}}\, V_{tb}\,V_{ts}^{\ast}\,
   \big\{ (a_{4}-\frac{1}{2}\,a_{10})\,
  {\cal A}_{ab}^{LL}(\overline{K},{\rho})
  + (C_{3}-\frac{1}{2}\,C_{9})\,
  {\cal A}_{cd}^{LL}(\overline{K},{\rho})
   \nonumber \\ & & \hspace{0.12\textwidth}
   + (a_{4}+a_{10})\,
  {\cal A}_{ef}^{LL}(\overline{K},{\rho})
  + (C_{3}+C_{9})\,
  {\cal A}_{gh}^{LL}(\overline{K},{\rho})
   \nonumber \\ & & \hspace{0.12\textwidth}
   + (a_{6}-\frac{1}{2}\,a_{8})\,
  {\cal A}_{ab}^{SP}(\overline{K},{\rho})
   + (C_{5}-\frac{1}{2}\,C_{7})\,
  {\cal A}_{cd}^{SP}(\overline{K},{\rho})
   \nonumber \\ & & \hspace{0.12\textwidth}
   + (a_{6}+a_{8})\,
  {\cal A}_{ef}^{SP}(\overline{K},{\rho})
   + (C_{5}+C_{7})\,
  {\cal A}_{gh}^{SP}(\overline{K},{\rho}) \big\}
   \label{amp-kz-rhom},
   \end{eqnarray}
   \begin{eqnarray} & &
  {\cal A}(B_{u}^{-}{\to}K^{-}{\phi})
   \nonumber \\ &=&
   \frac{G_{F}}{\sqrt{2}}\, V_{ub}\,V_{us}^{\ast}\, \big\{
     a_{1}\, {\cal A}_{ef}^{LL}({\phi},\overline{K})
    +C_{2}\, {\cal A}_{gh}^{LL}({\phi},\overline{K}) \big\}
   \nonumber \\ &-&
     \frac{G_{F}}{\sqrt{2}}\, V_{tb}\,V_{ts}^{\ast}\, \big\{
     ( a_{3}+a_{4}+a_{5}-\frac{1}{2}\,a_{7}
     -\frac{1}{2}\,a_{9}-\frac{1}{2}\,a_{10} )\,
     {\cal A}_{ab}^{LL}({\phi},\overline{K})
   \nonumber \\ & & \hspace{0.12\textwidth}
  +  ( a_{4}+a_{10} )\,{\cal A}_{ef}^{LL}({\phi},\overline{K})
  +  ( a_{6}+a_{8} )\,{\cal A}_{ef}^{SP}({\phi},\overline{K})
   \nonumber \\ & & \hspace{0.12\textwidth}
   + ( C_{3}+C_{4}-\frac{1}{2}\,C_{9}-\frac{1}{2}\,C_{10} )\,
     {\cal A}_{cd}^{LL}({\phi},\overline{K})
     \nonumber \\ & & \hspace{0.12\textwidth}
   + ( C_{6}-\frac{1}{2}\,C_{8} )\,
     {\cal A}_{cd}^{LR}({\phi},\overline{K})
   + ( C_{5}-\frac{1}{2}\,C_{7} )\,
     {\cal A}_{cd}^{SP}({\phi},\overline{K})
   \nonumber \\ & & \hspace{0.12\textwidth}
  +  ( C_{3}+C_{9} )\,{\cal A}_{gh}^{LL}({\phi},\overline{K})
  +  ( C_{5}+C_{7} )\,{\cal A}_{gh}^{SP}({\phi},\overline{K})  \big\}
   \label{km-phi-amp},
   \end{eqnarray}
   \begin{eqnarray} & &
  {\cal A}(\overline{B}_{d}^{0}{\to}{\rho}^{-}{\pi}^{+})
   \nonumber \\ &=&
   \frac{G_{F}}{\sqrt{2}}\, V_{ub}\,V_{ud}^{\ast}\, \big\{
    a_{1}\, {\cal A}_{ab}^{LL}({\rho},{\pi})
  + C_{2}\, {\cal A}_{cd}^{LL}({\rho},{\pi})
  + a_{2}\, {\cal A}_{ef}^{LL}({\pi},{\rho})
  + C_{1}\, {\cal A}_{gh}^{LL}({\pi},{\rho}) \big\}
   \nonumber \\ &-&
   \frac{G_{F}}{\sqrt{2}}\, V_{tb}\,V_{td}^{\ast}\, \big\{
    (a_{4}+a_{10})\, {\cal A}_{ab}^{LL}({\rho},{\pi})
  + (C_{3}+C_{9})\,  {\cal A}_{cd}^{LL}({\rho},{\pi})
   \nonumber \\ & & \hspace{0.12\textwidth}
   +(a_{3}+a_{4}-a_{5}+\frac{1}{2}\,a_{7}
    -\frac{1}{2}\,a_{9}-\frac{1}{2}\,a_{10} )\,
    {\cal A}_{ef}^{LL}({\rho},{\pi})
   \nonumber \\ & & \hspace{0.12\textwidth}
   +(C_{3}+C_{4}-\frac{1}{2}\,C_{9}-\frac{1}{2}\,C_{10} )\,
    {\cal A}_{gh}^{LL}({\rho},{\pi})
   +(C_{6}-\frac{1}{2}\,C_{8} )\,
    {\cal A}_{gh}^{LR}({\rho},{\pi})
   \nonumber \\ & & \hspace{0.12\textwidth}
   + (a_{6}-\frac{1}{2}\,a_{8}) \,
     {\cal A}_{ef}^{SP}({\rho},{\pi})
   + (C_{5}-\frac{1}{2}\,C_{7}) \,
     {\cal A}_{gh}^{SP}({\rho},{\pi})
   \nonumber \\ & & \hspace{0.12\textwidth}
    + (a_{3}-a_{5}-a_{7}+a_{9})\,
     {\cal A}_{ef}^{LL}({\pi},{\rho})
    + (C_{4}+C_{10})\,
     {\cal A}_{gh}^{LL}({\pi},{\rho})
   \nonumber \\ & & \hspace{0.12\textwidth}
   + (C_{6}+C_{8})\,
     {\cal A}_{gh}^{LR}({\pi},{\rho})
    + (C_{5}+C_{7})\,
     {\cal A}_{cd}^{SP}({\rho},{\pi}) \big\}
   \label{pip-rhom-amp},
   \end{eqnarray}
   \begin{eqnarray} & &
  {\cal A}(\overline{B}_{d}^{0}{\to}{\pi}^{-}{\rho}^{+})
   \nonumber \\ &=&
   \frac{G_{F}}{\sqrt{2}}\, V_{ub}\,V_{ud}^{\ast}\, \big\{
    a_{1}\, {\cal A}_{ab}^{LL}({\pi},{\rho})
  + C_{2}\, {\cal A}_{cd}^{LL}({\pi},{\rho})
  + a_{2}\, {\cal A}_{ef}^{LL}({\rho},{\pi})
  + C_{1}\, {\cal A}_{gh}^{LL}({\rho},{\pi}) \big\}
   \nonumber \\ &-&
   \frac{G_{F}}{\sqrt{2}}\, V_{tb}\,V_{td}^{\ast}\, \big\{
    (a_{4}+a_{10})\, {\cal A}_{ab}^{LL}({\pi},{\rho})
  + (C_{3}+C_{9})\,  {\cal A}_{cd}^{LL}({\pi},{\rho})
   \nonumber \\ & & \hspace{0.12\textwidth}
  + (a_{6}+a_{8})\, {\cal A}_{ab}^{SP}({\pi},{\rho})
  + (C_{5}+C_{7})\, {\cal A}_{cd}^{SP}({\pi},{\rho})
   \nonumber \\ & & \hspace{0.12\textwidth}
   +(a_{3}+a_{4}-a_{5}+\frac{1}{2}\,a_{7}
    -\frac{1}{2}\,a_{9}-\frac{1}{2}\,a_{10} )\,
    {\cal A}_{ef}^{LL}({\pi},{\rho})
   \nonumber \\ & & \hspace{0.12\textwidth}
   +(C_{3}+C_{4}-\frac{1}{2}\,C_{9}-\frac{1}{2}\,C_{10} )\,
    {\cal A}_{gh}^{LL}({\pi},{\rho})
   +(C_{6}-\frac{1}{2}\,C_{8} )\,
    {\cal A}_{gh}^{LR}({\pi},{\rho})
   \nonumber \\ & & \hspace{0.12\textwidth}
   + (a_{6}-\frac{1}{2}\,a_{8}) \,
     {\cal A}_{ef}^{SP}({\pi},{\rho})
   + (C_{5}-\frac{1}{2}\,C_{7}) \,
     {\cal A}_{gh}^{SP}({\pi},{\rho})
   \nonumber \\ & & \hspace{0.12\textwidth}
    + (a_{3}-a_{5}-a_{7}+a_{9})\,
     {\cal A}_{ef}^{LL}({\rho},{\pi})
    + (C_{4}+C_{10})\,
     {\cal A}_{gh}^{LL}({\rho},{\pi})
   \nonumber \\ & & \hspace{0.12\textwidth}
   + (C_{6}+C_{8})\,
     {\cal A}_{gh}^{LR}({\rho},{\pi})
     \big\}
   \label{pim-rhop-amp},
   \end{eqnarray}
   \begin{eqnarray} & &
   {\cal A}(\overline{B}_{d}^{0}{\to}{\pi}^{0}{\rho}^{0})
   \nonumber \\ &=&
   \frac{G_{F}}{2\,\sqrt{2}}\, V_{ub}\,V_{ud}^{\ast}\,
   \big\{ a_{2}\, \big[
  -{\cal A}_{ab}^{LL}({\pi},{\rho})
  -{\cal A}_{ab}^{LL}({\rho},{\pi})
  +{\cal A}_{ef}^{LL}({\pi},{\rho})
  +{\cal A}_{ef}^{LL}({\rho},{\pi}) \big]
   \nonumber \\ & & \hspace{0.13\textwidth}
   + C_{1}\, \big[
  -{\cal A}_{cd}^{LL}({\pi},{\rho})
  -{\cal A}_{cd}^{LL}({\rho},{\pi})
  +{\cal A}_{gh}^{LL}({\pi},{\rho})
  +{\cal A}_{gh}^{LL}({\rho},{\pi}) \big] \big\}
   \nonumber \\ &-&
   \frac{G_{F}}{2\,\sqrt{2}}\, V_{tb}\,V_{td}^{\ast}\, \big\{
    ( a_{4}-\frac{3}{2}\,a_{9}-\frac{1}{2}\,a_{10} )\, \big[
   {\cal A}_{ab}^{LL}({\pi},{\rho})
  +{\cal A}_{ab}^{LL}({\rho},{\pi}) \big]
  + \frac{3}{2}\,a_{7}\, \big[
   {\cal A}_{ab}^{LL}({\pi},{\rho})
  -{\cal A}_{ab}^{LL}({\rho},{\pi}) \big]
   \nonumber \\ & &\hspace{0.1\textwidth}
  +( C_{3}-\frac{1}{2}\,C_{9}-\frac{3}{2}\,C_{10} )\, \big[
   {\cal A}_{cd}^{LL}({\pi},{\rho})
  +{\cal A}_{cd}^{LL}({\rho},{\pi}) \big]
  - \frac{3}{2}\,C_{8}\, \big[
   {\cal A}_{cd}^{LR}({\pi},{\rho})
  +{\cal A}_{cd}^{LR}({\rho},{\pi}) \big]
   \nonumber \\ & &\hspace{0.1\textwidth}
  + (a_{6}-\frac{1}{2}\,a_{8})\, \big[
    {\cal A}_{ab}^{SP}({\pi},{\rho})
  + {\cal A}_{ef}^{SP}({\pi},{\rho})
  + {\cal A}_{ef}^{SP}({\rho},{\pi}) \big]
   \nonumber \\ & &\hspace{0.1\textwidth}
  + (C_{5}-\frac{1}{2}\,C_{7})\, \big[
    {\cal A}_{cd}^{SP}({\pi},{\rho})
  + {\cal A}_{cd}^{SP}({\rho},{\pi})
  + {\cal A}_{gh}^{SP}({\pi},{\rho})
  + {\cal A}_{gh}^{SP}({\rho},{\pi}) \big]
   \nonumber \\ & &\hspace{0.1\textwidth}
  +(2\,a_{3}+a_{4}-2\,a_{5}-\frac{1}{2}\,a_{7}
   +\frac{1}{2}\,a_{9}-\frac{1}{2}\,a_{10} )\, \big[
   {\cal A}_{ef}^{LL}({\pi},{\rho})
  +{\cal A}_{ef}^{LL}({\rho},{\pi}) \big]
   \nonumber \\ & &\hspace{0.1\textwidth}
  +(C_{3}+2\,C_{4}-\frac{1}{2}\,C_{9}+\frac{1}{2}\,C_{10} )\, \big[
   {\cal A}_{gh}^{LL}({\pi},{\rho})
  +{\cal A}_{gh}^{LL}({\rho},{\pi}) \big]
   \nonumber \\ & &\hspace{0.11\textwidth}
  + ( 2\,C_{6}+\frac{1}{2}\,C_{8} )\, \big[
   {\cal A}_{gh}^{LR}({\pi},{\rho})
  +{\cal A}_{gh}^{LR}({\rho},{\pi}) \big] \big\}
   \label{piz-rhoz-amp},
   \end{eqnarray}
   \begin{eqnarray} & &
   {\cal A}(\overline{B}_{d}^{0}{\to}{\pi}^{0}{\omega})
   \nonumber \\ &=&
   \frac{G_{F}}{2\,\sqrt{2}}\, V_{ub}\,V_{ud}^{\ast}\,
   \big\{ a_{2}\, \big[
   {\cal A}_{ab}^{LL}({\pi},{\omega})
  -{\cal A}_{ab}^{LL}({\omega},{\pi})
  +{\cal A}_{ef}^{LL}({\pi},{\omega})
  +{\cal A}_{ef}^{LL}({\omega},{\pi}) \big]
   \nonumber \\ & & \hspace{0.13\textwidth}
   + C_{1}\, \big[
   {\cal A}_{cd}^{LL}({\pi},{\omega})
  -{\cal A}_{cd}^{LL}({\omega},{\pi})
  +{\cal A}_{gh}^{LL}({\pi},{\omega})
  +{\cal A}_{gh}^{LL}({\omega},{\pi}) \big] \big\}
   \nonumber \\ &-&
   \frac{G_{F}}{2\,\sqrt{2}}\, V_{tb}\,V_{td}^{\ast}\, \big\{
    -( 2\,a_{3}+a_{4}+2\,a_{5}+\frac{1}{2}\,a_{7}
      + \frac{1}{2}\,a_{9} -\frac{1}{2}\,a_{10} )\,
   {\cal A}_{ab}^{LL}({\omega},{\pi})
   \nonumber \\ & &\hspace{0.1\textwidth}
  - ( C_{3}+2\,C_{4}-\frac{1}{2}\,C_{9} +\frac{1}{2}\,C_{10} )\,
   {\cal A}_{cd}^{LL}({\omega},{\pi})
  - ( 2\,C_{6}+\frac{1}{2}\,C_{8})\,
   {\cal A}_{cd}^{LR}({\omega},{\pi})
   \nonumber \\ & &\hspace{0.1\textwidth}
  - (a_{4}+\frac{3}{2}\,a_{7}-\frac{3}{2}\,a_{9}
    - \frac{1}{2}\,a_{10} )\, \big[
   {\cal A}_{ab}^{LL}({\pi},{\omega})
  +{\cal A}_{ef}^{LL}({\pi},{\omega})
  +{\cal A}_{ef}^{LL}({\omega},{\pi}) \big]
    \nonumber \\ & &\hspace{0.1\textwidth}
  - (C_{3}- \frac{1}{2}\,C_{9}-\frac{3}{2}\,C_{10} )\, \big[
   {\cal A}_{cd}^{LL}({\pi},{\omega})
  +{\cal A}_{gh}^{LL}({\pi},{\omega})
  +{\cal A}_{gh}^{LL}({\omega},{\pi}) \big]
    \nonumber \\ & &\hspace{0.1\textwidth}
  + \frac{3}{2}\,C_{8}\, \big[
   {\cal A}_{cd}^{LR}({\pi},{\omega})
  +{\cal A}_{gh}^{LR}({\pi},{\omega})
  +{\cal A}_{gh}^{LR}({\omega},{\pi}) \big]
    \nonumber \\ & &\hspace{0.1\textwidth}
  - (a_{6}-\frac{1}{2}\,a_{8})\, \big[
   {\cal A}_{ab}^{SP}({\pi},{\omega})
  +{\cal A}_{ef}^{SP}({\pi},{\omega})
  +{\cal A}_{ef}^{SP}({\omega},{\pi}) \big]
   \nonumber \\ & &\hspace{0.1\textwidth}
  -( C_{5}-\frac{1}{2}\,C_{7})\, \big[
   {\cal A}_{cd}^{SP}({\pi},{\omega})
  +{\cal A}_{cd}^{SP}({\omega},{\pi})
  +{\cal A}_{gh}^{SP}({\pi},{\omega})
  +{\cal A}_{gh}^{SP}({\omega},{\pi}) \big] \big\}
   \label{piz-w-amp},
   \end{eqnarray}
   \begin{eqnarray} & &
   {\cal A}(\overline{B}_{d}^{0}{\to}\overline{K}^{0}K^{{\ast}0})
   \nonumber \\ &=& -
   \frac{G_{F}}{\sqrt{2}}\, V_{tb}\,V_{td}^{\ast}\, \big\{
    ( a_{4}-\frac{1}{2}\,a_{10} )\, \big[
    {\cal A}_{ab}^{LL}(K^{\ast},\overline{K})
  + {\cal A}_{ef}^{LL}(K^{\ast},\overline{K}) \big]
    \nonumber \\ & &\hspace{0.13\textwidth}
  + ( C_{3}-\frac{1}{2}\,C_{9} )\, \big[
    {\cal A}_{cd}^{LL}(K^{\ast},\overline{K})
  + {\cal A}_{gh}^{LL}(K^{\ast},\overline{K}) \big]
    \nonumber \\ & &\hspace{0.13\textwidth}
  + ( a_{6}-\frac{1}{2}\,a_{8} )\,
    {\cal A}_{ef}^{SP}(K^{\ast},\overline{K})
    \nonumber \\ & &\hspace{0.13\textwidth}
   + ( C_{5}-\frac{1}{2}\,C_{7} )\, \big[
    {\cal A}_{cd}^{SP}(K^{\ast},\overline{K})
  + {\cal A}_{gh}^{SP}(K^{\ast},\overline{K}) \big]
    \nonumber \\ & &\hspace{0.13\textwidth}
  + ( a_{3}-a_{5}+\frac{1}{2}\,a_{7}-\frac{1}{2}\,a_{9} )\, \big[
    {\cal A}_{ef}^{LL}(K^{\ast},\overline{K})
  + {\cal A}_{ef}^{LL}(\overline{K},K^{\ast}) \big]
   \nonumber \\ & &\hspace{0.13\textwidth}
  + ( C_{4}-\frac{1}{2}\,C_{10} \big)\, \big[
    {\cal A}_{gh}^{LL}(K^{\ast},\overline{K})
  + {\cal A}_{gh}^{LL}(\overline{K},K^{\ast}) \big]
   \nonumber \\ & &\hspace{0.13\textwidth}
  + ( C_{6}-\frac{1}{2}\,C_{8} \big)\, \big[
    {\cal A}_{gh}^{LR}(K^{\ast},\overline{K})
  + {\cal A}_{gh}^{LR}(\overline{K},K^{\ast}) \big] \big\}
   \label{kzb-kvz-amp},
   \end{eqnarray}
   \begin{eqnarray} & &
   {\cal A}(\overline{B}_{d}^{0}{\to}\overline{K}^{{\ast}0}K^{0})
   \nonumber \\ &=& -
   \frac{G_{F}}{\sqrt{2}}\, V_{tb}\,V_{td}^{\ast}\, \big\{
    ( a_{4}-\frac{1}{2}\,a_{10} )\, \big[
    {\cal A}_{ab}^{LL}(K,\overline{K}^{\ast})
  + {\cal A}_{ef}^{LL}(K,\overline{K}^{\ast}) \big]
    \nonumber \\ & &\hspace{0.13\textwidth}
  + ( C_{3}-\frac{1}{2}\,C_{9} )\, \big[
    {\cal A}_{cd}^{LL}(K,\overline{K}^{\ast})
  + {\cal A}_{gh}^{LL}(K,\overline{K}^{\ast}) \big]
    \nonumber \\ & &\hspace{0.13\textwidth}
  + ( a_{6}-\frac{1}{2}\,a_{8} )\, \big[
    {\cal A}_{ab}^{SP}(K,\overline{K}^{\ast})
  + {\cal A}_{ef}^{SP}(K,\overline{K}^{\ast}) \big]
    \nonumber \\ & &\hspace{0.13\textwidth}
  + ( C_{5}-\frac{1}{2}\,C_{7} )\, \big[
    {\cal A}_{cd}^{SP}(K,\overline{K}^{\ast})
  + {\cal A}_{gh}^{SP}(K,\overline{K}^{\ast}) \big]
    \nonumber \\ & &\hspace{0.13\textwidth}
  + ( a_{3}-a_{5}+\frac{1}{2}\,a_{7}-\frac{1}{2}\,a_{9} )\, \big[
    {\cal A}_{ef}^{LL}(K,\overline{K}^{\ast})
  + {\cal A}_{ef}^{LL}(\overline{K}^{\ast},K) \big]
   \nonumber \\ & &\hspace{0.13\textwidth}
  + ( C_{4}-\frac{1}{2}\,C_{10} \big)\, \big[
    {\cal A}_{gh}^{LL}(K,\overline{K}^{\ast})
  + {\cal A}_{gh}^{LL}(\overline{K}^{\ast},K) \big]
   \nonumber \\ & &\hspace{0.13\textwidth}
  + ( C_{6}-\frac{1}{2}\,C_{8} \big)\, \big[
    {\cal A}_{gh}^{LR}(K,\overline{K}^{\ast})
  + {\cal A}_{gh}^{LR}(\overline{K}^{\ast},K) \big] \big\}
   \label{kz-kvz-amp},
   \end{eqnarray}
   \begin{eqnarray} & &
   {\cal A}(\overline{B}_{d}^{0}{\to}{\pi}^{0}{\phi})
   \nonumber \\ &=&
   \frac{G_{F}}{2}\, V_{tb}\,V_{td}^{\ast}\, \big\{
    ( a_{3}+a_{5}-\frac{1}{2}\,a_{7}-\frac{1}{2}\,a_{9} )\,
   {\cal A}_{ab}^{LL}({\phi},{\pi})
   \nonumber \\ & &
  + ( C_{4}-\frac{1}{2}\,C_{10} )\,
   {\cal A}_{cd}^{LL}({\phi},{\pi})
  + ( C_{6}-\frac{1}{2}\,C_{8})\,
   {\cal A}_{cd}^{LR}({\phi},{\pi}) \big\}
   \label{piz-phi-amp},
   \end{eqnarray}
   \begin{eqnarray} & &
   {\cal A}(\overline{B}_{d}^{0}{\to}K^{{\ast}-}{\pi}^{+})
   \nonumber \\ &=&
   \frac{G_{F}}{\sqrt{2}}\, V_{ub}\,V_{us}^{\ast}\, \big\{
     a_{1}\, {\cal A}_{ab}^{LL}(\overline{K}^{\ast},{\pi})
    +C_{2}\, {\cal A}_{cd}^{LL}(\overline{K}^{\ast},{\pi}) \big\}
   \nonumber \\ &-&
   \frac{G_{F}}{\sqrt{2}}\, V_{tb}\,V_{ts}^{\ast}\, \big\{
     ( a_{4}+a_{10})\,
     {\cal A}_{ab}^{LL}(\overline{K}^{\ast},{\pi})
   + ( C_{3}+C_{9} )\,
     {\cal A}_{cd}^{LL}(\overline{K}^{\ast},{\pi})
   \nonumber \\ & & \hspace{0.1\textwidth}
   + ( a_{4}-\frac{1}{2}\,a_{10})\,
     {\cal A}_{ef}^{LL}(\overline{K}^{\ast},{\pi})
   + ( C_{3}-\frac{1}{2}\,C_{9} )\,
     {\cal A}_{gh}^{LL}(\overline{K}^{\ast},{\pi})
   \nonumber \\ & & \hspace{0.1\textwidth}
   + ( a_{6}-\frac{1}{2}\,a_{8} )\,
     {\cal A}_{ef}^{SP}(\overline{K}^{\ast},{\pi})
   + ( C_{5}-\frac{1}{2}\,C_{7} )\,
     {\cal A}_{gh}^{SP}(\overline{K}^{\ast},{\pi})
   \nonumber \\ & & \hspace{0.1\textwidth}
   + ( C_{5}+C_{7} )\,
     {\cal A}_{cd}^{SP}(\overline{K}^{\ast},{\pi}) \big\}
   \label{pip-kvm-amp},
   \end{eqnarray}
   \begin{eqnarray} & &
   {\cal A}(\overline{B}_{d}^{0}{\to}K^{-}{\rho}^{+})
   \nonumber \\ &=&
   \frac{G_{F}}{\sqrt{2}}\, V_{ub}\,V_{us}^{\ast}\, \big\{
     a_{1}\, {\cal A}_{ab}^{LL}(\overline{K},{\rho})
    +C_{2}\, {\cal A}_{cd}^{LL}(\overline{K},{\rho}) \big\}
   \nonumber \\ &-&
   \frac{G_{F}}{\sqrt{2}}\, V_{tb}\,V_{ts}^{\ast}\, \big\{
   ( a_{4}+a_{10})\,
     {\cal A}_{ab}^{LL}(\overline{K},{\rho})
   + ( C_{3}+C_{9} )\,
     {\cal A}_{cd}^{LL}(\overline{K},{\rho})
   \nonumber \\ & & \hspace{0.1\textwidth}
   + ( a_{6}+a_{8} )\,
     {\cal A}_{ab}^{SP}(\overline{K},{\rho})
   + ( C_{5}+C_{7} )\,
     {\cal A}_{cd}^{SP}(\overline{K},{\rho})
   \nonumber \\ & & \hspace{0.1\textwidth}
   + ( a_{4}-\frac{1}{2}\,a_{10})\,
     {\cal A}_{ef}^{LL}(\overline{K},{\rho})
   + ( C_{3}-\frac{1}{2}\,C_{9} )\,
     {\cal A}_{gh}^{LL}(\overline{K},{\rho})
   \nonumber \\ & & \hspace{0.1\textwidth}
   + ( a_{6}-\frac{1}{2}\,a_{8} )\,
     {\cal A}_{ef}^{SP}(\overline{K},{\rho})
   + ( C_{5}-\frac{1}{2}\,C_{7} )\,
     {\cal A}_{gh}^{SP}(\overline{K},{\rho})  \big\}
   \label{km-rhop-amp},
   \end{eqnarray}
   \begin{eqnarray} & &
   {\cal A}(\overline{B}_{d}^{0}{\to}\overline{K}^{{\ast}0}{\pi}^{0})
   \nonumber \\ &=&
   \frac{G_{F}}{2}\, V_{ub}\,V_{us}^{\ast}\, \big\{
     a_{2}\, {\cal A}_{ab}^{LL}({\pi},\overline{K}^{\ast})
   + C_{1}\, {\cal A}_{cd}^{LL}({\pi},\overline{K}^{\ast}) \big\}
   \nonumber \\ &+&
   \frac{G_{F}}{2}\, V_{tb}\,V_{ts}^{\ast}\, \big\{
     (a_{4}-\frac{1}{2}\,a_{10})\, \big[
     {\cal A}_{ab}^{LL}(\overline{K}^{\ast},{\pi})
   + {\cal A}_{ef}^{LL}(\overline{K}^{\ast},{\pi}) \big]
     \nonumber \\ & & \hspace{0.11\textwidth}
   + (C_{3}-\frac{1}{2}\,C_{9})\, \big[
     {\cal A}_{cd}^{LL}(\overline{K}^{\ast},{\pi})
   + {\cal A}_{gh}^{LL}(\overline{K}^{\ast},{\pi}) \big]
   \nonumber \\ & & \hspace{0.11\textwidth}
   + \frac{3}{2}\, (a_{7}-a_{9})\,
     {\cal A}_{ab}^{LL}({\pi},\overline{K}^{\ast})
   - \frac{3}{2}\, C_{8}\,
     {\cal A}_{cd}^{LR}({\pi},\overline{K}^{\ast})
  \nonumber \\ & & \hspace{0.11\textwidth}
   - \frac{3}{2}\, C_{10}\,
     {\cal A}_{cd}^{LL}({\pi},\overline{K}^{\ast})
   + (a_{6}-\frac{1}{2}\,a_{8})\,
     {\cal A}_{ef}^{SP}(\overline{K}^{\ast},{\pi})
  \nonumber \\ & & \hspace{0.11\textwidth}
   + (C_{5}-\frac{1}{2}\,C_{7})\, \big[
     {\cal A}_{cd}^{SP}(\overline{K}^{\ast},{\pi})
   + {\cal A}_{gh}^{SP}(\overline{K}^{\ast},{\pi}) \big] \big\}
   \label{piz-kvz-amp},
   \end{eqnarray}
   \begin{eqnarray} & &
   {\cal A}(\overline{B}_{d}^{0}{\to}\overline{K}^{0}{\rho}^{0})
   \nonumber \\ &=&
   \frac{G_{F}}{2}\, V_{ub}\,V_{us}^{\ast}\, \big\{
     a_{2}\, {\cal A}_{ab}^{LL}({\rho},\overline{K})
   + C_{1}\, {\cal A}_{cd}^{LL}({\rho},\overline{K}) \big\}
   \nonumber \\ &+&
   \frac{G_{F}}{2}\, V_{tb}\,V_{ts}^{\ast}\, \big\{
     (a_{4}-\frac{1}{2}\,a_{10})\, \big[
     {\cal A}_{ab}^{LL}(\overline{K},{\rho})
   + {\cal A}_{ef}^{LL}(\overline{K},{\rho}) \big]
     \nonumber \\ & & \hspace{0.11\textwidth}
   + (C_{3}-\frac{1}{2}\,C_{9})\, \big[
     {\cal A}_{cd}^{LL}(\overline{K},{\rho})
   + {\cal A}_{gh}^{LL}(\overline{K},{\rho}) \big]
   \nonumber \\ & & \hspace{0.11\textwidth}
   + (a_{6}-\frac{1}{2}\,a_{8})\, \big[
     {\cal A}_{ab}^{SP}(\overline{K},{\rho})
   + {\cal A}_{ef}^{SP}(\overline{K},{\rho}) \big]
   \nonumber \\ & & \hspace{0.11\textwidth}
    + (C_{5}-\frac{1}{2}\,C_{7})\, \big[
     {\cal A}_{cd}^{SP}(\overline{K},{\rho})
   + {\cal A}_{gh}^{SP}(\overline{K},{\rho}) \big]
   \nonumber \\ & & \hspace{0.11\textwidth}
   - \frac{3}{2}\, (a_{7}+a_{9})\,
     {\cal A}_{ab}^{LL}({\rho},\overline{K})
   - \frac{3}{2}\, C_{8}\,
     {\cal A}_{cd}^{LR}({\rho},\overline{K})
  \nonumber \\ & & \hspace{0.11\textwidth}
   - \frac{3}{2}\, C_{10}\,
     {\cal A}_{cd}^{LL}({\rho},\overline{K})  \big\}
   \label{kz-rhoz-amp},
   \end{eqnarray}
   \begin{eqnarray} & &
   {\cal A}(\overline{B}_{d}^{0}{\to}\overline{K}^{0}{\omega})
   \nonumber \\ &=&
   \frac{G_{F}}{2}\, V_{ub}\,V_{us}^{\ast}\, \big\{
     a_{2}\, {\cal A}_{ab}^{LL}({\omega},\overline{K})
   + C_{1}\, {\cal A}_{cd}^{LL}({\omega},\overline{K}) \big\}
   \nonumber \\ &-&
   \frac{G_{F}}{2}\, V_{tb}\,V_{ts}^{\ast}\, \big\{
     (a_{4}-\frac{1}{2}\,a_{10})\, \big[
     {\cal A}_{ab}^{LL}(\overline{K},{\omega})
   + {\cal A}_{ef}^{LL}(\overline{K},{\omega}) \big]
     \nonumber \\ & & \hspace{0.11\textwidth}
   + (C_{3}-\frac{1}{2}\,C_{9})\, \big[
     {\cal A}_{cd}^{LL}(\overline{K},{\omega})
   + {\cal A}_{gh}^{LL}(\overline{K},{\omega}) \big]
   \nonumber \\ & & \hspace{0.11\textwidth}
   + (a_{6}-\frac{1}{2}\,a_{8})\, \big[
     {\cal A}_{ab}^{SP}(\overline{K},{\omega})
   + {\cal A}_{ef}^{SP}(\overline{K},{\omega}) \big]
   \nonumber \\ & & \hspace{0.11\textwidth}
    + (C_{5}-\frac{1}{2}\,C_{7})\, \big[
     {\cal A}_{cd}^{SP}(\overline{K},{\omega})
   + {\cal A}_{gh}^{SP}(\overline{K},{\omega}) \big]
   \nonumber \\ & & \hspace{0.11\textwidth}
   + (2\,a_{3}+2\,a_{5}+\frac{1}{2}\,a_{7}
     +\frac{1}{2}\,a_{9})\,
     {\cal A}_{ab}^{LL}({\omega},\overline{K})
   \nonumber \\ & & \hspace{0.11\textwidth}
  + (2\,C_{4}+\frac{1}{2}\,C_{10})\,
     {\cal A}_{cd}^{LL}({\omega},\overline{K})
  + (2\,C_{6}+\frac{1}{2}\,C_{8})\,
     {\cal A}_{cd}^{LR}({\omega},\overline{K}) \big\}
   \label{kz-w-amp},
   \end{eqnarray}
   \begin{eqnarray} & &
   {\cal A}(\overline{B}_{d}^{0}{\to}\overline{K}^{0}{\phi})
   \nonumber \\ &=& -
   \frac{G_{F}}{\sqrt{2}}\, V_{tb}\,V_{ts}^{\ast}\, \big\{
     (a_{3}+a_{4}+a_{5}-\frac{1}{2}\,a_{7}
     -\frac{1}{2}\,a_{9}-\frac{1}{2}\,a_{10})\,
     {\cal A}_{ab}^{LL}({\phi},\overline{K})
     \nonumber \\ & &
    +(a_{4}-\frac{1}{2}\,a_{10})\,
    {\cal A}_{ef}^{LL}({\phi},\overline{K})
    +(a_{6}-\frac{1}{2}\,a_{8})\,
    {\cal A}_{ef}^{SP}({\phi},\overline{K})
     \nonumber \\ & &
   + (C_{3}+C_{4}-\frac{1}{2}\,C_{9}-\frac{1}{2}\,C_{10})\,
     {\cal A}_{cd}^{LL}({\phi},\overline{K})
   + (C_{6}-\frac{1}{2}\,C_{8})\,
     {\cal A}_{cd}^{LR}({\phi},\overline{K})
     \nonumber \\ & &
   + (C_{5}-\frac{1}{2}\,C_{7})\, \big[
     {\cal A}_{cd}^{SP}({\phi},\overline{K})
    +{\cal A}_{gh}^{SP}({\phi},\overline{K}) \big]
   + (C_{3}-\frac{1}{2}\,C_{9})\,
     {\cal A}_{gh}^{LL}({\phi},\overline{K}) \big\}
   \label{kz-phi-amp},
   \end{eqnarray}
   \begin{eqnarray} & &
   {\cal A}(\overline{B}_{d}^{0}{\to}K^{{\ast}-}K^{+})
   \nonumber \\ &=&
   \frac{G_{F}}{\sqrt{2}}\, V_{ub}\,V_{ud}^{\ast}\, \big\{
     a_{2}\,{\cal A}_{ef}^{LL}(K,\overline{K}^{\ast})
   + C_{1}\,{\cal A}_{gh}^{LL}(K,\overline{K}^{\ast}) \big\}
   \nonumber \\ &-&
   \frac{G_{F}}{\sqrt{2}}\, V_{tb}\,V_{td}^{\ast}\, \big\{
     ( a_{3}-a_{5}-a_{7}+a_{9} )\,
     {\cal A}_{ef}^{LL}(K,\overline{K}^{\ast})
   \nonumber \\ & & \hspace{0.11\textwidth}
   + ( C_{4}+C_{10} )\,
     {\cal A}_{gh}^{LL}(K,\overline{K}^{\ast})
   + ( C_{6}+C_{8} )\,
     {\cal A}_{gh}^{LR}(K,\overline{K}^{\ast})
     \nonumber \\ & & \hspace{0.11\textwidth}
   + ( a_{3}-a_{5}+\frac{1}{2}\,a_{7}-\frac{1}{2}\,a_{9} )\,
     {\cal A}_{ef}^{LL}(\overline{K}^{\ast},K)
   \nonumber \\ & & \hspace{0.11\textwidth}
   + ( C_{4}-\frac{1}{2}\,C_{10} )\,
     {\cal A}_{gh}^{LL}(\overline{K}^{\ast},K)
   + ( C_{6}-\frac{1}{2}\,C_{8} )\,
     {\cal A}_{gh}^{LR}(\overline{K}^{\ast},K) \big\}
   \label{kp-kvm-amp},
   \end{eqnarray}
   \begin{eqnarray} & &
   {\cal A}(\overline{B}_{d}^{0}{\to}K^{-}K^{{\ast}+})
   \nonumber \\ &=&
   \frac{G_{F}}{\sqrt{2}}\, V_{ub}\,V_{ud}^{\ast}\, \big\{
     a_{2}\,{\cal A}_{ef}^{LL}(K^{\ast},\overline{K})
   + C_{1}\,{\cal A}_{gh}^{LL}(K^{\ast},\overline{K}) \big\}
   \nonumber \\ &-&
   \frac{G_{F}}{\sqrt{2}}\, V_{tb}\,V_{td}^{\ast}\, \big\{
     ( a_{3}-a_{5}-a_{7}+a_{9} )\,
     {\cal A}_{ef}^{LL}(K^{\ast},\overline{K})
   \nonumber \\ & & \hspace{0.11\textwidth}
   + ( C_{4}+C_{10} )\,
     {\cal A}_{gh}^{LL}(K^{\ast},\overline{K})
   + ( C_{6}+C_{8} )\,
     {\cal A}_{gh}^{LR}(K^{\ast},\overline{K})
     \nonumber \\ & & \hspace{0.11\textwidth}
   + ( a_{3}-a_{5}+\frac{1}{2}\,a_{7}-\frac{1}{2}\,a_{9} )\,
     {\cal A}_{ef}^{LL}(\overline{K},K^{\ast})
   \nonumber \\ & & \hspace{0.11\textwidth}
   + ( C_{4}-\frac{1}{2}\,C_{10} )\,
     {\cal A}_{gh}^{LL}(\overline{K},K^{\ast})
   + ( C_{6}-\frac{1}{2}\,C_{8} )\,
     {\cal A}_{gh}^{LR}(\overline{K},K^{\ast}) \big\}
   \label{km-kvp-amp}.
   \end{eqnarray}
   The shorthands are
   \begin{equation}
   a_{i}\, =\, \left\{ \begin{array}{lll}
   \displaystyle C_{i}+\frac{1}{N_{c}}C_{i+1},
   & \quad & \text{for odd }i; \\ ~ \\
   \displaystyle C_{i}+\frac{1}{N_{c}}C_{i-1},
   & \quad & \text{for even }i,
   \end{array} \right.
   \label{wilson-ai}
   \end{equation}
   \begin{equation}
   C_{m}\,{\cal A}_{ij}^{k}(M_{1},M_{2})\, =\,
   {\cal A}_{i}^{k}(C_{m},M_{1},M_{2})
  +{\cal A}_{j}^{k}(C_{m},M_{1},M_{2})
   \label{mode-factor-01},
   \end{equation}
   where the explicit expressions of the amplitude building blocks
   ${\cal A}_{i}^{k}(C_{m},M_{1},M_{2})$ including
   contributions from the $B$ mesonic WFs ${\phi}_{B2}$ are
   given in Appendix \ref{sec:block}.

   \section{the amplitude building blocks}
   \label{sec:block}
   For the sake of simplification and convenience,
   shorthand is used for the amplitude building blocks.
   \begin{equation}
  {\phi}_{B1,B2}\, =\, {\phi}_{B1,B2}(x_{1},b_{1})\,e^{-S_{B}}
   \label{shorthand-phi-b1},
   \end{equation}
   \begin{equation}
  {\phi}_{P}^{a}\, =\, {\phi}_{P}^{a}(x_{2})\,e^{-S_{P}}
   \label{shorthand-phi-pa},
   \end{equation}
   \begin{equation}
  {\phi}_{P}^{p,t}\, =\,
   r_{P}\, {\phi}_{P}^{p,t}(x_{2})\,e^{-S_{P}}
   \label{shorthand-phi-pp-pt},
   \end{equation}
   \begin{equation}
  {\phi}_{V}^{v}\, =\, f_{V}^{\parallel}\,
  {\phi}_{V}^{v}(x_{3})\,e^{-S_{V}}
   \label{shorthand-phi-vv},
   \end{equation}
   \begin{equation}
  {\phi}_{V}^{t,s}\, =\,
   r_{V}\, f_{V}^{\perp}\,
  {\phi}_{V}^{t,s}(x_{3})\,e^{-S_{V}}
   \label{shorthand-phi-vt-vs},
   \end{equation}
   \begin{equation}
  {\cal C}\, =\,
   \frac{{\pi}\,C_{F}}{N_{c}^{2}}\,m_{B}^{4}\, f_{B}\,f_{P}
   \label{shorthand-coefficient},
   \end{equation}
  where $r_{P}$ $=$ ${\mu}_{P}/m_{B}$ and $r_{V}$ $=$ $m_{V}/m_{B}$.
  For the amplitude building block ${\cal A}_{i}^{j}(M_{1},M_{2})$,
  the subscript $i$ corresponds to the indices of
  Fig. \ref{fig:feynman}, and the superscript
  $j$ refers to the three possible Dirac structures
  ${\Gamma}_{1}{\otimes}{\Gamma}_{2}$ of the operator
  $(\bar{q}_{1}q_{2})_{{\Gamma}_{1}}(\bar{q}_{3}q_{4})_{{\Gamma}_{2}}$,
  namely $j$ $=$ $LL$ for $(V-A){\otimes}(V-A)$, $j$ $=$ $LR$
  for $(V-A){\otimes}(V+A)$ and $j$ $=$ $SP$ for $
  -2\,(S-P){\otimes}(S+P)$.
  The explicit expressions of ${\cal A}_{i}^{j}(M_{1},M_{2})$
  up to the order of $r_{P}$ and $r_{V}$
  are written as follows.
   \begin{eqnarray}
  {\cal A}^{LL}_{a}(P,V) &=& {\cal C}\,
  {\int} \mathbbm{d}x_{1}\,\mathbbm{d}x_{3}\,
         \mathbbm{d}b_{1}\,\mathbbm{d}b_{3}\,
  H_{ab}({\alpha}_{g}^{V},{\beta}_{a}^{V},b_{1},b_{3})\,
  {\alpha}_{s}(t_{a}^{V})\, C_{i}(t_{a}^{V})
   \nonumber \\ & &
   \big\{ {\phi}_{B1}\, \big[ {\phi}_{V}^{v}\, (1+x_{3})
   + \big( {\phi}_{V}^{t}+{\phi}_{V}^{s} \big)\,
   (\bar{x}_{3}-x_{3}) \big]
   \nonumber \\ & &
  -{\phi}_{B2}\, \big[ {\phi}_{V}^{v}
   -\big( {\phi}_{V}^{t}+{\phi}_{V}^{s} \big)\,
    x_{3} \big] \big\}\, S_{t}(x_{3})
   \label{amp-apv-left},
   \end{eqnarray}
   \begin{equation}
  {\cal A}^{LR}_{a}(P,V)\, =\, -{\cal A}^{LL}_{a}(P,V)
   \label{amp-apv-right},
   \end{equation}
   \begin{eqnarray}
  {\cal A}^{SP}_{a}(P,V) &=& {\cal C}\,
  {\int} \mathbbm{d}x_{1}\,\mathbbm{d}x_{3}\,
         \mathbbm{d}b_{1}\,\mathbbm{d}b_{3}\,
  H_{ab}({\alpha}_{g}^{V},{\beta}_{a}^{V},b_{1},b_{3})\,
  {\alpha}_{s}(t_{a}^{V})\, C_{i}(t_{a}^{V})
   \nonumber \\ & & 2\,r_{P}\,
   \big\{ {\phi}_{B1}\, \big[ -{\phi}_{V}^{v}
   +{\phi}_{V}^{t}\,x_{3}
   -{\phi}_{V}^{s}\,(2+x_{3}) \big]
   \nonumber \\ & &
  +{\phi}_{B2}\, \big[ {\phi}_{V}^{v}
  -{\phi}_{V}^{t}+{\phi}_{V}^{s} \big] \big\}\, S_{t}(x_{3})
   \label{amp-apv-sp},
   \end{eqnarray}
   \begin{eqnarray}
  {\cal A}^{LL}_{a}(V,P) &=&  {\cal C}\, f_{V}^{\parallel}\,
  {\int} \mathbbm{d}x_{1}\,\mathbbm{d}x_{2}\,
         \mathbbm{d}b_{1}\,\mathbbm{d}b_{2}\,
  H_{ab}({\alpha}_{g}^{P},{\beta}_{a}^{P},b_{1},b_{2})\,
  {\alpha}_{s}(t_{a}^{P})
   \nonumber \\ & &
   \big\{ {\phi}_{B1}\, \big[ {\phi}_{P}^{a}\, (1+x_{2})
   + \big( {\phi}_{P}^{p}+{\phi}_{P}^{t} \big)\,
   (\bar{x}_{2}-x_{2}) \big]
   \nonumber \\ & &
  -{\phi}_{B2}\, \big[ {\phi}_{P}^{a}
   -\big( {\phi}_{P}^{p}+{\phi}_{P}^{t} \big)\,
    x_{2} \big] \big\}\, C_{i}(t_{a}^{P}) \, S_{t}(x_{2})
   \label{amp-avp-left},
   \end{eqnarray}
   \begin{equation}
  {\cal A}^{LR}_{a}(V,P)\, =\, {\cal A}^{LL}_{a}(V,P)
   \label{amp-avp-right},
   \end{equation}
   \begin{equation}
  {\cal A}^{SP}_{a}(V,P) \, =\, 0
   \label{amp-avp-sp},
   \end{equation}
   \begin{equation}
  {\cal A}^{LL}_{b}(P,V)\, =\, 2\,{\cal C}\,
  {\int} \mathbbm{d}x_{1}\,\mathbbm{d}x_{3}\,
         \mathbbm{d}b_{1}\,\mathbbm{d}b_{3}\,
   H_{ab}({\alpha}_{g}^{V},{\beta}_{b}^{V},b_{3},b_{1})\,
  {\alpha}_{s}(t_{b}^{V})\, C_{i}(t_{b}^{V})\,
  S_{t}(x_{1})\, {\phi}_{B1}\, {\phi}_{V}^{s}
   \label{amp-bpv-left},
   \end{equation}
   \begin{equation}
  {\cal A}^{LR}_{b}(P,V)\, =\, -{\cal A}^{LL}_{b}(P,V)
   \label{amp-bpv-right},
   \end{equation}
   \begin{eqnarray}
  {\cal A}^{SP}_{b}(P,V) &=& -{\cal C}\,
  {\int} \mathbbm{d}x_{1}\,\mathbbm{d}x_{3}\,
         \mathbbm{d}b_{1}\,\mathbbm{d}b_{3}\,
   H_{ab}({\alpha}_{g}^{V},{\beta}_{b}^{V},b_{3},b_{1})\,
  {\alpha}_{s}(t_{b}^{V})\, C_{i}(t_{b}^{V})\, S_{t}(x_{1})
   \nonumber \\ & & 2\,r_{P}\,
   \big\{ {\phi}_{B1}\, \big[ {\phi}_{V}^{v}\,x_{1}
   +2\,{\phi}_{V}^{s} \bar{x}_{1} \big]
   +2\, {\phi}_{B2}\, {\phi}_{V}^{s}\,x_{1} \big] \big\}
   \label{amp-bpv-sp},
   \end{eqnarray}
   \begin{equation}
  {\cal A}^{LL}_{b}(V,P)\, =\,
  2\,{\cal C}\, f_{V}^{\parallel}\,
  {\int} \mathbbm{d}x_{1}\,\mathbbm{d}x_{2}\,
         \mathbbm{d}b_{1}\,\mathbbm{d}b_{2}\,
   H_{ab}({\alpha}_{g}^{P},{\beta}_{b}^{P},b_{2},b_{1})\,
  {\alpha}_{s}(t_{b}^{P})\, C_{i}(t_{b}^{P})\,
  S_{t}(x_{1})\, {\phi}_{B1}\, {\phi}_{P}^{p}
   \label{amp-bvp-left},
   \end{equation}
   \begin{equation}
  {\cal A}^{LR}_{b}(V,P)\, =\, {\cal A}^{LL}_{b}(V,P)
   \label{amp-bvp-right},
   \end{equation}
   \begin{equation}
  {\cal A}^{SP}_{b}(V,P) \, =\, 0
   \label{amp-bvp-sp},
   \end{equation}
   \begin{eqnarray}
  {\cal A}^{LL}_{c}(P,V) &=& {\cal C}\,
  {\int} \mathbbm{d}x_{1}\,\mathbbm{d}x_{2}\,\mathbbm{d}x_{3}\,
         \mathbbm{d}b_{1}\,\mathbbm{d}b_{2}\,
   H_{cd}({\alpha}_{g}^{V},{\beta}_{c}^{V},b_{1},b_{2})\,
  {\alpha}_{s}(t_{c}^{V})\, C_{i}(t_{c}^{V})\, S_{t}(x_{3})
   \nonumber \\ & &
  {\phi}_{P}^{a}\, \big\{
   \big( {\phi}_{B1}-{\phi}_{B2} \big)\,
  {\phi}_{V}^{v}\,(\bar{x}_{2}-x_{1})
  +{\phi}_{B1}\, \big( {\phi}_{V}^{t}-{\phi}_{V}^{s}
   \big)\,x_{3} \big\}_{b_{1}=b_{3}}
   \label{amp-cpv-left},
   \end{eqnarray}
   \begin{eqnarray}
  {\cal A}^{LR}_{c}(P,V) &=& {\cal C}\,
  {\int} \mathbbm{d}x_{1}\,\mathbbm{d}x_{2}\,\mathbbm{d}x_{3}\,
         \mathbbm{d}b_{1}\,\mathbbm{d}b_{2}\,
   H_{cd}({\alpha}_{g}^{V},{\beta}_{c}^{V},b_{1},b_{2})\,
  {\alpha}_{s}(t_{c}^{V})\, C_{i}(t_{c}^{V})\, S_{t}(x_{3})
   \nonumber \\ & & \hspace{-0.05\textwidth}
   {\phi}_{P}^{a}\, \big\{ \big( {\phi}_{B1}-{\phi}_{B2} \big)\,
   \big[ {\phi}_{V}^{v}\,(x_{1}-\bar{x}_{2})
  +\big( {\phi}_{V}^{t}+{\phi}_{V}^{s} \big)\,x_{3} \big]
  -{\phi}_{B1}\, {\phi}_{V}^{v}\,x_{3} \big\}_{b_{1}=b_{3}}
   \label{amp-cpv-right},
   \end{eqnarray}
   \begin{eqnarray}
  {\cal A}^{SP}_{c}(P,V) &=&  {\cal C}\,
  {\int} \mathbbm{d}x_{1}\,\mathbbm{d}x_{2}\,\mathbbm{d}x_{3}\,
         \mathbbm{d}b_{1}\,\mathbbm{d}b_{2}\,
   H_{cd}({\alpha}_{g}^{V},{\beta}_{c}^{V},b_{1},b_{2})\,
  {\alpha}_{s}(t_{c}^{V})\, C_{i}(t_{c}^{V})\, S_{t}(x_{3})
   \nonumber \\ & &
   \big\{ \big( {\phi}_{B1}-{\phi}_{B2} \big)\,
   \big( {\phi}_{P}^{p}+{\phi}_{P}^{t} \big)\,
   \big( {\phi}_{V}^{v}-{\phi}_{V}^{t}+{\phi}_{V}^{s} \big)\,
   (x_{1}-\bar{x}_{2})
   \nonumber \\ & &
   -{\phi}_{B1}\, \big( {\phi}_{P}^{p}-{\phi}_{P}^{t} \big)\,
    \big( {\phi}_{V}^{t}+{\phi}_{V}^{s} \big)\, x_{3}
    \big\}_{b_{1}=b_{3}}
   \label{amp-cpv-sp},
   \end{eqnarray}
   \begin{eqnarray}
  {\cal A}^{LL}_{c}(V,P) &=& {\cal C}\,
  {\int} \mathbbm{d}x_{1}\,\mathbbm{d}x_{2}\,\mathbbm{d}x_{3}\,
         \mathbbm{d}b_{1}\,\mathbbm{d}b_{3}\,
   H_{cd}({\alpha}_{g}^{P},{\beta}_{c}^{P},b_{1},b_{3})\,
  {\alpha}_{s}(t_{c}^{P})\, C_{i}(t_{c}^{P})\, S_{t}(x_{2})
   \nonumber \\ & &
  {\phi}_{V}^{v}\, \big\{ \big( {\phi}_{B1}-{\phi}_{B2} \big)\,
  {\phi}_{P}^{a}\, (\bar{x}_{3}-x_{1}) -{\phi}_{B1}\,
   \big({\phi}_{P}^{p}-{\phi}_{P}^{t}\big)\,x_{2}
   \big\}_{b_{1}=b_{2}}
   \label{amp-cvp-left},
   \end{eqnarray}
   \begin{eqnarray}
  {\cal A}^{LR}_{c}(V,P) &=& {\cal C}\,
  {\int} \mathbbm{d}x_{1}\,\mathbbm{d}x_{2}\,\mathbbm{d}x_{3}\,
         \mathbbm{d}b_{1}\,\mathbbm{d}b_{3}\,
   H_{cd}({\alpha}_{g}^{P},{\beta}_{c}^{P},b_{1},b_{3})\,
  {\alpha}_{s}(t_{c}^{P})\, C_{i}(t_{c}^{P})\, S_{t}(x_{2})
   \nonumber \\ & & \hspace{-0.05\textwidth}
  {\phi}_{V}^{v}\, \big\{ \big( {\phi}_{B1}-{\phi}_{B2} \big)\,
   \big[ {\phi}_{P}^{a}\, (\bar{x}_{3}-x_{1})
  -\big( {\phi}_{P}^{p}+{\phi}_{P}^{t}\big)\,x_{2} \big]
  +{\phi}_{B1}\,{\phi}_{P}^{a}\,x_{2} \big\}_{b_{1}=b_{2}}
   \label{amp-cvp-right},
   \end{eqnarray}
   \begin{eqnarray}
  {\cal A}^{SP}_{c}(V,P) &=&  {\cal C}\,
  {\int} \mathbbm{d}x_{1}\,\mathbbm{d}x_{2}\,\mathbbm{d}x_{3}\,
         \mathbbm{d}b_{1}\,\mathbbm{d}b_{3}\,
   H_{cd}({\alpha}_{g}^{P},{\beta}_{c}^{P},b_{1},b_{3})\,
  {\alpha}_{s}(t_{c}^{P})\, C_{i}(t_{c}^{P})\, S_{t}(x_{2})
   \nonumber \\ & &
   \big\{ \big( {\phi}_{B1}-{\phi}_{B2} \big)\,
   \big( {\phi}_{P}^{a}+{\phi}_{P}^{p}-{\phi}_{P}^{t} \big)\,
   \big( {\phi}_{V}^{t}+{\phi}_{V}^{s} \big)\,
   (\bar{x}_{3}-x_{1})
   \nonumber \\ & &
  -{\phi}_{B1}\,\big( {\phi}_{P}^{p}+{\phi}_{P}^{t}\big)\,
   \big( {\phi}_{V}^{t}-{\phi}_{V}^{s} \big)\,x_{2}
   \big\}_{b_{1}=b_{2}}
   \label{amp-cvp-sp},
   \end{eqnarray}
   \begin{eqnarray}
  {\cal A}^{LL}_{d}(P,V) &=& {\cal C}\,
  {\int} \mathbbm{d}x_{1}\,\mathbbm{d}x_{2}\,\mathbbm{d}x_{3}\,
         \mathbbm{d}b_{1}\,\mathbbm{d}b_{2}\,
   H_{cd}({\alpha}_{g}^{V},{\beta}_{d}^{V},b_{1},b_{2})\,
  {\alpha}_{s}(t_{d}^{V})\, C_{i}(t_{d}^{V})\, S_{t}(x_{3})
   \nonumber \\ & & \hspace{-0.05\textwidth}
   {\phi}_{p}^{a}\, \big\{ \big( {\phi}_{B1}-{\phi}_{B2} \big)\,
   \big[ {\phi}_{V}^{v}\,(x_{1}-x_{2})
  +\big( {\phi}_{V}^{t}+{\phi}_{V}^{s}\big)\,x_{3} \big]
  -{\phi}_{B1}\, {\phi}_{V}^{v}\,x_{3} \big\}_{b_{1}=b_{3}}
   \label{amp-dpv-left},
   \end{eqnarray}
   \begin{eqnarray}
  {\cal A}^{LR}_{d}(P,V) &=& {\cal C}\,
  {\int} \mathbbm{d}x_{1}\,\mathbbm{d}x_{2}\,\mathbbm{d}x_{3}\,
         \mathbbm{d}b_{1}\,\mathbbm{d}b_{2}\,
   H_{cd}({\alpha}_{g}^{V},{\beta}_{d}^{V},b_{1},b_{2})\,
  {\alpha}_{s}(t_{d}^{V})\, C_{i}(t_{d}^{V})\, S_{t}(x_{3})
   \nonumber \\ & &
  {\phi}_{p}^{a}\, \big\{ \big( {\phi}_{B1}-{\phi}_{B2} \big)\,
  {\phi}_{V}^{v}\, (x_{2}-x_{1}) + {\phi}_{B1}\,
   \big( {\phi}_{V}^{t}-{\phi}_{V}^{s} \big)\,x_{3} \big\}_{b_{1}=b_{3}}
   \label{amp-dpv-right},
   \end{eqnarray}
   \begin{eqnarray}
  {\cal A}^{SP}_{d}(P,V) &=& {\cal C}\,
  {\int} \mathbbm{d}x_{1}\,\mathbbm{d}x_{2}\,\mathbbm{d}x_{3}\,
         \mathbbm{d}b_{1}\,\mathbbm{d}b_{2}\,
   H_{cd}({\alpha}_{g}^{V},{\beta}_{d}^{V},b_{1},b_{2})\,
  {\alpha}_{s}(t_{d}^{V})\, C_{i}(t_{d}^{V})\, S_{t}(x_{3})
   \nonumber \\ & &
   \big\{ \big( {\phi}_{B1}-{\phi}_{B2} \big)\,
   \big( {\phi}_{P}^{p}-{\phi}_{P}^{t}\big)\,
   \big( {\phi}_{V}^{v}-{\phi}_{V}^{t}+{\phi}_{V}^{s} \big)\,
   (x_{2}-x_{1})
   \nonumber \\ & &
   +{\phi}_{B1}\, \big( {\phi}_{P}^{p}+{\phi}_{P}^{t} \big)\,
    \big( {\phi}_{V}^{t}+{\phi}_{V}^{s} \big)\, x_{3}
    \big\}_{b_{1}=b_{3}}
   \label{amp-dpv-sp},
   \end{eqnarray}
   \begin{eqnarray}
  {\cal A}^{LL}_{d}(V,P) &=& {\cal C}\,
  {\int} \mathbbm{d}x_{1}\,\mathbbm{d}x_{2}\,\mathbbm{d}x_{3}\,
         \mathbbm{d}b_{1}\,\mathbbm{d}b_{3}\,
   H_{cd}({\alpha}_{g}^{P},{\beta}_{d}^{P},b_{1},b_{3})\,
  {\alpha}_{s}(t_{d}^{P})\, C_{i}(t_{d}^{P})\, S_{t}(x_{2})
   \nonumber \\ & & \hspace{-0.05\textwidth}
  {\phi}_{V}^{v}\, \big\{ \big( {\phi}_{B1}-{\phi}_{B2} \big)\,
   \big[ {\phi}_{P}^{a}\, (x_{1}-x_{3})
  +\big( {\phi}_{P}^{p}+{\phi}_{P}^{t}\big)\,x_{2} \big]
  -{\phi}_{B1}\,{\phi}_{P}^{a}\,x_{2} \big\}_{b_{1}=b_{2}}
   \label{amp-dvp-left},
   \end{eqnarray}
   \begin{eqnarray}
  {\cal A}^{LR}_{d}(V,P) &=& {\cal C}\,
  {\int} \mathbbm{d}x_{1}\,\mathbbm{d}x_{2}\,\mathbbm{d}x_{3}\,
         \mathbbm{d}b_{1}\,\mathbbm{d}b_{3}\,
   H_{cd}({\alpha}_{g}^{P},{\beta}_{d}^{P},b_{1},b_{3})\,
  {\alpha}_{s}(t_{d}^{P})\, C_{i}(t_{d}^{P})\, S_{t}(x_{2})
   \nonumber \\ & &
  {\phi}_{V}^{v}\, \big\{ \big( {\phi}_{B1}-{\phi}_{B2} \big)\,
  {\phi}_{P}^{a}\, (x_{1}-x_{3}) +{\phi}_{B1}\,
   \big({\phi}_{P}^{p}-{\phi}_{P}^{t}\big)\,x_{2}
   \big\}_{b_{1}=b_{2}}
   \label{amp-dvp-right},
   \end{eqnarray}
   \begin{eqnarray}
  {\cal A}^{SP}_{d}(V,P) &=& {\cal C}\,
  {\int} \mathbbm{d}x_{1}\,\mathbbm{d}x_{2}\,\mathbbm{d}x_{3}\,
         \mathbbm{d}b_{1}\,\mathbbm{d}b_{3}\,
   H_{cd}({\alpha}_{g}^{P},{\beta}_{d}^{P},b_{1},b_{3})\,
  {\alpha}_{s}(t_{d}^{P})\, C_{i}(t_{d}^{P})\, S_{t}(x_{2})
   \nonumber \\ & &
   \big\{ \big( {\phi}_{B1}-{\phi}_{B2} \big)\,
   \big( {\phi}_{P}^{a}+{\phi}_{P}^{p}-{\phi}_{P}^{t} \big)\,
   \big( {\phi}_{V}^{t}-{\phi}_{V}^{s} \big)\,
   (x_{3}-x_{1})
   \nonumber \\ & &
  -{\phi}_{B1}\,\big( {\phi}_{P}^{p}+{\phi}_{P}^{t}\big)\,
   \big( {\phi}_{V}^{t}+{\phi}_{V}^{s} \big)\,x_{2}
    \big\}_{b_{1}=b_{2}}
   \label{amp-dvp-sp},
   \end{eqnarray}
   \begin{eqnarray}
  {\cal A}^{LL}_{e}(P,V) &=& {\cal C}\,
  {\int} \mathbbm{d}x_{2}\,\mathbbm{d}x_{3}\,
         \mathbbm{d}b_{2}\,\mathbbm{d}b_{3}\,
   H_{ef}({\alpha}_{a}^{V},{\beta}_{e}^{V},b_{2},b_{3})\,
   {\alpha}_{s}(t_{e}^{V})\,C_{i}(t_{e}^{V})
   \nonumber \\ & &
   \big\{ 2\,{\phi}_{P}^{p}\,\big[ {\phi}_{V}^{t}\,x_{3}
    + {\phi}_{V}^{s}\,(1+\bar{x}_{3}) \big]
    - {\phi}_{P}^{a}\,{\phi}_{V}^{v}\, \bar{x}_{3}
   \big\}\, S_{t}(\bar{x}_{3})
   \label{amp-epv-left},
   \end{eqnarray}
   \begin{equation}
  {\cal A}^{LR}_{e}(P,V)\, =\, -{\cal A}^{LL}_{e}(P,V)
   \label{amp-epv-right},
   \end{equation}
   \begin{eqnarray}
  {\cal A}^{SP}_{e}(P,V) &=& 2\,{\cal C}\,
  {\int} \mathbbm{d}x_{2}\,\mathbbm{d}x_{3}\,
         \mathbbm{d}b_{2}\,\mathbbm{d}b_{3}\,
   H_{ef}({\alpha}_{a}^{V},{\beta}_{e}^{V},b_{2},b_{3})\,
   {\alpha}_{s}(t_{e}^{V})
   \nonumber \\ & &
   C_{i}(t_{e}^{V})\, S_{t}(\bar{x}_{3})\,
   \big\{ {\phi}_{P}^{a}\, \big( {\phi}_{V}^{t}
    + {\phi}_{V}^{s} \big)\, \bar{x}_{3}
    -2\, {\phi}_{P}^{p}\,{\phi}_{V}^{v} \big\}
   \label{amp-epv-sp},
   \end{eqnarray}
   \begin{eqnarray}
  {\cal A}^{LL}_{e}(V,P) &=& -{\cal C}\,
  {\int} \mathbbm{d}x_{2}\,\mathbbm{d}x_{3}\,
         \mathbbm{d}b_{2}\,\mathbbm{d}b_{3}\,
   H_{ef}({\alpha}_{a}^{P},{\beta}_{e}^{P},b_{3},b_{2})\,
   {\alpha}_{s}(t_{e}^{P})\,C_{i}(t_{e}^{P})
   \nonumber \\ & &
   \big\{ {\phi}_{P}^{a}\,{\phi}_{V}^{v}\,\bar{x}_{2}
   +2\,{\phi}_{V}^{s}\,\big[ {\phi}_{P}^{p}\,(1+\bar{x}_{2})
   +{\phi}_{P}^{t}\,x_{2} \big] \big\}\, S_{t}(\bar{x}_{2})
   \label{amp-evp-left},
   \end{eqnarray}
   \begin{equation}
  {\cal A}^{LR}_{e}(V,P)\, =\, -{\cal A}^{LL}_{e}(V,P)
   \label{amp-evp-right},
   \end{equation}
   \begin{eqnarray}
  {\cal A}^{SP}_{e}(V,P) &=& 2\,{\cal C}\,
  {\int} \mathbbm{d}x_{2}\,\mathbbm{d}x_{3}\,
         \mathbbm{d}b_{2}\,\mathbbm{d}b_{3}\,
   H_{ef}({\alpha}_{a}^{P},{\beta}_{e}^{P},b_{3},b_{2})\,
   {\alpha}_{s}(t_{e}^{P})
   \nonumber \\ & &
   C_{i}(t_{e}^{P})\, S_{t}(\bar{x}_{2})\,
   \big\{ 2\, {\phi}_{P}^{a}\, {\phi}_{V}^{s}
   + {\phi}_{V}^{v}\, \big( {\phi}_{P}^{p}
    + {\phi}_{P}^{t} \big)\, \bar{x}_{2} \big\}
   \label{amp-evp-sp},
   \end{eqnarray}
   \begin{eqnarray}
  {\cal A}^{LL}_{f}(P,V) &=& {\cal C}\,
  {\int} \mathbbm{d}x_{2}\,\mathbbm{d}x_{3}\,
         \mathbbm{d}b_{2}\,\mathbbm{d}b_{3}\,
   H_{ef}({\alpha}_{a}^{V},{\beta}_{f}^{V},b_{3},b_{2})\,
   {\alpha}_{s}(t_{f}^{V})\,C_{i}(t_{f}^{V})
   \nonumber \\ & &
   \big\{ {\phi}_{P}^{a}\,{\phi}_{V}^{v}\, x_{2}
   -2\, {\phi}_{V}^{s}\, \big[ {\phi}_{P}^{p}\,(1+x_{2})
   -{\phi}_{P}^{t}\,\bar{x}_{2} \big] \big\}\,  S_{t}(x_{2})
   \label{amp-fpv-left},
   \end{eqnarray}
   \begin{equation}
  {\cal A}^{LR}_{f}(P,V) \, =\, -{\cal A}^{LL}_{f}(P,V)
   \label{amp-fpv-right},
   \end{equation}
   \begin{eqnarray}
  {\cal A}^{SP}_{f}(P,V) &=& 2\, {\cal C}\,
  {\int} \mathbbm{d}x_{2}\,\mathbbm{d}x_{3}\,
         \mathbbm{d}b_{2}\,\mathbbm{d}b_{3}\,
   H_{ef}({\alpha}_{a}^{V},{\beta}_{f}^{V},b_{3},b_{2})\,
   {\alpha}_{s}(t_{f}^{V})
   \nonumber \\ & & C_{i}(t_{f}^{V})\,  S_{t}(x_{2})\,
   \big\{ 2\,{\phi}_{P}^{a}\,{\phi}_{V}^{s}\,
  -\big( {\phi}_{P}^{p}-{\phi}_{P}^{t} \big)\,
   {\phi}_{V}^{v}\, x_{2} \big\}
   \label{amp-fpv-sp},
   \end{eqnarray}
   \begin{eqnarray}
  {\cal A}^{LL}_{f}(V,P) &=& {\cal C}\,
  {\int} \mathbbm{d}x_{2}\,\mathbbm{d}x_{3}\,
         \mathbbm{d}b_{2}\,\mathbbm{d}b_{3}\,
   H_{ef}({\alpha}_{a}^{P},{\beta}_{f}^{P},b_{2},b_{3})\,
   {\alpha}_{s}(t_{f}^{P})\,C_{i}(t_{f}^{P})
   \nonumber \\ & &
   \big\{ {\phi}_{P}^{a}\,{\phi}_{V}^{v}\, x_{3}
   -2\,{\phi}_{P}^{p}\, \big[ {\phi}_{V}^{t}\,
   \bar{x}_{3} - {\phi}_{V}^{s}\, (1+x_{3}) \big]
   \big\}\, S_{t}(x_{3})
   \label{amp-fvp-left},
   \end{eqnarray}
   \begin{equation}
  {\cal A}^{LR}_{f}(V,P) \, =\, -{\cal A}^{LL}_{f}(V,P)
   \label{amp-fvp-right},
   \end{equation}
   \begin{eqnarray}
  {\cal A}^{SP}_{f}(V,P) &=& 2\, {\cal C}\,
  {\int} \mathbbm{d}x_{2}\,\mathbbm{d}x_{3}\,
         \mathbbm{d}b_{2}\,\mathbbm{d}b_{3}\,
   H_{ef}({\alpha}_{a}^{P},{\beta}_{f}^{P},b_{2},b_{3})\,
   {\alpha}_{s}(t_{f}^{P})
   \nonumber \\ & & C_{i}(t_{f}^{P})\,  S_{t}(x_{3})\,
   \big\{ 2\,{\phi}_{P}^{p}\,{\phi}_{V}^{v}\,
  -{\phi}_{P}^{a}\, \big( {\phi}_{V}^{t}-{\phi}_{V}^{s}
    \big)\, x_{3} \big\}
   \label{amp-fvp-sp},
   \end{eqnarray}
   \begin{eqnarray}
  {\cal A}^{LL}_{g}(P,V) &=& {\cal C}\,
  {\int} \mathbbm{d}x_{1}\,\mathbbm{d}x_{2}\,\mathbbm{d}x_{3}\,
         \mathbbm{d}b_{1}\,\mathbbm{d}b_{2}\,
   H_{gh}({\alpha}_{a}^{V},{\beta}_{g}^{V},b_{1},b_{2})\,
   {\alpha}_{s}(t_{g}^{V})\,C_{i}(t_{g}^{V})
   \nonumber \\ & &
   \big\{ {\phi}_{B1}\, \big[
   \big( {\phi}_{P}^{p}\,{\phi}_{V}^{t}-
  {\phi}_{P}^{t}\,{\phi}_{V}^{s} \big)\,(\bar{x}_{3}-x_{2})
  - {\phi}_{P}^{a}\,{\phi}_{V}^{v}\,(x_{1}+x_{2})
   \nonumber \\ & & \qquad
   + \big( {\phi}_{P}^{p}\,{\phi}_{V}^{s}-
  {\phi}_{P}^{t}\,{\phi}_{V}^{t} \big)\,
  (x_{2}+\bar{x}_{3}-2\,\bar{x}_{1})
  +4\, {\phi}_{P}^{p}\,{\phi}_{V}^{s} \big]
   \nonumber \\ & & \hspace{-0.07\textwidth}
   + {\phi}_{B2}\, \big[
   \big( {\phi}_{P}^{p}-{\phi}_{P}^{t} \big)\,
   \big( {\phi}_{V}^{t}+{\phi}_{V}^{s} \big)\, (x_{3}-x_{1})
   -2\,\big( {\phi}_{P}^{p}-{\phi}_{P}^{t} \big)\,
   {\phi}_{V}^{s} \big] \big\}_{b_{2}=b_{3}}
   \label{amp-gpv-left},
   \end{eqnarray}
   \begin{eqnarray}
  {\cal A}^{LR}_{g}(P,V) &=&  {\cal C}\,
  {\int} \mathbbm{d}x_{1}\,\mathbbm{d}x_{2}\,\mathbbm{d}x_{3}\,
         \mathbbm{d}b_{1}\,\mathbbm{d}b_{2}\,
   H_{gh}({\alpha}_{a}^{V},{\beta}_{g}^{V},b_{1},b_{2})\,
   {\alpha}_{s}(t_{g}^{V})\,C_{i}(t_{g}^{V})
   \nonumber \\ & &
   \big\{ {\phi}_{B1}\, \big[
   \big( {\phi}_{P}^{p}\,{\phi}_{V}^{t}-
  {\phi}_{P}^{t}\,{\phi}_{V}^{s} \big)\,(\bar{x}_{3}-x_{2})
  + {\phi}_{P}^{a}\,{\phi}_{V}^{v}\,(x_{1}+\bar{x}_{3})
   \nonumber \\ & &  \qquad
  - \big( {\phi}_{P}^{p}\,{\phi}_{V}^{s}-
  {\phi}_{P}^{t}\,{\phi}_{V}^{t} \big)\,
  (x_{2}+\bar{x}_{3}-2\,\bar{x}_{1})
  -4\, {\phi}_{P}^{p}\,{\phi}_{V}^{s} \big]
   \nonumber \\ & &
   + {\phi}_{B2}\, \big[
   \big( {\phi}_{P}^{p}+{\phi}_{P}^{t} \big)\,
   \big( {\phi}_{V}^{t}-{\phi}_{V}^{s} \big)\, (x_{3}-x_{1})
   \nonumber \\ & & \qquad
   -2\,{\phi}_{P}^{p}\,
    \big( {\phi}_{V}^{t}-{\phi}_{V}^{s} \big)
   - {\phi}_{P}^{a}\,{\phi}_{V}^{v}\,(x_{1}+\bar{x}_{3})
    \big] \big\}_{b_{2}=b_{3}}
   \label{amp-gpv-right},
   \end{eqnarray}
   \begin{eqnarray}
  {\cal A}^{SP}_{g}(P,V) &=& {\cal C}\,
  {\int} \mathbbm{d}x_{1}\,\mathbbm{d}x_{2}\,\mathbbm{d}x_{3}\,
         \mathbbm{d}b_{1}\,\mathbbm{d}b_{2}\,
   H_{gh}({\alpha}_{a}^{V},{\beta}_{g}^{V},b_{1},b_{2})\,
   {\alpha}_{s}(t_{g}^{V})\,C_{i}(t_{g}^{V})
   \nonumber \\ & &
   \big\{ \big( {\phi}_{B1}-{\phi}_{B2} \big)\,
   \big[ {\phi}_{P}^{a}\, \big( {\phi}_{V}^{t}-{\phi}_{V}^{s} \big)\,
   (x_{3}-x_{1})- \big( {\phi}_{P}^{p}+{\phi}_{P}^{t} \big)\,
   {\phi}_{V}^{v} \big]
   \nonumber \\ & &
   + {\phi}_{B1}\, \big[  {\phi}_{P}^{a}\,
   \big( {\phi}_{V}^{t}-{\phi}_{V}^{s} \big)
   + \big( {\phi}_{P}^{p}+{\phi}_{P}^{t} \big)\,
  {\phi}_{V}^{v}\, (x_{2}-\bar{x}_{1}) \big]
   \big\}_{b_{2}=b_{3}}
   \label{amp-gpv-sp},
   \end{eqnarray}
   \begin{eqnarray}
  {\cal A}^{LL}_{g}(V,P) &=& {\cal C}\,
  {\int} \mathbbm{d}x_{1}\,\mathbbm{d}x_{2}\,\mathbbm{d}x_{3}\,
         \mathbbm{d}b_{1}\,\mathbbm{d}b_{3}\,
   H_{gh}({\alpha}_{a}^{P},{\beta}_{g}^{P},b_{1},b_{3})\,
  {\alpha}_{s}(t_{g}^{P})\,C_{i}(t_{g}^{P})
   \nonumber \\ & &
   \big\{ {\phi}_{B1}\, \big[
   \big( {\phi}_{P}^{p}\,{\phi}_{V}^{t}-
  {\phi}_{P}^{t}\,{\phi}_{V}^{s} \big)\,(\bar{x}_{2}-x_{3})
  - {\phi}_{P}^{a}\,{\phi}_{V}^{v}\,(x_{1}+x_{3})
   \nonumber \\ & & \qquad
   - \big( {\phi}_{P}^{p}\,{\phi}_{V}^{s}-
  {\phi}_{P}^{t}\,{\phi}_{V}^{t} \big)\,
  (\bar{x}_{2}+x_{3}-2\,\bar{x}_{1})
  -4\, {\phi}_{P}^{p}\,{\phi}_{V}^{s} \big]
   \nonumber \\ & &
   + {\phi}_{B2}\, \big[
   \big( {\phi}_{P}^{p}-{\phi}_{P}^{t} \big)\,
   \big( {\phi}_{V}^{t}+{\phi}_{V}^{s} \big)\,
   (x_{3}-\bar{x}_{1})
   \nonumber \\ & & \qquad
   +2\,{\phi}_{P}^{p}\, \big( {\phi}_{V}^{t}+{\phi}_{V}^{s} \big)
   + {\phi}_{P}^{a}\,{\phi}_{V}^{v}\,(x_{1}+x_{3})
    \big] \big\}_{b_{2}=b_{3}}
   \label{amp-gvp-left},
   \end{eqnarray}
   \begin{eqnarray}
  {\cal A}^{LR}_{g}(V,P) &=&  {\cal C}\,
  {\int} \mathbbm{d}x_{1}\,\mathbbm{d}x_{2}\,\mathbbm{d}x_{3}\,
         \mathbbm{d}b_{1}\,\mathbbm{d}b_{3}\,
   H_{gh}({\alpha}_{a}^{P},{\beta}_{g}^{P},b_{1},b_{3})\,
   {\alpha}_{s}(t_{g}^{P})\,C_{i}(t_{g}^{P})
   \nonumber \\ & &
   \big\{ {\phi}_{B1}\, \big[
   \big( {\phi}_{P}^{p}\,{\phi}_{V}^{t}-
  {\phi}_{P}^{t}\,{\phi}_{V}^{s} \big)\,(\bar{x}_{2}-x_{3})
  + {\phi}_{P}^{a}\,{\phi}_{V}^{v}\,(x_{1}+\bar{x}_{2})
   \nonumber \\ & &  \qquad
  + \big( {\phi}_{P}^{p}\,{\phi}_{V}^{s}-
  {\phi}_{P}^{t}\,{\phi}_{V}^{t} \big)\,
  (\bar{x}_{2}+x_{3}-2\,\bar{x}_{1})
  +4\, {\phi}_{P}^{p}\,{\phi}_{V}^{s} \big]
   \nonumber \\ & & \hspace{-0.07\textwidth}
   + {\phi}_{B2}\, \big[
   \big( {\phi}_{P}^{p}+{\phi}_{P}^{t} \big)\,
   \big( {\phi}_{V}^{t}-{\phi}_{V}^{s} \big)\,
   (x_{3}-\bar{x}_{1})
   -2\,\big( {\phi}_{P}^{p}+{\phi}_{P}^{t} \big)\,
    {\phi}_{V}^{s} \big] \big\}_{b_{2}=b_{3}}
   \label{amp-gvp-right},
   \end{eqnarray}
   \begin{eqnarray}
  {\cal A}^{SP}_{g}(V,P) &=& {\cal C}\,
  {\int} \mathbbm{d}x_{1}\,\mathbbm{d}x_{2}\,\mathbbm{d}x_{3}\,
         \mathbbm{d}b_{1}\,\mathbbm{d}b_{3}\,
   H_{gh}({\alpha}_{a}^{P},{\beta}_{g}^{P},b_{1},b_{3})\,
   {\alpha}_{s}(t_{g}^{P})\,C_{i}(t_{g}^{P})
   \nonumber \\ & &
   \big\{ \big( {\phi}_{B1}-{\phi}_{B2} \big)\,
   \big[ {\phi}_{P}^{a}\, \big( {\phi}_{V}^{t}+{\phi}_{V}^{s} \big)\,
   (\bar{x}_{1}-x_{3})- \big( {\phi}_{P}^{p}-{\phi}_{P}^{t} \big)\,
   {\phi}_{V}^{v} \big]
   \nonumber \\ & &
   + {\phi}_{B1}\, \big[  {\phi}_{P}^{a}\,
   \big( {\phi}_{V}^{t}+{\phi}_{V}^{s} \big)
   + \big( {\phi}_{P}^{p}-{\phi}_{P}^{t} \big)\,
  {\phi}_{V}^{v}\, (x_{1}-x_{2}) \big]
   \big\}_{b_{2}=b_{3}}
   \label{amp-gvp-sp},
   \end{eqnarray}
   \begin{eqnarray}
  {\cal A}^{LL}_{h}(P,V) &=&  {\cal C}\,
  {\int} \mathbbm{d}x_{1}\,\mathbbm{d}x_{2}\,\mathbbm{d}x_{3}\,
         \mathbbm{d}b_{1}\,\mathbbm{d}b_{2}\,
   H_{gh}({\alpha}_{a}^{V},{\beta}_{h}^{V},b_{1},b_{2})\,
   {\alpha}_{s}(t_{h}^{V})\,C_{i}(t_{h}^{V})
   \nonumber \\ & &
   \big\{ {\phi}_{B1}\, \big[
   \big( {\phi}_{P}^{p}\,{\phi}_{V}^{t}-
  {\phi}_{P}^{t}\,{\phi}_{V}^{s} \big)\,(\bar{x}_{3}-x_{2})
  +{\phi}_{P}^{a}\,{\phi}_{V}^{v}\,(\bar{x}_{3}-x_{1})
   \nonumber \\ & & \qquad
   - \big( {\phi}_{P}^{p}\,{\phi}_{V}^{s}-
  {\phi}_{P}^{t}\,{\phi}_{V}^{t} \big)\,
  (x_{2}+\bar{x}_{3}-2\,x_{1}) \big]
   \nonumber \\ & &
   + {\phi}_{B2}\, \big[ {\phi}_{P}^{a}\,{\phi}_{V}^{v}
   + \big( {\phi}_{P}^{p}+{\phi}_{P}^{t} \big)\,
   \big( {\phi}_{V}^{t}-{\phi}_{V}^{s} \big)\,
   \big]\, (x_{1}-\bar{x}_{3}) \big\}_{b_{2}=b_{3}}
   \label{amp-hpv-left},
   \end{eqnarray}
   \begin{eqnarray}
  {\cal A}^{LR}_{h}(P,V) &=& {\cal C}\,
  {\int} \mathbbm{d}x_{1}\,\mathbbm{d}x_{2}\,\mathbbm{d}x_{3}\,
         \mathbbm{d}b_{1}\,\mathbbm{d}b_{2}\,
   H_{gh}({\alpha}_{a}^{V},{\beta}_{h}^{V},b_{1},b_{2})\,
   {\alpha}_{s}(t_{h}^{V})\,C_{i}(t_{h}^{V})
   \nonumber \\ & &
   \big\{ {\phi}_{B1}\, \big[
   \big( {\phi}_{P}^{p}\,{\phi}_{V}^{t}-
  {\phi}_{P}^{t}\,{\phi}_{V}^{s} \big)\,(\bar{x}_{3}-x_{2})
  +{\phi}_{P}^{a}\,{\phi}_{V}^{v}\,(x_{1}-x_{2})
   \nonumber \\ & & \qquad
   + \big( {\phi}_{P}^{p}\,{\phi}_{V}^{s}-
  {\phi}_{P}^{t}\,{\phi}_{V}^{t} \big)\,
  (x_{2}+\bar{x}_{3}-2\,x_{1}) \big]
   \nonumber \\ & &
   + {\phi}_{B2}\, \big( {\phi}_{P}^{p}-{\phi}_{P}^{t} \big)\,
   \big( {\phi}_{V}^{t}+{\phi}_{V}^{s} \big)\,
   (x_{1}-\bar{x}_{3}) \big\}_{b_{2}=b_{3}}
   \label{amp-hpv-right},
   \end{eqnarray}
   \begin{eqnarray}
  {\cal A}^{SP}_{h}(P,V) &=& {\cal C}\,
  {\int} \mathbbm{d}x_{1}\,\mathbbm{d}x_{2}\,\mathbbm{d}x_{3}\,
         \mathbbm{d}b_{1}\,\mathbbm{d}b_{2}\,
   H_{gh}({\alpha}_{a}^{V},{\beta}_{h}^{V},b_{1},b_{2})\,
   {\alpha}_{s}(t_{h}^{V})\,C_{i}(t_{h}^{V})
   \nonumber \\ & &
   \big\{ \big( {\phi}_{B1}-{\phi}_{B2} \big)\,
   \big( {\phi}_{P}^{p}+{\phi}_{P}^{t} \big)\,
   {\phi}_{V}^{v}\, (x_{1}-x_{2})
   \nonumber \\ & &
   + {\phi}_{B1}\, {\phi}_{P}^{a}\,
     \big( {\phi}_{V}^{t}-{\phi}_{V}^{s} \big)
     (\bar{x}_{3}-x_{1}) \big\}_{b_{2}=b_{3}}
   \label{amp-hpv-sp},
   \end{eqnarray}
   \begin{eqnarray}
  {\cal A}^{LL}_{h}(V,P) &=&  {\cal C}\,
  {\int} \mathbbm{d}x_{1}\,\mathbbm{d}x_{2}\,\mathbbm{d}x_{3}\,
         \mathbbm{d}b_{1}\,\mathbbm{d}b_{3}\,
   H_{gh}({\alpha}_{a}^{P},{\beta}_{h}^{P},b_{1},b_{3})\,
  {\alpha}_{s}(t_{h}^{P})\,C_{i}(t_{h}^{P})
   \nonumber \\ & &
   \big\{ {\phi}_{B1}\, \big[
   \big( {\phi}_{P}^{p}\,{\phi}_{V}^{t}-
  {\phi}_{P}^{t}\,{\phi}_{V}^{s} \big)\,(\bar{x}_{2}-x_{3})
  +{\phi}_{P}^{a}\,{\phi}_{V}^{v}\,(\bar{x}_{2}-x_{1})
   \nonumber \\ & & \qquad
   + \big( {\phi}_{P}^{p}\,{\phi}_{V}^{s}-
  {\phi}_{P}^{t}\,{\phi}_{V}^{t} \big)\,
  (\bar{x}_{2}+x_{3}-2\,x_{1}) \big]
   \nonumber \\ & &
   + {\phi}_{B2}\, \big( {\phi}_{P}^{p}+{\phi}_{P}^{t} \big)\,
   \big( {\phi}_{V}^{t}-{\phi}_{V}^{s} \big)\, (x_{3}-x_{1})
   \big\}_{b_{2}=b_{3}}
   \label{amp-hvp-left},
   \end{eqnarray}
   \begin{eqnarray}
  {\cal A}^{LR}_{h}(V,P) &=& {\cal C}\,
  {\int} \mathbbm{d}x_{1}\,\mathbbm{d}x_{2}\,\mathbbm{d}x_{3}\,
         \mathbbm{d}b_{1}\,\mathbbm{d}b_{3}\,
   H_{gh}({\alpha}_{a}^{P},{\beta}_{h}^{P},b_{1},b_{3})\,
  {\alpha}_{s}(t_{h}^{P})\,C_{i}(t_{h}^{P})
   \nonumber \\ & &
   \big\{ {\phi}_{B1}\, \big[
   \big( {\phi}_{P}^{p}\,{\phi}_{V}^{t}-
  {\phi}_{P}^{t}\,{\phi}_{V}^{s} \big)\,(\bar{x}_{2}-x_{3})
  +{\phi}_{P}^{a}\,{\phi}_{V}^{v}\,(x_{1}-x_{3})
  \nonumber \\ & & \qquad
   - \big( {\phi}_{P}^{p}\,{\phi}_{V}^{s}-
  {\phi}_{P}^{t}\,{\phi}_{V}^{t} \big)\,
  (\bar{x}_{2}+x_{3}-2\,x_{1}) \big]
   \nonumber \\ & &
   + {\phi}_{B2}\, \big[ {\phi}_{P}^{a}\,{\phi}_{V}^{v}
   + \big( {\phi}_{P}^{p}-{\phi}_{P}^{t} \big)\,
   \big( {\phi}_{V}^{t}+{\phi}_{V}^{s} \big)\,
   \big]\, (x_{3}-x_{1}) \big\}_{b_{2}=b_{3}}
   \label{amp-hvp-right},
   \end{eqnarray}
   \begin{eqnarray}
  {\cal A}^{SP}_{h}(V,P) &=& {\cal C}\,
  {\int} \mathbbm{d}x_{1}\,\mathbbm{d}x_{2}\,\mathbbm{d}x_{3}\,
         \mathbbm{d}b_{1}\,\mathbbm{d}b_{3}\,
   H_{gh}({\alpha}_{a}^{P},{\beta}_{h}^{P},b_{1},b_{3})\,
  {\alpha}_{s}(t_{h}^{P})\,C_{i}(t_{h}^{P})
   \nonumber \\ & &
   \big\{ \big( {\phi}_{B1}-{\phi}_{B2} \big)\,
   \big( {\phi}_{P}^{p}-{\phi}_{P}^{t} \big)\,
   {\phi}_{V}^{v}\, (x_{1}-\bar{x}_{2})
   \nonumber \\ & &
   + {\phi}_{B1}\, {\phi}_{P}^{a}\,
      \big( {\phi}_{V}^{t}+{\phi}_{V}^{s} \big)
     (x_{3}-x_{1}) \big\}_{b_{2}=b_{3}}
   \label{amp-hvp-sp},
   \end{eqnarray}
   \begin{equation}
  {\alpha}_{g}^{V}\, =\, m_{B}^{2}\,x_{1}\,x_{3}
   \label{gluon-epv},
   \end{equation}
   \begin{equation}
  {\alpha}_{g}^{P}\, =\, m_{B}^{2}\,x_{1}\,x_{2}
   \label{gluon-evp},
   \end{equation}
   \begin{equation}
  {\alpha}_{a}^{V}\, =\, m_{B}^{2}\,x_{2}\,\bar{x}_{3}
   \label{gluon-apv},
   \end{equation}
   \begin{equation}
  {\alpha}_{a}^{P}\, =\, m_{B}^{2}\,\bar{x}_{2}\,x_{3}
   \label{gluon-avp},
   \end{equation}
   \begin{equation}
  {\beta}_{a}^{V}\, =\, m_{B}^{2}\,x_{3}
   \label{quark-apv},
   \end{equation}
   \begin{equation}
  {\beta}_{a}^{P}\, =\, m_{B}^{2}\,x_{2}
   \label{quark-avp},
   \end{equation}
   \begin{equation}
  {\beta}_{b}^{V}\, =\,
  {\beta}_{b}^{P}\, =\, m_{B}^{2}\,x_{1}
   \label{quark-bpv},
   \end{equation}
   \begin{equation}
  {\beta}_{c}^{V}\, =\, m_{B}^{2}\,x_{3}\,(x_{1}-\bar{x}_{2})
   \label{quark-cpv},
   \end{equation}
   \begin{equation}
  {\beta}_{c}^{P}\, =\, m_{B}^{2}\,x_{2}\,(x_{1}-\bar{x}_{3})
   \label{quark-cvp},
   \end{equation}
   \begin{equation}
  {\beta}_{d}^{V}\, =\, m_{B}^{2}\,x_{3}\,(x_{1}-x_{2})
   \label{quark-dpv},
   \end{equation}
   \begin{equation}
  {\beta}_{d}^{P}\, =\, m_{B}^{2}\,x_{2}\,(x_{1}-x_{3})
   \label{quark-dvp},
   \end{equation}
   \begin{equation}
  {\beta}_{e}^{V}\, =\, m_{B}^{2}\,\bar{x}_{3}
   \label{quark-epv},
   \end{equation}
   \begin{equation}
  {\beta}_{e}^{P}\, =\, m_{B}^{2}\,\bar{x}_{2}
   \label{quark-evp},
   \end{equation}
   \begin{equation}
  {\beta}_{f}^{V}\, =\, m_{B}^{2}\, x_{2}
   \label{quark-fpv},
   \end{equation}
   \begin{equation}
  {\beta}_{f}^{P}\, =\, m_{B}^{2}\, x_{3}
   \label{quark-fvp},
   \end{equation}
   \begin{equation}
  {\beta}_{g}^{V} \, =\, {\alpha}_{a}^{V}
   -m_{B}^{2}\,\bar{x}_{1}\,(x_{2}+\bar{x}_{3})
   \label{quark-gpv},
   \end{equation}
   \begin{equation}
  {\beta}_{g}^{P} \, =\, {\alpha}_{a}^{P}
  -m_{B}^{2}\,\bar{x}_{1}\,(\bar{x}_{2}+x_{3})
   \label{quark-gvp},
   \end{equation}
   \begin{equation}
  {\beta}_{h}^{V} \, =\, {\alpha}_{a}^{V}
  -m_{B}^{2}\,x_{1}\,(x_{2}+\bar{x}_{3})
   \label{quark-hpv},
   \end{equation}
   \begin{equation}
  {\beta}_{h}^{P} \, =\, {\alpha}_{a}^{P}
  -m_{B}^{2}\,x_{1}\,(\bar{x}_{2}+x_{3})
   \label{quark-hvp},
   \end{equation}
   \begin{equation}
   t_{a,b}^{V} \, =\, {\max}({\alpha}_{g}^{V},{\beta}_{a,b}^{V},b_{1},b_{3})
   \label{scale-apv},
   \end{equation}
   \begin{equation}
   t_{a,b}^{P} \, =\, {\max}({\alpha}_{g}^{P},{\beta}_{a,b}^{P},b_{1},b_{2})
   \label{scale-avp},
   \end{equation}
   \begin{equation}
   t_{c,d}^{i} \, =\, {\max}({\alpha}_{g}^{i},{\beta}_{c,d}^{i},b_{2},b_{3})
   \label{scale-cpv},
   \end{equation}
   \begin{equation}
   t_{e,f}^{i} \, =\, {\max}({\alpha}_{a}^{i},{\beta}_{e,f}^{i},b_{2},b_{3})
   \label{scale-evp},
   \end{equation}
   \begin{equation}
   t_{g,h}^{i} \, =\, {\max}({\alpha}_{a}^{i},{\beta}_{g,h}^{i},b_{1},b_{2})
   \label{scale-gpv},
   \end{equation}
  \begin{equation}
  H_{ab}({\alpha},{\beta},b_{i},b_{j}) \, =\,
  b_{i}\,b_{j}\, K_{0}\big(b_{i}\,\sqrt{\alpha}\big)\,
  \big\{ {\theta}\big(b_{i}-b_{j}\big)\,
  K_{0}\big(b_{i}\,\sqrt{\beta}\big)\,
  I_{0}\big(b_{j}\,\sqrt{\beta}\big)
  +\big(b_{i}\,{\leftrightarrow}\,b_{j}\big) \big\}
  \label{amp-ab-denominator},
  \end{equation}
  \begin{eqnarray}
   N_{c}\, H_{cd}({\alpha},{\beta},b_{1},b_{i}) &=&
   b_{1}\,b_{i}\, \big\{ {\theta}(b_{1}-b_{2})\,
   K_{0}\big(b_{1}\sqrt{{\alpha}}\big)\,
   I_{0}\big(b_{i}\sqrt{{\alpha}}\big)
   +\big(b_{1}\,{\leftrightarrow}\,b_{i}\big)\big\}
   \nonumber \\ & & \hspace{-0.125\textwidth}
   \big\{ {\theta}({\beta})\, K_{0}\big(b_{i}\sqrt{\beta}\big)
   +i\,\frac{\pi}{2}\,{\theta}(-{\beta})\,\big[
   J_{0}\big(b_{i}\sqrt{-{\beta}}\big)+i\,
   Y_{0}\big(b_{i}\sqrt{-{\beta}}\big) \big] \big\}
   \label{amp-cd-denominator},
   \end{eqnarray}
   \begin{eqnarray} & &
   H_{ef}({\alpha},{\beta},b_{i},b_{j})\, =\,
   -\frac{{\pi}^{2}}{4}\,b_{i}\,b_{j}\,
    \big\{ J_{0}\big(b_{i}\,\sqrt{\alpha} \big)
       +i\,Y_{0}\big(b_{i}\,\sqrt{\alpha} \big) \big\}
    \nonumber \\ & &
    \big\{ {\theta}\big(b_{i}-b_{j}\big)\, \big[
      J_{0} \big(b_{i}\,\sqrt{\beta} \big)
  +i\,Y_{0} \big(b_{i}\,\sqrt{\beta} \big) \big]\,
      J_{0} \big(b_{j}\,\sqrt{\beta} \big)
    +\big(b_{i}\,{\leftrightarrow}\,b_{j}\big) \big\}
    \label{amp-ef-denominator},
    \end{eqnarray}
   \begin{eqnarray} & &
   N_{c}\,H_{gh}({\alpha},{\beta},b_{1},b_{i})
   \nonumber \\ &=&
   b_{1}\, b_{i}\, \big\{ \frac{i\,{\pi}}{2}\,
   {\theta} \big( {\beta} \big)\, \big[
     J_{0} \big( b_{1}\,\sqrt{ {\beta} }\big)
   +i\,Y_{0} \big( b_{1}\,\sqrt{ {\beta} }\big) \big]
   + {\theta} \big( -{\beta} \big)\,
    K_{0} \big( b_{1}\,\sqrt{-{\beta} }\big) \big\}
    \nonumber \\ & &
    \frac{i\,{\pi}}{2}\, \big\{ {\theta}\big(b_{1}-b_{i}\big)\,
    \big[ J_{0} \big( b_{1}\,\sqrt{\alpha} \big)
    +i\, Y_{0} \big( b_{1}\,\sqrt{\alpha} \big) \big]\,
    J_{0} \big( b_{i}\,\sqrt{\alpha} \big)
    +\big(b_{1}\,{\leftrightarrow}\,b_{i}\big) \big\}
    \label{amp-gh-denominator}.
    \end{eqnarray}

   \end{appendix}
   \end{widetext}


\begin{thebibliography}{99}
  \bibitem{1808.10567}
  \href{https://doi.org/10.1093/ptep/ptaa008}
       {K. Kou {\em et al.},
        Prog. Theor. Exp. Phys. 2019, 123C01 (2019);}
  \href{https://doi.org/10.1093/ptep/ptz106}
       {2020, 029201 (2020)(E).}
  \bibitem{1808.08865}
  \href{https://arxiv.org/abs/1808.08865}
       {I. Bediaga {\em et al.} (LHCb Collaboration),
        arXiv:1808.08865.}
  \bibitem{1811.10545}
  \href{https://arxiv.org/abs/1811.10545}
       {J. Costa, {\em et al.} IHEP-CEPC-DR-2018-02,
        arXiv:1811.10545.}
  \bibitem{epjc.79.474}
  \href{https://doi.org/10.1140/epjc/s10052-019-6904-3}
       {A. Abada, {\em et al.},
        Eur. Phys. J. C 79, 474 (2019).}
  \bibitem{pdg2020}
  \href{https://doi.org/10.1093/ptep/ptaa104}
       {P. Zyla {\em et al.} (Particle Data Group),
        Prog. Theor. Exp. Phys. 2020, 083C01 (2020).}
  \bibitem{plb.576.29}
  \href{https://doi.org/10.1016/j.physletb.2003.09.070}
       {J. Abdallah {\em et al.} (DELPHI Collaboration),
        Phys. Lett. B 576, 29 (2003).}

  \bibitem{PhysRevLett.10.531}
  \href{https://doi.org/10.1103/PhysRevLett.10.531}
       {N. Cabibbo, Phys. Rev. Lett. 10, 531 (1963).}
  \bibitem{PTP.49.652}
  \href{https://doi.org/10.1143/PTP.49.652}
       {M. Kobayashi, T. Maskawa,
        Prog. Theor. Phys. 49, 652 (1973).}
  \bibitem{PhysRevD.22.2157}
  \href{https://doi.org/10.1103/PhysRevD.22.2157}
       {G. Lepage, S. Brodsky,
        Phys. Rev. D 22, 2157 (1980).}
  \bibitem{PhysRevLett.74.4388}
  \href{https://doi.org/10.1103/PhysRevLett.74.4388}
       {H. Li, H. Yu, Phys. Rev. Lett. 74, 4388 (1995).}
  \bibitem{plb.348.597}
  \href{https://doi.org/10.1016/0370-2693(95)00174-J}
       {H. Li, Phys. Lett. B 348, 597 (1995).}
  \bibitem{PhysRevD.52.3958}
  \href{https://doi.org/10.1103/PhysRevD.52.3958}
       {H. Li, Phys. Rev. D 52, 3958 (1995).}
  \bibitem{PhysRevD.63.074006}
  \href{https://doi.org/10.1103/PhysRevD.63.074006}
       {Y. Keum, H. Li,
        Phys. Rev. D 63, 074006 (2001).}
  \bibitem{PhysRevD.63.054008}
  \href{https://doi.org/10.1103/PhysRevD.63.054008}
       {Y. Keum, H. Li, A. Sanda,
        Phys. Rev. D 63, 054008 (2001).}
  \bibitem{PhysRevD.63.074009}
  \href{https://doi.org/10.1103/PhysRevD.63.074009}
       {C. L\"{u}, K. Ukai, M. Yang,
        Phys. Rev. D 63, 074009 (2001).}
  \bibitem{plb.555.197}
  \href{https://doi.org/10.1016/S0370-2693(03)00049-2}
       {H. Li, K. Ukai,
        Phys. Lett. B 555, 197 (2003).}
  \bibitem{PhysRevLett.83.1914}
  \href{https://doi.org/10.1103/PhysRevLett.83.1914}
       {M. Beneke, G. Buchalla, M. Neubert, C. Sachrajda,
        Phys. Rev. Lett. 83, 1914 (1999).}
  \bibitem{npb.591.313}
  \href{https://doi.org/10.1016/S0550-3213(00)00559-9}
       {M. Beneke, G. Buchalla, M. Neubert, C. Sachrajda,
        Nucl. Phys. B 591, 313 (2000).}
  \bibitem{npb.606.245}
  \href{https://doi.org/10.1016/S0550-3213(01)00251-6}
       {M. Beneke, G. Buchalla, M. Neubert, C. Sachrajda,
        Nucl. Phys. B 606, 245 (2001).}
  \bibitem{plb.488.46}
  \href{https://doi.org/10.1016/S0370-2693(00)00854-6}
       {D. Du, D. Yang, G. Zhu,
        Phys. Lett. B 488, 46 (2000).}
  \bibitem{plb.509.263}
  \href{https://doi.org/10.1016/S0370-2693(01)00398-7}
       {D. Du, D. Yang, G. Zhu,
        Phys. Lett. B 509, 263 (2001).}
  \bibitem{PhysRevD.64.014036}
  \href{https://doi.org/10.1103/PhysRevD.64.014036}
       {D. Du, D. Yang, G. Zhu,
        Phys. Rev. D 64, 014036 (2001).}
  \bibitem{epjc.36.365}
  \href{https://doi.org/10.1140/epjc/s2004-01945-7}
       {Z. Song, C. Meng, K. Chao,
        Eur. Phys. J. C 36, 365 (2004).}
  \bibitem{PhysRevD.69.054009}
  \href{https://doi.org/10.1103/PhysRevD.69.054009}
       {Z. Song, C. Meng, Y. Gao, K. Chao,
        Phys. Rev. D 69, 054009 (2004).}
  \bibitem{npb.774.64}
  \href{https://doi.org/10.1016/j.nuclphysb.2007.03.020}
       {M. Beneke, J. Rohrer, D. Yang,
        Nucl. Phys. B 774, 64 (2007).}
  \bibitem{PhysRevD.77.074013}
  \href{https://doi.org/10.1103/PhysRevD.77.074013}
       {J. Sun, G. Xue, Y. Yang {\em et al.},
        Phys. Rev. D 77, 074013 (2008).}
  \bibitem{PhysRevD.63.014006}
  \href{https://doi.org/10.1103/PhysRevD.63.014006}
       {C. Bauer, S. Fleming, M. Luke,
        Phys. Rev. D 63, 014006 (2000).}
  \bibitem{PhysRevD.63.114020}
  \href{https://doi.org/10.1103/PhysRevD.63.114020}
       {C. Bauer {\em et al.},
        Phys. Rev. D 63, 114020 (2001).}
  \bibitem{plb.516.134}
  \href{https://doi.org/10.1016/S0370-2693(01)00902-9}
       {C. Bauer, I. Stewart,
        Phys. Lett. B 516, 134 (2001).}
  \bibitem{PhysRevD.65.054022}
  \href{https://doi.org/10.1103/PhysRevD.65.054022}
       {C. Bauer, D. Pirjol, I. Stewart,
        Phys. Rev. D 65, 054022 (2002).}
  \bibitem{PhysRevD.66.014017}
  \href{https://doi.org/10.1103/PhysRevD.66.014017}
       {C. Bauer {\em et al.},
        Phys. Rev. D 66, 014017 (2002).}
  \bibitem{npb.643.431}
  \href{https://doi.org/10.1016/S0550-3213(02)00687-9}
       {M. Beneke {\em et al.},
        Nucl. Phys. B 643, 431 (2002).}
  \bibitem{plb.553.267}
  \href{https://doi.org/10.1016/S0370-2693(02)03204-5}
       {M. Beneke, T. Feldmann,
        Phys. Lett. B 553, 267 (2003).}
  \bibitem{npb.685.249}
  \href{https://doi.org/10.1016/j.nuclphysb.2004.02.033}
       {M. Beneke, T. Feldmann,
        Nucl. Phys. B 685, 249 (2004).}
  \bibitem{npb.794.154}
  \href{https://doi.org/10.1016/j.nuclphysb.2007.10.028}
       {V. Pilipp, Nucl. Phys. B 794, 154 (2008).}
  \bibitem{npb.832.109}
  \href{https://doi.org/10.1016/j.nuclphysb.2010.02.002}
       {M. Beneke, T. Huber, X. Li ,
        Nucl. Phys. B 832,  109 (2010).}
  \bibitem{plb.750.348}
  \href{https://doi.org/10.1016/j.physletb.2015.09.037}
       {G. Bell, M. Beneke, T. Huber, Tobias, X. Li,
        Phys. Lett. B 750, 348 (2015).}
  \bibitem{jhep.2016.09.112}
  \href{https://doi.org/10.1007/JHEP09(2016)112}
       {T. Huber,  S. Kr\"{a}nkl, X. Li,
        JHEP 09, 112 (2016).}
  \bibitem{jhep.2020.04.055}
  \href{https://doi.org/10.1007/JHEP04(2020)055}
       {G. Bell, M. Beneke, T. Hubera, X. Li,
        JHEP 04, 055 (2020).}
  \bibitem{PhysRevD.71.034018}
  \href{https://doi.org/10.1103/PhysRevD.71.034018}
       {T. Huang, X. Wu,
        Phys. Rev. D 71, 034018 (2005).}
  \bibitem{PhysRevD.103.056006} 
  \href{https://doi.org/10.1103/PhysRevD.103.056006}
       {Y. Yang, L. Lang, X. Zhao {\em et al.},
        Phys. Rev. D 103, 056006 (2021).}
  \bibitem{PhysRevD.74.014027}
  \href{https://doi.org/10.1103/PhysRevD.74.014027}
       {T. Kurimoto,
        Phys. Rev. D 74, 014027 (2006).}
  \bibitem{epjc.28.515}
  \href{https://doi.org/10.1140/epjc/s2003-01199-y}
       {C. L\"{u}, M. Yang,
        Eur. Phys. J. C 28, 515 (2003).}
  \bibitem{npb.625.239}
  \href{https://doi.org/10.1016/S0550-3213(02)00017-2}
       {S. Descotes-Genon, C. Sachrajda,
        Nucl. Phys. B 625, 239 (2002).}
  \bibitem{npb.642.263}
  \href{https://doi.org/10.1016/S0550-3213(02)00623-5}
       {Z. Wei, M. Yang,
        Nucl. Phys. B 642, 263 (2002).}
  \bibitem{PhysRevD.89.094004}
  \href{https://doi.org/10.1103/PhysRevD.89.094004}
       {S. Cheng, Y. Fan, X. Yu {\em et al.},
        Phys. Rev. D 89, 094004 (2012).}
  \bibitem{front.phys.16.24201}
  \href{https://doi.org/10.1007/s11467-020-1036-7}
       {S. Cheng, Z. Xiao,
        Front. Phys. 16, 24201 (2021).}
  \bibitem{PhysRevD.64.112002} 
  \href{https://doi.org/10.1103/PhysRevD.64.112002}
       {C. Chen, Y. Keum, H. Li,
        Phys. Rev. D 64, 112002 (2001).}
  \bibitem{epjc.23.275} 
  \href{https://doi.org/10.1007/s100520100878}
       {C. L\"{u}, M. Yang,
        Eur. Phys. J. C 23, 275 (2002).}
  \bibitem{epjc.72.1923} 
  \href{https://doi.org/10.1140/epjc/s10052-012-1923-3}
       {R. Zhou, X. Gao, C. L\"{u},
        Eur. Phys. J. C 72, 1923 (2012).}
  \bibitem{PhysRevD.90.074018} 
  \href{https://doi.org/10.1103/PhysRevD.90.074018}
       {H. Li, S. Mishima,
        Phys. Rev. D 90, 074018 (2014).}
  \bibitem{PhysRevD.75.014019} 
  \href{https://doi.org/10.1103/PhysRevD.75.014019}
       {L. Guo, Q. Xu, Z. Xiao,
        Phys. Rev. D 75, 014019 (2007).}
  \bibitem{epjc.59.49} 
  \href{https://doi.org/10.1140/epjc/s10052-008-0805-1}
       {Z. Zhang, Z. Xiao,
        Eur. Phys. J. C 59, 49 (2009).}
  \bibitem{PhysRevD.104.016025} 
  \href{https://doi.org/10.1103/PhysRevD.104.016025}
       {J. Hua, H. Li, C. L\"{u} {\em et al.},
        Phys. Rev. D 104, 016025 (2021).}
  \bibitem{PhysRevD.74.094020} 
  \href{https://doi.org/10.1103/PhysRevD.74.094020}
       {H. Li, S. Mishima,
        Phys. Rev. D 74, 094020 (2006).}
  \bibitem{RevModPhys.68.1125}
  \href{https://doi.org/10.1103/RevModPhys.68.1125}
       {G. Buchalla, A. Buras, M. Lautenbacher,
        Rev. Mod. Phys. 68, 1125, (1996).}
  \bibitem{PhysRevD.55.272}
  \href{https://doi.org/10.1103/PhysRevD.55.272}
       {A. Grozin, M. Neubert,
        Phys. Rev. D 55, 272 (1997).}
  \bibitem{npb.592.3}
  \href{https://doi.org/10.1016/S0550-3213(00)00585-X}
       {M. Beneke, Th. Feldmann,
        Nucl. Phys. B 592, 3 (2001).}
  \bibitem{jhep.1999.01.010}
  \href{https://doi.org/10.1088/1126-6708/1999/01/010}
       {P. Ball,
        JHEP 01, 010 (1999).}
  \bibitem{PhysRevD.65.014007}
  \href{https://doi.org/10.1103/PhysRevD.65.014007}
       {T. Kurimoto, H. Li, A. Sanda,
        Phys. Rev. D 65, 014007 (2001).}
  \bibitem{jhep.2006.05.004}
  \href{https://doi.org/10.1088/1126-6708/2006/05/004}
       {P. Ball, V. Braun, A. Lenz,
        JHEP 05, 004 (2006).}
  \bibitem{jhep.2007.03.069}
  \href{https://doi.org/10.1088/1126-6708/2007/03/069}
       {P. Ball, G. Jones,
        JHEP 03, 069 (2007).}
  \bibitem{PhysRevD.59.014005}
  \href{https://doi.org/10.1103/PhysRevD.59.014005}
       {A. Ali, K. Kramer, C. L\"{u},
        Phys. Rev. D 59, 014005 (1998).}
  \bibitem{PhysRevD.65.094025}
  \href{https://doi.org/10.1103/PhysRevD.65.094025}
       {D. Du, H. Gong, J. Sun {\em et al.},
        Phys. Rev. D 65, 094025 (2002).}

  \end{thebibliography}
   \end{document}